\documentclass[aca4,a4paper]{jacow}
\usepackage{pdfpages,multirow,ragged2e} %
\usepackage{hyperref}

\makeatletter%
\ifboolexpr{bool{xetex}}
{\renewcommand{\Gin@extensions}{.pdf,%
		.png,.jpg,.bmp,.pict,.tif,.psd,.mac,.sga,.tga,.gif,%
		.eps,.ps,%
}}{}
\makeatother

\ifboolexpr{bool{xetex} or bool{luatex}} 
{}                                      
{\usepackage[utf8]{inputenc}}           

\usepackage[USenglish]{babel}
\usepackage{siunitx}
\usepackage{wasysym}

\usepackage{pdfpages,multirow,ragged2e}
\usepackage{graphics} 
\usepackage{graphicx}   
\ifboolexpr{bool{jacowbiblatex}}%
{%
	\addbibresource{jacow-test.bib}
	\addbibresource{biblatex-examples.bib}
}{}
\makeatletter
\newcommand{\setlabel}[1]{\edef\@currentlabel{#1}\label}
\makeatother

\begin{document}
	
\setlength{\titleblockheight}{33mm}

\title{PREDOMINANTLY ELECTRIC STORAGE RING WITH NUCLEAR SPIN CONTROL CAPABILITY}
	
\author{Richard Talman\thanks{richard.talman@cornell.edu}, Laboratory for Elementary-Particle Physics, Cornell University, Ithaca, NY, USA\\
and John Talman, UAL Consultants, Ithaca, NY, USA}
	
\maketitle

\tableofcontents

\abstract
A predominantly electric E\&m storage ring, with weak superimposed magnetic bending, is shown to be capable of storing two different particle type bunches, such as helion (h) and deuteron (d), or h and electron ($e^-$), co-traveling with different velocities on the same central orbit. Rear-end collisions occurring periodically in a full acceptance particle detector/polarimeter, allow the (previously inaccessible) direct measurement of the spin dependence of nuclear transmutation for center of mass (CM) kinetic energies (KE) ranging from hundreds of keV up toward pion production thresholds.  With the nuclear process occurring in a semi-relativistic moving frame, all initial and final state particles have convenient laboratory frame KEs in the tens to hundreds of MeV.
The rear-end collisions occur as faster stored bunches pass through slower bunches. An inexpensive facility capable of meeting these requirements is described, with several nuclear channels as examples.  Especially noteworthy are the $e^{+/-}$-induced weak interaction triton (t) $\beta$-decay processes, t + $e^+ \rightarrow$ h + $\nu$ and  h + $e^- \rightarrow$ t + $\nu$.  Experimental capability of measurement of the spin dependence of the induced triton case is emphasized.  For cosmological nuclear physics, the experimental improvement will be produced by the storage ring's capability to investigate the spin dependence of nuclear transmutation processes at reduced kinetic energies compared to what can be obtained with fixed target geometry.

\section{Introduction}\label{sec:Introduction}

The proton is the only stable elementary particle for which no experimentally testable fundamental theory predictions exist!  
Direct $p,p$ and $p,n$ coupling is too strong for their interactions to be calculable using relativistic quantum field theory. 
Next-best: the meson-nucleon perturbation parameter (roughly 1/5) is small enough for standard model theory, 
with its quarks and gluons, to be based, numerically, 
predominantly on $\pi$ meson, nucleon scattering.  This \emph{finesses} complications associated with finite size, internal structure, 
and compound nucleus formation.

These issues should be addressed experimentally, but this is seriously impeded by the absence of nuclear physics measurement, 
especially concerning spin dependence, for particle kinetic energies (KE) in the range from \SI{100}{keV} to several MeV, comparable with 
Coulomb potential barrier heights.  Even though multi-keV scale energies are easily produced in vacuum, until now spin measurement 
in this region  has been prevented by space charge and negligibly short particle ranges in matter.  In this energy range, negligible 
compared to all nucleon rest masses, the lab frame and the CM frame coincide. 

To study spin dependence in nuclear scattering, one must cause the scattering to occur in what is (at least a weakly relativistic) 
moving frame of reference.  This is possible using ``rear-end'' collisions in a predominantly electric E\&m storage ring.  Superimposed 
weak magnetic bending makes it possible for two beams of different velocity to circulate in the same direction, at the same time, 
in the same storage ring.  ``Rear-end'' collisions occurring during the passage of faster bunches through slower bunches can be used 
to study spin dependence on nucleon-nucleon collisions in a moving coordinate frame.  

Such ``rear-end'' collisions allow the CM KEs to 
be in the several \SI{100}{keV} range, while all incident and scattered particles have convenient laboratory KEs, two orders of magnitude 
higher, in the tens of MeV range.  Multi-MeV scale incident beams can then be established in pure spin states and the momenta and 
polarizations of all final state particles can be measured with high analyzing power and high efficiency.  In this way the storage 
ring satisfies the condition that all nuclear collisions take place in a coordinate frame moving at convenient semi-relativistic 
speed in the laboratory, with CM KEs comparable with Coulomb barrier heights.

\subsection{Philosophical digression\label{sec:Philosophical}}\mbox{}

The term \emph{finesse'} was used in a metaphorical sense in the introduction, to suggest a strategy of proceeding without the need for fundamental
understanding.  In other words, having looked at nuclei from both low energy and high energy, we still ``do not understand'' nuclei.  
This is in contrast to atomic physics, for which quantum mechanics, by now axiomatized, explains everything, to 
every ones satisfaction, from all sides.  

One conjectures that certain historical figures from the past, say Rutherford, Bohr, and Einstein, for example, if alive today, might 
be disappointed by the progress that has been made in our understanding of nuclear physics.  

Once convinced of the existence of pions and muons, Rutherford would have had no trouble understanding the need for incorporating some 
probabilistic description into his classical mechanics.  In a two body collision, once two nucleons have ``captured each other'' into a 
compound nucleus, temporarily converting their kinetic energy into rotational energy, it becomes quite complicated to redistribute the 
rotational energy back into a few final state particles.  

With nuclear sizes small and the number of particles in a beam bunch large, say $10^8$ particles, indistinguishable from each other,
there is no detectable difference between one-by-one conservation (to one part in $10^4$) of angular momentum and conservation of net bunch 
angular momentum. In each actual particle-particle collision it is not determinable whether the particles have missed, left-right, for example, 
or in any other orientation.  Certainly, quantum mechanics, with its treatment of wave/particle duality, introduces probabilistic description 
into this classical mechanics visualization.  This uncertainty could reasonably be interpreted as an extension of the Heisenberg 
uncertainly principle.

Next best, is to conserve angular momentum ``on the average''.  De Broglie\cite{deBroglie}, in a 1927 paper attempting to reconcile 
Heisenberg and Schroedinger interpretations, introduced the issue of the evolution of a ``swarm of particles'', with each particle
described by a Hamilton-Jacobi wave eikonal; a particle terminology resembling ``beam bunch'' in modern terminology and not very 
different from the wave terminology of ``wave packet''.  

Requiring conservation of energy and momentum, both linear and angular, at every instant in time, Rutherford could scarcely avoid the need 
for a temporary compound nucleus phase, as deBroglie (and Bohr as well) describe, in which two nuclei ``capture each other'' (like ice dancers) 
temporarily converting their kinetic energies into rotational energy.  Clearly, in the fullness of time, this energy has to be converted 
back into kinetic energy of a small number of point-like particles.  Some kind of a probabilistic description of the final state, as 
provided by quantum mechanical wave functions, with consistent spin and angular distribution, has become obligatory.  

\begin{figure*}[hbt]
\centering
\includegraphics[scale=0.40]{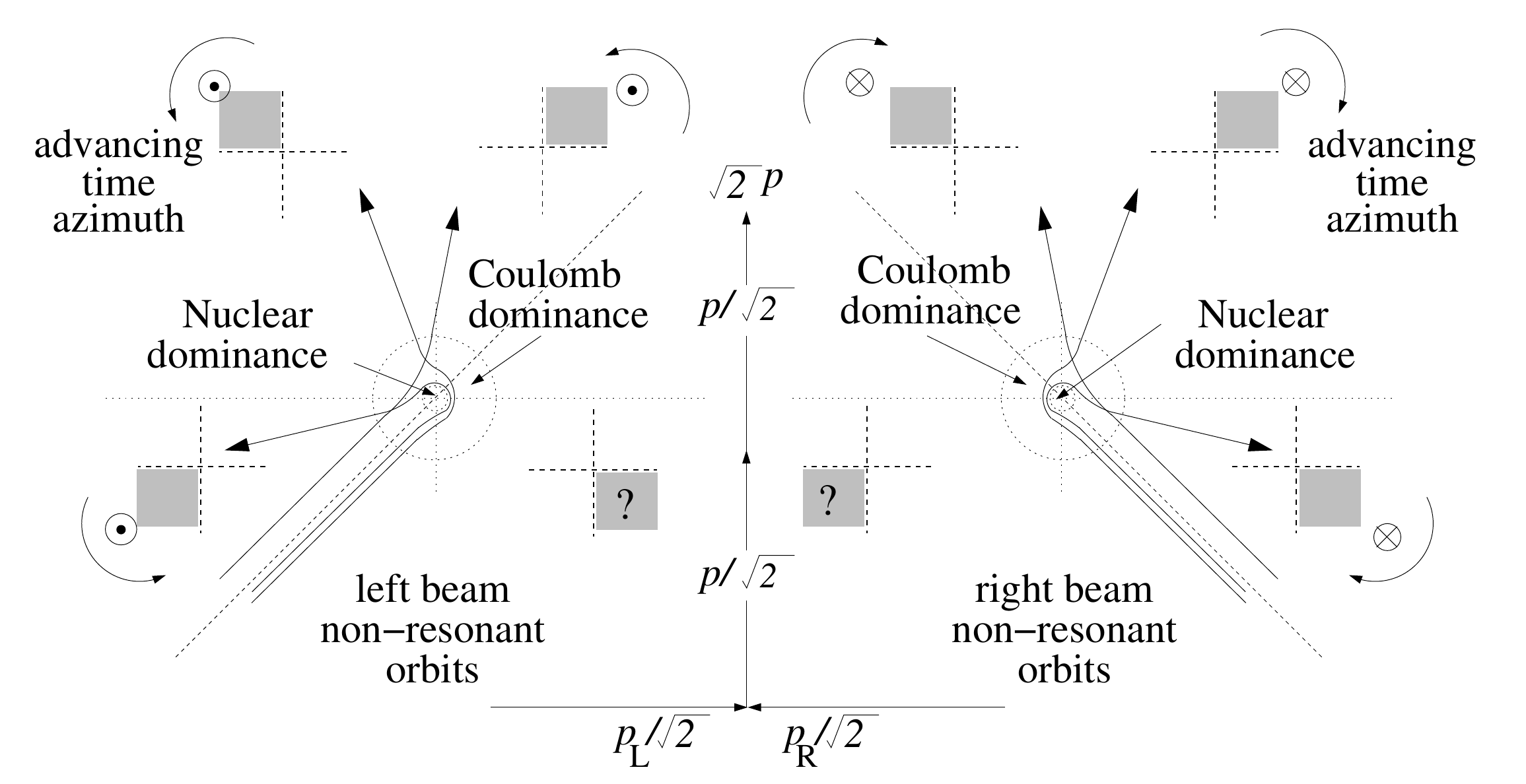}
\caption{\label{fig:EikonalTrajectories}Scattering of identical nuclear particles contained in beams viewed from a reference 
frame in which the beams are orthogonal.  Because the nuclear and electrical forces are inextricably linked, and the particle sizes
are too small to allow the orbits to be followed in detail, the scattering can only be described in a statistical sense, based,
for example, on their QM wave functions, in spite of the fact that, classically, their orbits are perfectly mirror symmetric.
At impact parameter large enough for the nuclear force to be negligible, initial and final state orbits are easily associated.
For smaller impact parameters a ``compound nucleus phase'' is inevitable, in which the subsequent evolution can only be described in a 
statistical sense.}
\end{figure*}

These ideas are pursued pictorially in Figure~\ref{fig:EikonalTrajectories}. 
''Prompt'' classical orbits (or wave eikonal curves in Hamilton-Jacobi terminology) representing 
a sequence of three  particles (from the same wave packet) participating in the mutual scattering of pairs of protons, 
one from each beam, one into each of three azimuthally advancing sectors. By the fourth sector it has become ambiguous 
(classically) or probabilistic (quantum mechanically) whether the particle is detectable immediately or has been  
``mutually captured'' by the other nucleus.  Semicircular arrows indicate the sense of azimuthal advance (which is necessarily 
exactly opposite for the two scattering particles.  Note though that ``spooky action at a distance, nowadays referred to as
quantum entanglement, is implied by Figure~\ref{fig:EikonalTrajectories}.

Figure~\ref{fig:EikonalTrajectories} demonstrates pictorially how angular momentum that is stochastically assigned locally (by QM) 
rather than being conserved, can be exactly compensated globally, by mirror symmetry.  But, with rotational energy recoverable 
as mass energy, this means also that energy can be ``teletransported'' over arbitrarily large distances.
\footnote{Etienne Forest, while
at the Central Design Group (CDG) at UC Berkeley, with his characteristic humility, while critiquing accelerator codes for which the 
design orbit did not quite close, by 1\,mm for example, was simply neglected, referred to restoring the continuity as ``beam me up, Scotty''
teletransportation.  Etienne, himself, was known to play Star Trek inspired computer games while being paid to design the Supercollider.}

The epoch represented in the figure begins just before and ends just after the actual collision.  Inset graphs show horizontal plane quadrants, 
viewed from above. The shadings represent the sequencing of events, by indicating the quadrants in which successive events 
terminate.  Individual orbits orbits can be labeled by ``azimuthal time'', with one ``typical'' prompt orbit shown in each of 
the first three quadrants after the collision.  

As regards Newtonian particle description this represents the 
``complete story''---the subsequent evolution is unpredictable in principle.  As regards ``quantum wave physics'' the figure 
represents evolution only up to the end of an initial time interval---the subsequent evolution is predictable in principal, but
only in a statistical sense---for example, the fourth scatter may, or may not, result in an event in the fourth quadrant.  
Calculation of the subsequent probabilities requires quantum mechanics, and is limited by the Heisenberg (time) uncertainty; or 
(equivalently) by limitations imposed by mathematical properties of the Fourier transforms used to calculate the probabilities.

Conservation of angular momentum complicates ones understanding of this conversion process. As with gymnasts, divers, and falling cats,
it is hard to keep track of the orientation of the total angular momentum.  There is unavoidable ``anomaly'' (or Berry phase)
by which the orientation of the total angular momentum cannot be recovered, even in principle.  The ``smallness'' of nuclei
and wave packet representation are not irrelevant to this issue.  In ``near miss'' collisions one cannot be sure of the 
azimuthal orientation of the impact parameter.  Any theoretical description needs to represent this anomalous behavior consistently.  In 
nuclear physics this requirement is associated with the existence of $G$, the anomalous magnetic moment (for which there is currently 
no fundamental theory). This topic is pursued (but only in practical measurement terms) in Appendix E.

Discounting the pion as transitory, and recognizing that satisfying the conservation laws would become harder and harder with 
increasing incident particle energies, Rutherford might reasonably have anticipated the need for a massive, spin 1/2, muon-like 
particle.   Also needed, would be a probabilistic wave mechanical theory, like quantum mechanics, to repackage the rotational energy
into the momentum vectors of a small number of point-like particles.  

Recognizing, also, that electric and nuclear forces were inseparably present in the proton, Rutherford might well have considered it
unnatural to treat electrical and nuclear forces individually.  In this respect he might have expected support from Bohr and
Einstein, concerning the issue of ``exchange potential'', a theoretical formalism introduced by Heisenberg, with later refinement
and support by Wigner, Bethe, and others.  

The exchange potential is needed to account  empirically for the ``saturation'' of the density of nuclear matter within nuclei.  
Yet, with nuclei being ``as big as a barn'' in Fermi language, the exchange potential seems like ``spooky action at a distance'' 
in Einstein language.  Bohr, on the other hand, though justifiably proud and supportive of the treatment of energy in the quantum 
mechanics of atoms, might eventually have become dubious about its uncritical acceptance for nuclear physics.

The purpose for this digression, has been to motivate the investigation of these nuclear physics issues experimentally.  
Predominantly electric E\&m storage rings make this both possible and inexpensive.

\subsection{Importance of anomalous nuclear MDM $G$-values\label{sec:AnomMDMs}}\mbox{}

One motivation for the E\&m storage ring being promoted in this paper centers on the careful study of
elastic or weakly inelastic nucleon scattering, and emphasizes the possible role played by the  anomalous 
MDM, $G$.  An essential feature of the rings being advocated here follows from their superimposed electric and magnetic 
bending, which provides the capability of simultaneously co- or counter-circulating frozen or pseudo-frozen spin 
beams of different particle type. 

The original motivation for the development of E\&m rings was to investigate time reversal violation in the form
of non-vanishing proton electric dipole moment (EDM), which has always been assumed to constrain the strong nuclear force.
But, in actuality, the electromagnetic and nuclear forces are inextricably connected in actual protons.  The influence of this 
marriage has been well accounted for, in both classical and quantum mechanics, for low energy Rutherford scattering differential 
scattering cross sections.  However, in p,p scattering, there is also proton spin precession caused by the 
(relativistically-implied) ${\bf B=v\times E}$ magnetic field (in the proton's rest frame) acting on the proton's 
anomalous magnetic moment\cite{RT-RAST} .

Appendix~E discusses the consistent treatment of g-factor $g$ and anomalous magnetic moment factor $G$ and the
conversion of $g$ to $G$, following the treatment in reference\ \cite{RT-ICFA}, which explains how 
E\&m storage rings can be utilized as ``MDM Comparators''.  In the present context, when ultrahigh frequency domain MDM
precision is required, it is appropriate to have runs long enough for spin orientations to complete an integral number of 
rotations after an integral number of turns.  For this purpose it is appropriate to express the anomalous MDM as a rational 
fraction, in order to determine the minimum number of turns required, and the exact number of turns required to produce 
an integral number of spin revolutions.  This approach is abbreviated to the phrase 
\emph{with frequency domain precision} in the sequel.

As explained in section ``\ref{sec:Set-Reset}'', the E\&m storage ring configuration is ideal for 
the precision measurement of anomalous nuclear MDM $G$-values. Such rings serve naturally for the function of 
``mutual co-magnetometry'' for precision experimental determination of $G$-values of nuclear particles.

In the present context there is an equally important need for knowing the MDMs of nuclear isotopes to the highest possible 
precision.  What needs to be explained is the way that storage ring steering can be set and reset 
to \emph{frequency domain precision} (i.e. with precision that would be unachievable by direct field strength control)
using the particle anomalous magnetic moments as ``magnetometric gyroscopes''.

For historical reasons, based probably on the great importance and successful application of the 
$g$-factor in atomic physics, the anomalous MDM parameter $G$, a fundamental measurable ratio of 
nucleus angular momentum (proportional to inertial mass $m$ of nucleon) to magnetic moment (proportional to charge of 
the same nucleus) is less systematically updated and made available than is $g$.  
With $Z$ and $A$ being dimensionless measures, the ratio of integers, $A/Z$, justifies regarding   
$g(A/Z)$ as being a function of $A$ and $Z$ only via the ratio $A/Z$.  To be ``anomalous'' the 
dimensionality of $G$ and $g$ must be the same: i.e. their ratio is dimensionless.  
For every nucleon, $Z$ is truly an integer multiple of (positive) proton charge $e$.  Regrettably, 
for example because of nuclear binding energy, nucleon mass ratio is only approximately given by the 
mass number $A$.

This discussion is relegated to Appendix~E, but not because it is unimportant;
in fact this paper provides further strong support for the precise measurement, and consistent
treatment of nuclear isotope MDMs and mass values. But the discussion is both technical and boring.  
This justifies treating Appendix~E as a self-contained discussion of the experimental and 
theoretical connections between $g$ and $G$\cite{NIST-background}\cite{NIST-isotope-abundance}.

\begin{figure}[hbt]
\centering
\includegraphics[scale=0.9]{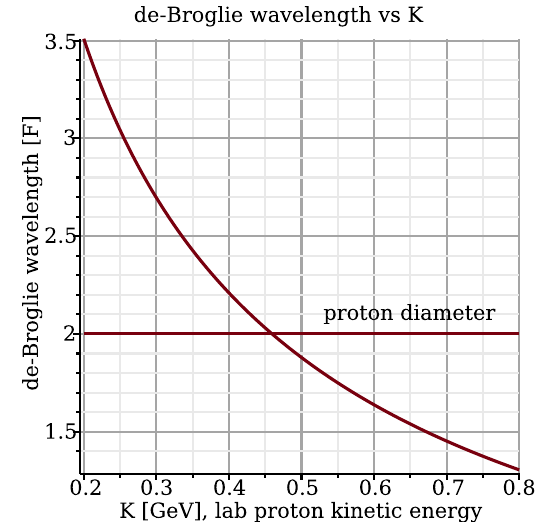}
\caption{\label{fig:deBroglie-lambda-vs-E_tot}Fermi length units plot of the deBroglie wavelength of an 
incident proton versus its laboratory kinetic energy.  The approximate match of deBroglie wavelength with proton 
size with the onset of resonant scattering behavior can scarcely be simple coincidence.  It seems natural for 
resonance-like behavior to set in when wavelengths in a wave picture match particle sizes in a particle picture.  
See Figure~\ref{fig:pp-total-crsctn-vs-KE}. Surely this represents a nominal transition point from particle 
to wave description of $p,p$ scattering; much like the transition from geometric to wave optics. 
New ``particles'' in this case pi-mesons, need to be massive, to account for the available energy, 
while conserving momentum at the local velocity.  Azimuthal symmetry is broken on a particle by particle basis, 
but is preserved on average. } 
\end{figure}

\subsection{Modern spin control; ancient nuclear physics\label{sec:Spin-control}}\mbox{}

The energy region emphasized in this paper, ``above'' Rutherford scattering, ``below'' meson threshold,
seems paradoxical in various ways.  Total $p,p$ cross sections are discussed and plotted in detail
in a heroic 1993 review by Lechanoine-LeLuc and F. Lehar\cite{LeLuc-Lehar}, containing seven pages of 
references, from an era in which a large experimental group had five members.

\begin{figure}[hbt]
\centering
\centerline{\includegraphics[scale=0.36]{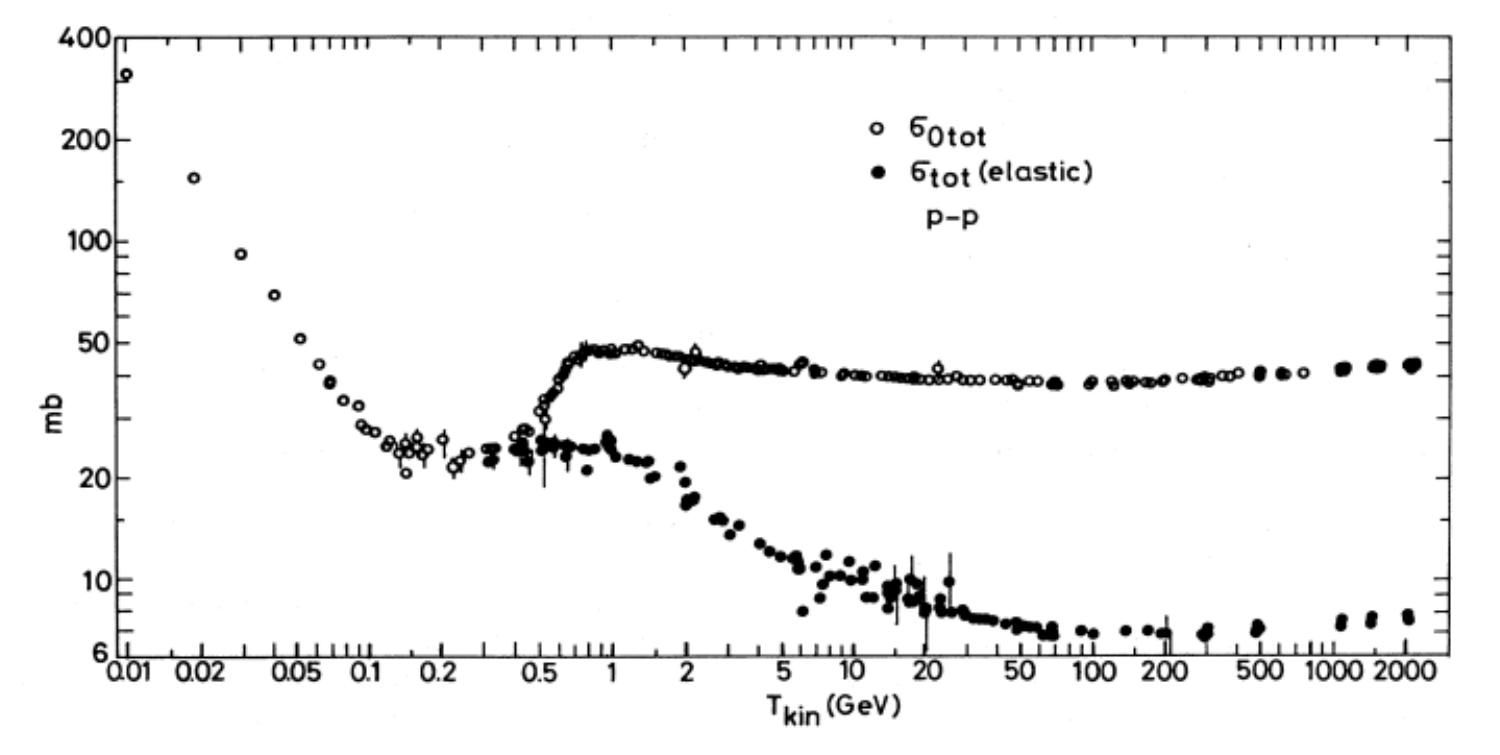}}
\centerline{\includegraphics[scale=0.38]{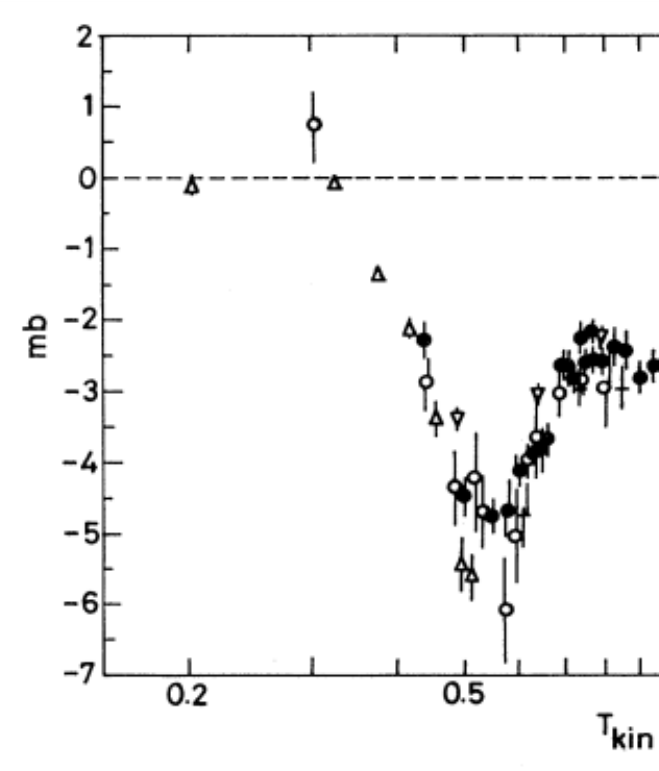}\includegraphics[scale=0.38]{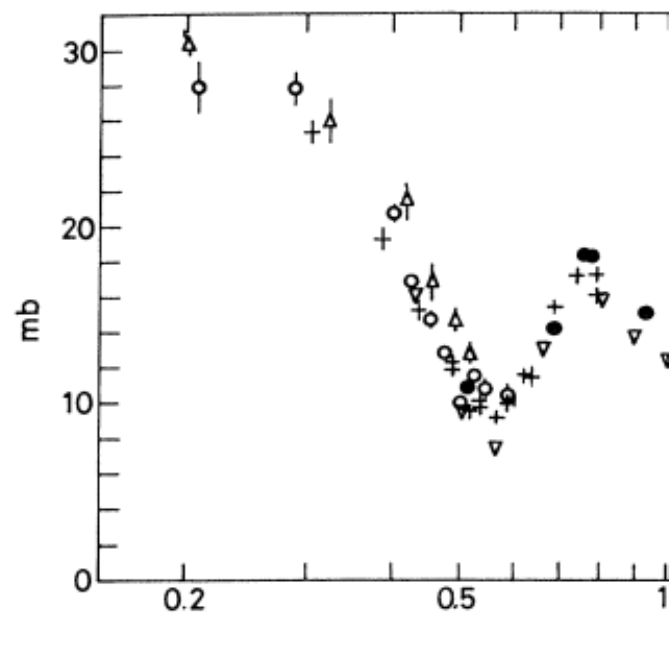}}
\caption{\label{fig:pp-total-crsctn-vs-KE}{(Copied from reference\ \cite{LeLuc-Lehar}) 
{\bf Above:\ }Energy dependence of $p,p$ spin-independent total cross section (open circles) and total 
elastic cross section (solid curves).}
{\bf Below left:\ }$\sigma_{1 tot}(p,p) = -(1/2)\Delta\sigma_T(p,p)$ energy dependence,
{\bf Below right:\ }$-\Delta\sigma_L(p,p)$ energy dependence on the incident state polarizations.
This figure should be compared with Figure~\ref{fig:deBroglie-lambda-vs-E_tot}.
}
\end{figure}

Figure~\ref{fig:pp-total-crsctn-vs-KE} (copied from LeLuc and Lehar) shows measured elastic and inelastic 
$p,p$ cross sections. Spin dependent $p,p$ cross sections, measured with polarized beams and polarized hydrogen 
target are plotted at the bottom of Figure~\ref{fig:pp-total-crsctn-vs-KE}. 

Though originally mysterious, the complicated long de Broglie wavelength behavior (i.e. low energy region below, 
say, 100\,MeV) quickly became well understood in terms of interference between Rutherford and nuclear 
amplitudes.  The same cannot be said for the short wavelength, higher energy region, above, 
say, 400\,MeV, where inelastic scattering quickly becomes dominant.  Notice, in Figure~\ref{fig:deBroglie-lambda-vs-E_tot},
that the dramatic change in $p,p$ scattering begins quite precisely when the deBroglie wavelength of the proton is
equal to the proton diameter. 
\hspace*{-1cm} 

\section{Storage ring with predominantly electric bending}\label{sec:Predom-elec}\mbox{}

\subsection {``Circular E\&m'' storage ring closed orbitry}\label{sec:Co-traveling-orbits}\mbox{}

This section, concerning the simultaneous storage of two different particle type beams in the  circular arcs
of a predominantly electric ``E\&m'' storage ring with superimposed magnetic bending, is supposed to be clear 
on its own.  To the extent this is not the case, it may be helpful to refer to Appendix~F, 
``Superimposed E\&M storage rings'', which is copied almost verbatim from reference~\cite{RT-Instrumentation-paper}.  

For simplicity the arcs are assumed to be perfect circles, of bending radius $r_0$, joined 
tangentially by bend-free straight sections of arbitrary length.  Without essential loss of generality, we assume 
the geometry has super-periodicity four, giving it the shape of a rounded square, or a squared-off circle.

Fractional bending coefficients $\eta_E$ and $\eta_m$ are defined by
\begin{equation}
\eta_E = \frac{qr_0}{pc/e}\,\frac{E_0}{\beta},\ 
\eta_M = \frac{qr_0}{pc/e}\,cB_0,
\label{eq:BendFrac.2}
\end{equation}
neither of which is necessarily positive.  These fractional
bending fractions satisfy
\begin{equation}
\eta_E + \eta_M = 1,\quad\hbox{and}\quad
\frac{\eta_E}{\eta_M} = \frac{E_0/\beta}{cB_0}.
\label{eq:BendFrac.2p}
\end{equation}
By symmetry, stable \emph{all-electric} storage ring orbits are forward/backward symmetric and there 
are continua of different orbit velocities and radii, one of which matches the design ring radius $r_0$
in each direction.  To represent the required bending force at radius $r_0$ being augmented by magnetic 
bending while preserving the orbit curvature we require
\begin{equation}
\eta_E + \eta_M=1, \hbox{\ where\ } |\eta_M/\eta_E| \apprle 0.5 \, .
\label{eq:bendingFractions} 
\end{equation}
The resulting magnetic force dependence on direction causes an $\eta_M>0$ (call this ``constructive'') 
or $\eta_M<0$ (``destructive'') perturbation to shift opposite direction orbit velocities (v) of the same radius, 
one up in radius and one down, resulting in two stable orbits in each direction.  For stored beams, any further 
$\Delta\eta_M \ne 0$ change causes beam velocities to ramp up in kinetic energy ($KE=\mathcal{E}-mc^2$) 
in one direction, down in the other.  Our proposed E\&m storage ring is ideal for investigating low-energy nuclear 
processes and, especially, their spin dependence at low energy.  

Consider the possible existence of a stable orbit particle pair (necessarily of different particle type) such 
as deuteron/proton ($d,p$) or deuteron/helion ($d,h$), each with laboratory kinetic energy (KE) in the tens of 
MeV range, and traveling simultaneously with different velocities in the same direction.  This periodically 
enables ``rear-end'' collision events whose CM KEs can be tuned into the several \SI{100}{keV} range by changing 
$\eta_M$.  

This description is not effective for ``same particle'' pairs, such as $p,p$ or $d,d$.  Their resultant 
co-traveling bunch velocities remain identical and no ``rear-end'' collisions ensue.  (Treatment 
of this fundamentally important case of identical particle scattering has to be deferred for now.)

With careful tuning of $E$ and $B$, such nucleon bunch pairs will have appropriately different charge, mass, and 
velocity for their kinematic rigidities to be identical.  Both beams can then co-circulate indefinitely, with 
different velocities.  

Depending on the sign of magnetic field $B$, either the lighter or the heavier particle bunches can be faster, 
``lapping'' the slower bunches periodically, and enabling ``rear-end'' nuclear collision events. (The only 
longitudinal complication introduced by dual beam operation is that the ``second'' beam  needs to be injected 
with accurate velocity, directly into stable RF buckets.)

Only in such a storage ring can ``rear-end'' collisions occur with heavier particle bunches
passing through lighter particle bunches, or vice versa.  From a relativistic perspective, treated as 
point particles, the two configurations just mentioned would be indistinguishable~\cite{Jaffe}.  As observed in the 
laboratory, to the extent the particles are composite, such collisions would classically be expected to be
quite different and easily distinguishable.  

Pavsic, in a 1973 paper reproduced in 2001~\cite{Pavsic}, develops a 
``mirror matter'' Hamiltonian formalism, distinguishing between ``external'' and ``internal'' symmetry.  
He points out, for example, that ``the existence of the anomalous proton or neutron magnetic moments indicates 
the asymmetric internal structure of two particles''; a comment that applies directly to the present paper.
Otherwise, Pavsic is agnostic, suggesting that his formalism provides only a parameterization for experiments
sensitive to internal structure, with possible implications concerning mirror matter.

\subsection{Storage ring PTR with E\&\lowercase{m} bending}\label{sec:PTR-ring}\mbox{}

First suggested by Koop~\cite{Koop}, (in the context of counter-rotating proton beams for proton EDM  measurement),
design of the E\&m configuration has been described in a series of papers by or including a present 
author~\cite{RT-ICFA}\cite{EDM-Challenge}.  The acronym PTR chosen in Ref.~\cite{CYR} to stand for ``prototype'' 
has been retained, in spite of the much altered rationale for its existence. 

It is possible, with superimposed electric and magnetic bending, for beam pairs of different particle 
type to co-circulate simultaneously.   This opens the possibility of ``rear-end'' collisions occurring 
while a fast bunch of one nuclear isotope type passes through a bunch of lighter, yet slower, isotope 
type (or vice versa).  The Pavsic formalism just mentioned seems well suited to the empirical experimental 
representation of measured differences between these two possibilities.

A schematic diagram of the proposed PTR storage ring is shown on the left in 
Fig.~\ref{fig:RTR-figures}, and its optimized beta functions are shown on the right.  PTR lattice description ``sxf'' 
files can be obtained at Ref.~\cite{web-JDT}.
\begin{figure*}
\centering
\includegraphics[scale=0.24]{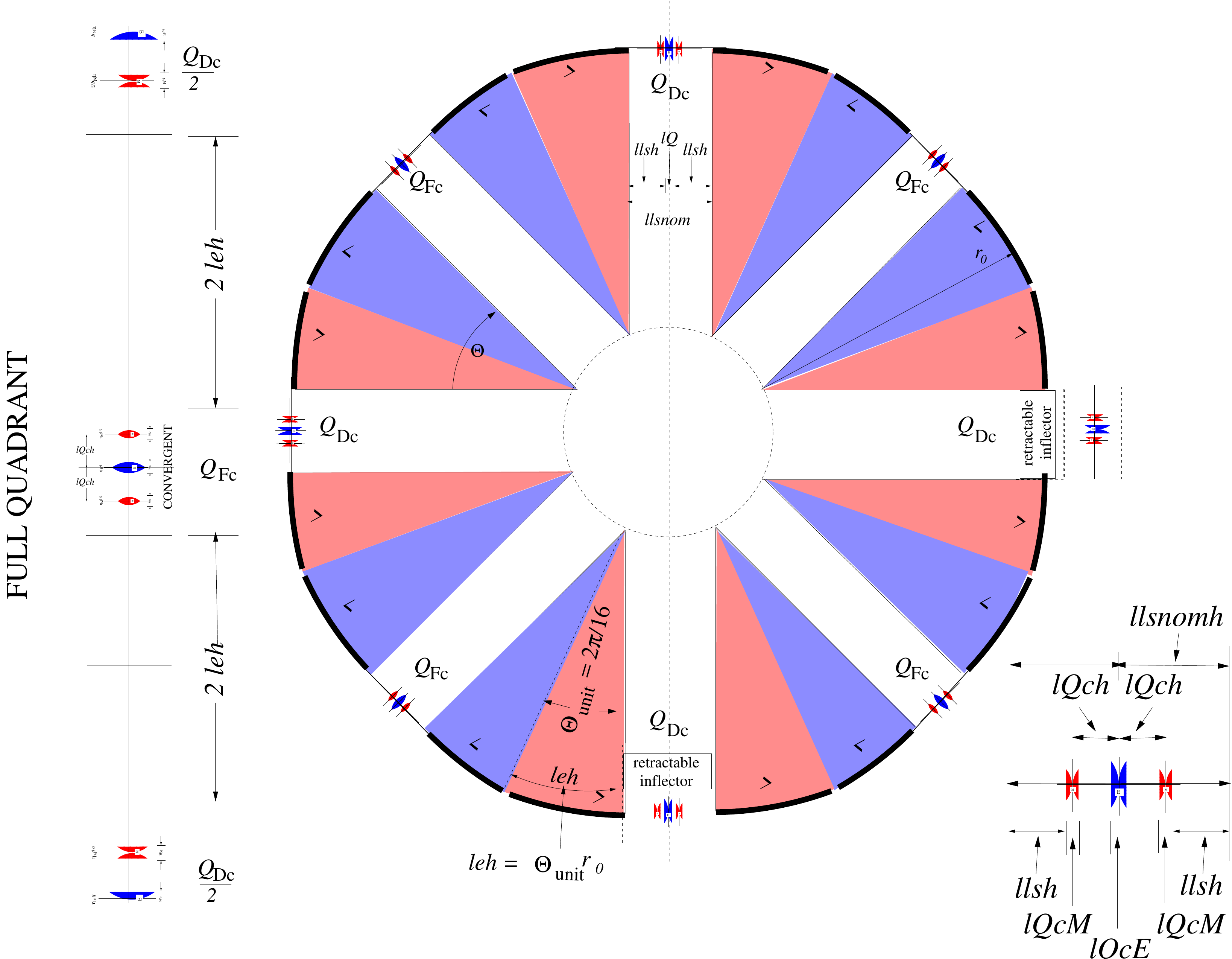}
\caption{Lattice layouts for PTR, the proposed prototype nuclear transmutation storage ring.
``Compromise'' quadrupoles~\cite{RT-Compromise} are shown lower right.  The need for a  m/2,E,m/2 quadrupole triplet to implement 
the E\&m superimposed focusing by a ``compromise quadrupole'' is explained in Appendix A.
The circumference is \SI{102.2}{m}.}
\label{fig:RTR-figures} 
\end{figure*}
\begin{figure*}
\centering
\includegraphics[scale=1.0]{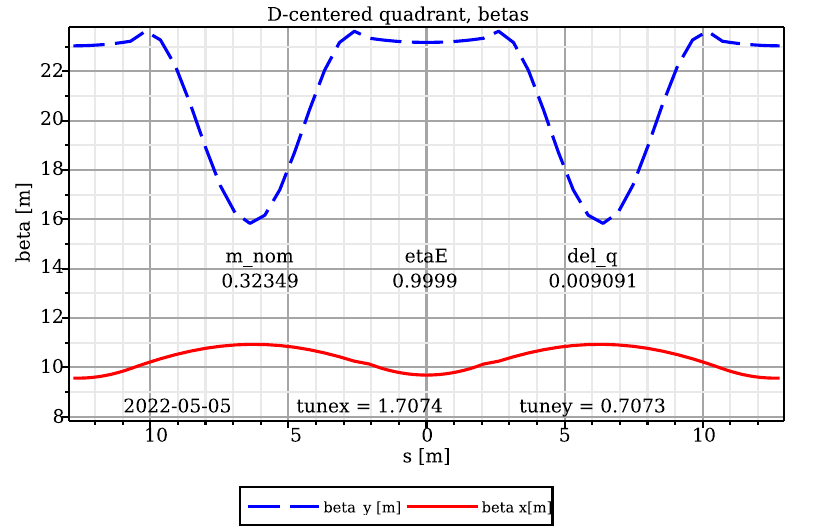}
\caption{Refined PTR tuning, with quad strengths and $m_{\rm nom.}$ (adjusted to 0.32349) for (distortion-free) equal-fractional-tune, 
$Q_x=Q_y+1$, operation on the difference resonance.  Not counting geometric horizontal focusing, thick lens pole shape horizontal and 
vertical focusing strengths are then identical~\cite{Haissinski} (mnemonic: $m_{\rm nom.}$=1/3).  The entire lattice design can be scaled, 
e.g. to reduce peak field requirements.}
\label{fig:RTR-figures} 
\end{figure*}

\begin{figure*}
\includegraphics[scale=0.41]{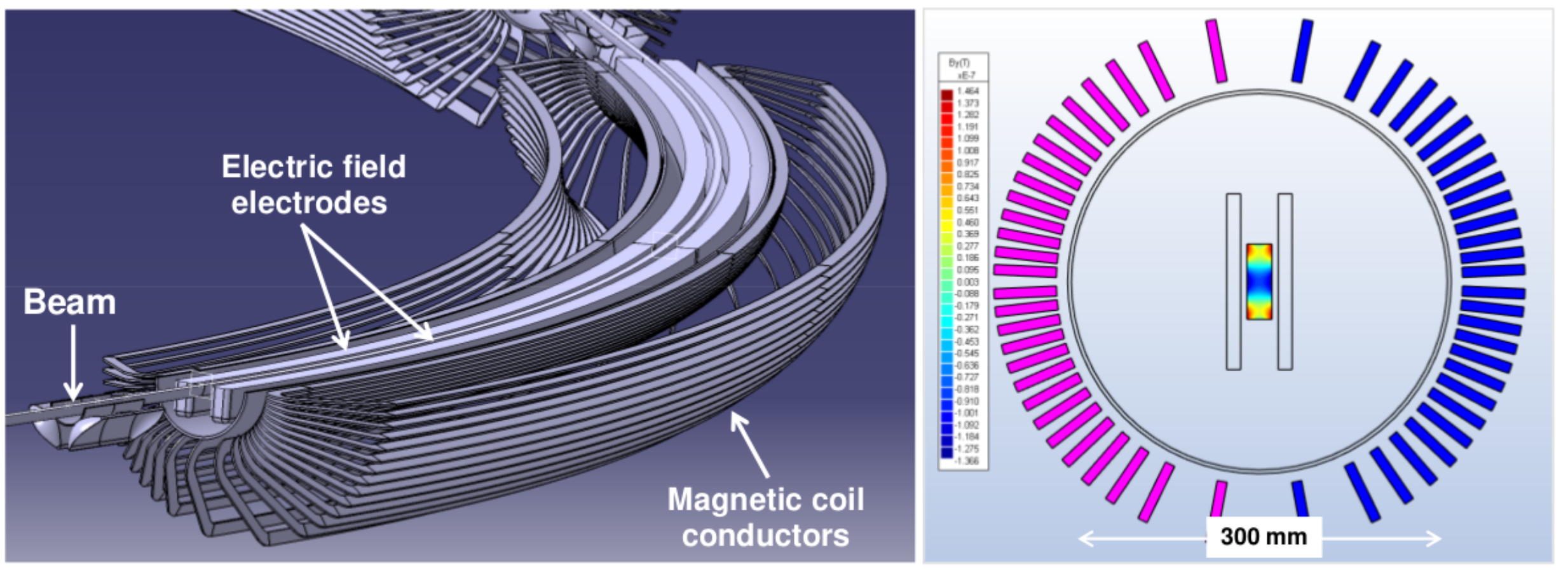}
\caption{{\bf Left:\ }Cutaway drawing of one sector of the PTR ring bending element. {\bf Right:\ } A transverse section showing an end view of the (inner legs of the) magnet coil, as well as a field map of the good field region.  The (brilliant) design is due to Helmut S\"oltner~\cite{CYR}. The magnet is ``air core'', limited to quite weak magnetic field, but sufficient for most applications. Current design maximum values for electric and magnetic fields are \SI{10}{MV/m} and \SI{30}{mT}.} 
\label{fig:SectorPerspective}
\end{figure*}
\begin{figure}
\includegraphics[scale=0.35]{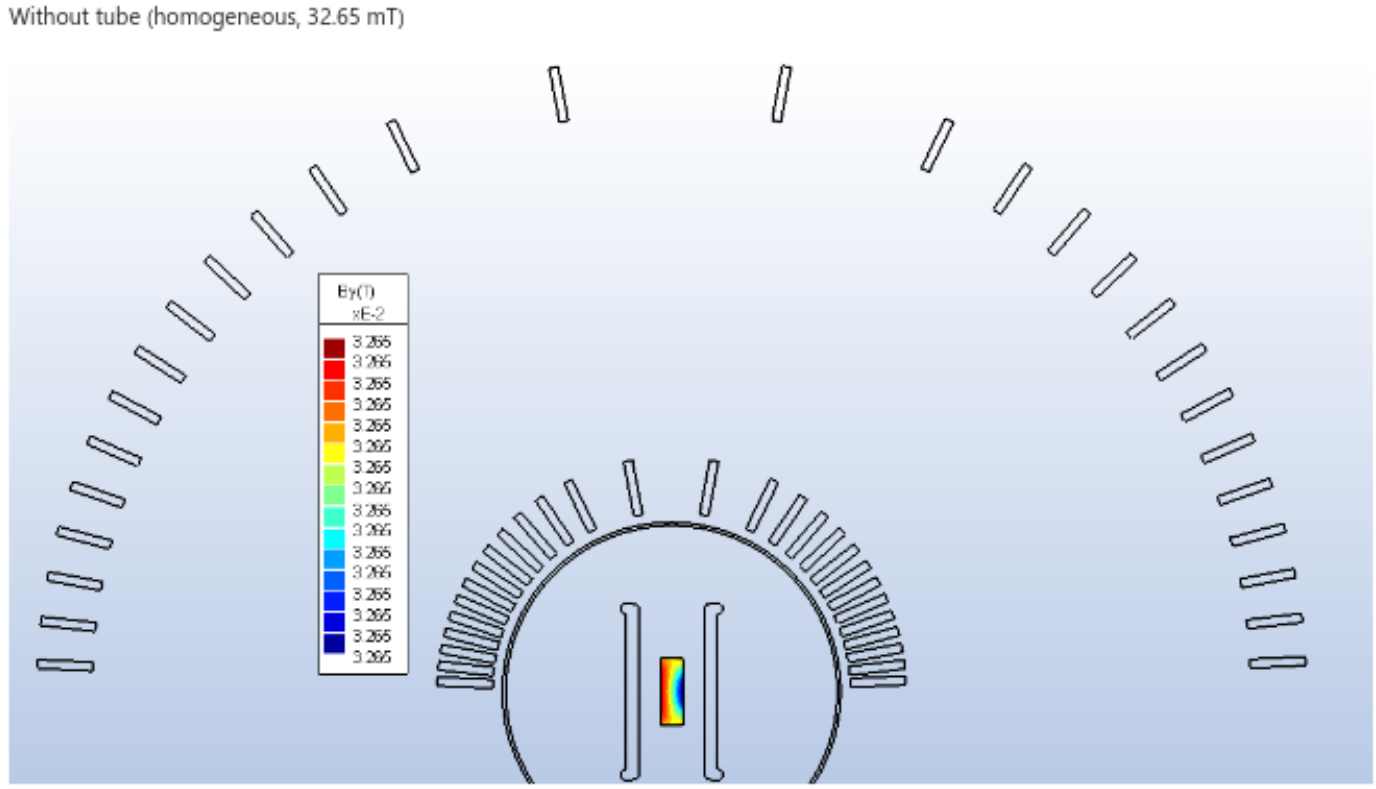}
\includegraphics[scale=0.35]{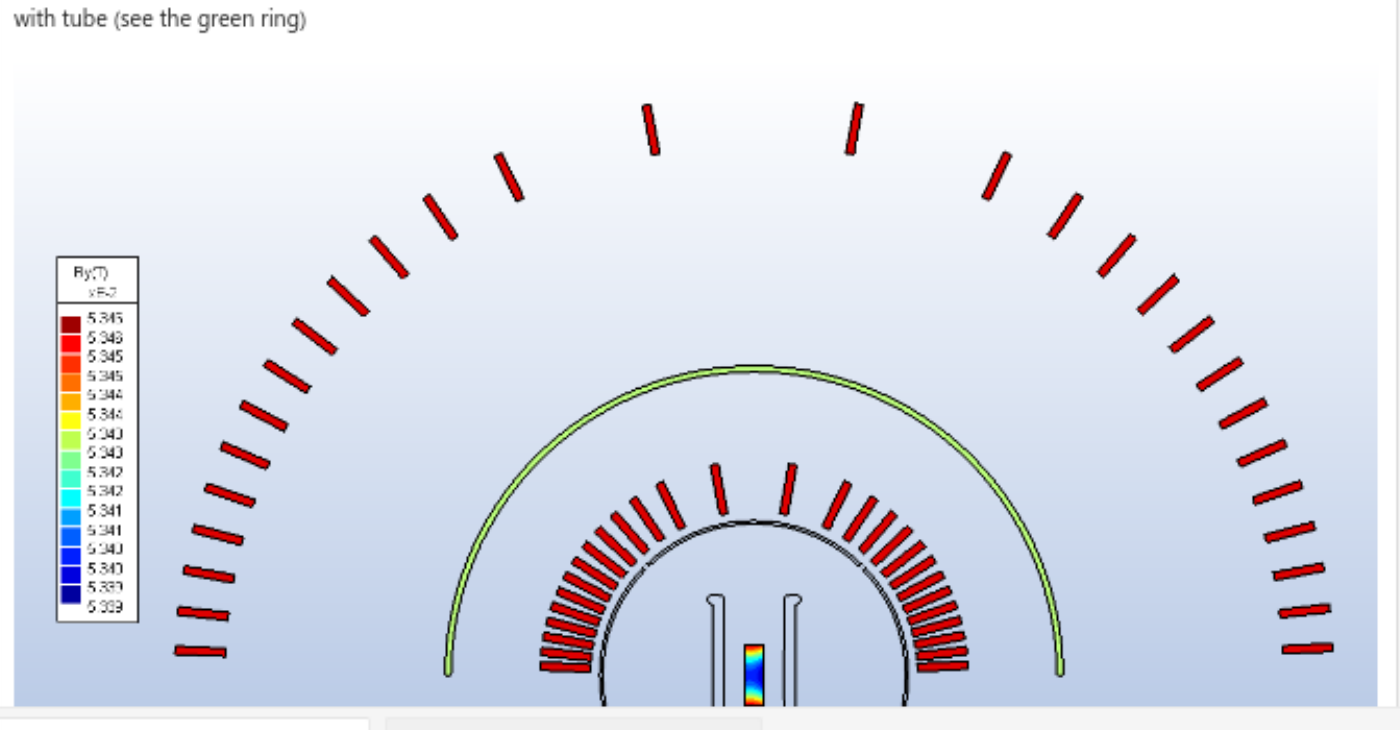}
\caption{{\bf Top:\ }Full transverse section of top half of the air core bending element shown on the right in Figure~\ref{fig:SectorPerspective}\,.
{\bf Bottom:\ } Full transverse section of top half of an iron-enhanced version of the bending element shown on the right in Figure~\ref{fig:SectorPerspective}\,.
For predominant magnetic, e\&M, bending, the lower design would be required. This design has been graciously provided by Helmut S\"oltner.}
\label{fig:Helmut-iron-enhancedment}
\end{figure}

\begin{figure*}
\centering
\includegraphics[scale=0.35]{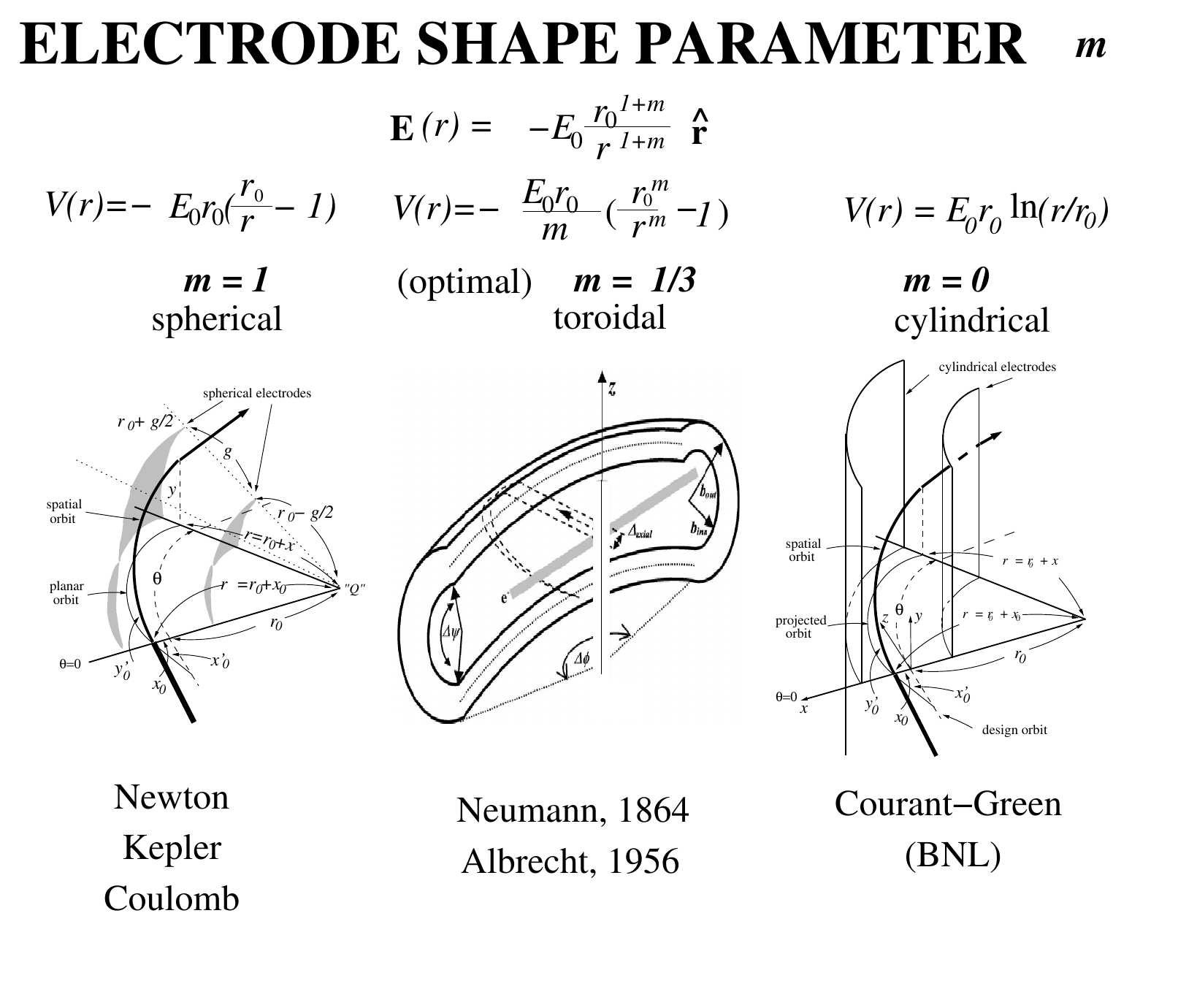}\includegraphics[scale=0.48]{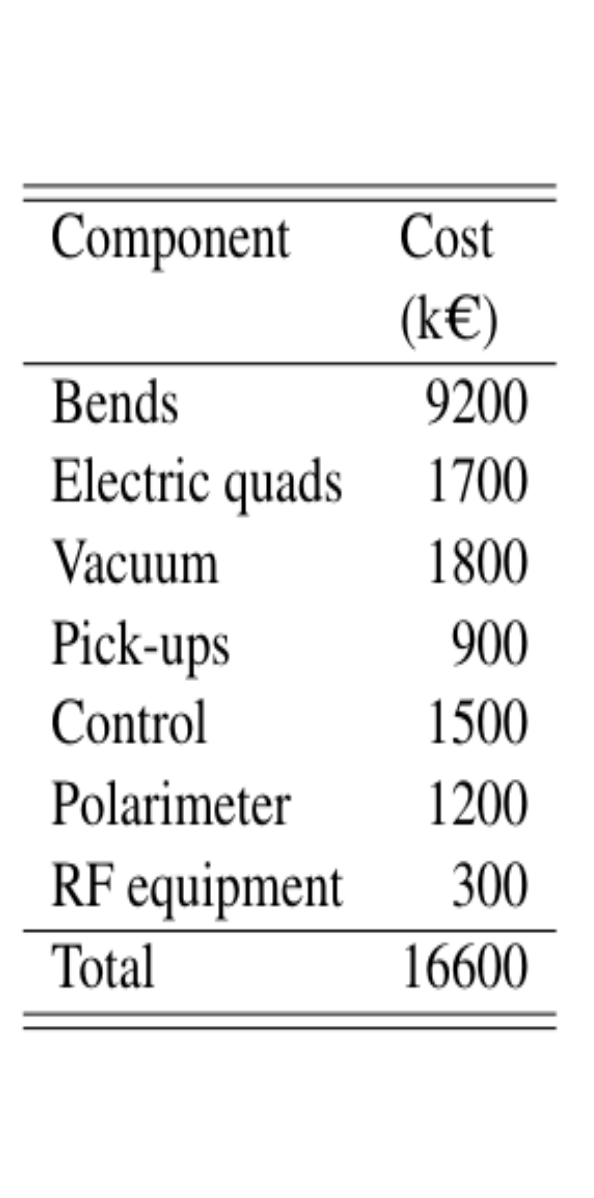}
\caption{Left: Electrode shapes are shown with their focusing strength parameters $m$, for spherical, 
(optimized) toroidal, and cylindrical bending fields~\cite{Albrecht} \cite{Plotkin} \cite{EDM-Challenge} \cite{RT-JT-AGS-Analogue}. 
Right: 2021 CERN Yellow report cost estimate~\cite{CYR} for the apparatus shown in Figure~\ref{fig:SectorPerspective}.} 
\label{fig:ElectrodeShappes_cost}
\end{figure*}

Though the quadrupole strengths are minimal (as can be seen by the vanishing entrance and exit slopes) in the figure on the
right of Fig.~\ref{fig:RTR-figures}, they have been trimmed
for ``equal'' \emph{fractional} $x,y$ tune values (0.7074, 0.7073).  

The optimal thick lens PTR optics (i.e. with quadrupoles essentially turned off, and functioning only for trimming) is 
uniquely determined, with $m_{\rm nom.}$ (defined by formulas in Figure~\ref{fig:ElectrodeShappes_cost}) being curiously 
close to 1/3, closer to $m=0$ (cylindrical) than to $m=1$ (spherical) electrode shape.  With obvious scaling changes, 
namely electric, $E_0/\beta$, and magnetic, $cB_0$, field strengths varying inversely with the factor $qr_0/p$, 
as given in Eq.~(\ref{eq:BendFrac.2}). \emph{The same focal relationship is valid at all scales, from microscopic to 
cosmological.}  For example, by doubling $r_0$ to \SI{22}{m}, the value of $E_0$ would be 
reduced from \SI{5.06}{MV/m} to \SI{2.53}{MV/m}.  See, for example, the central row of Table~\ref{tbl:BendParams4-hd}.

\subsection{Bunching of both beams by the same RF cavity}\label{sec:Beam-bunching}\mbox{}

\begin{figure*}
\centering
\includegraphics[width=0.80\textwidth]{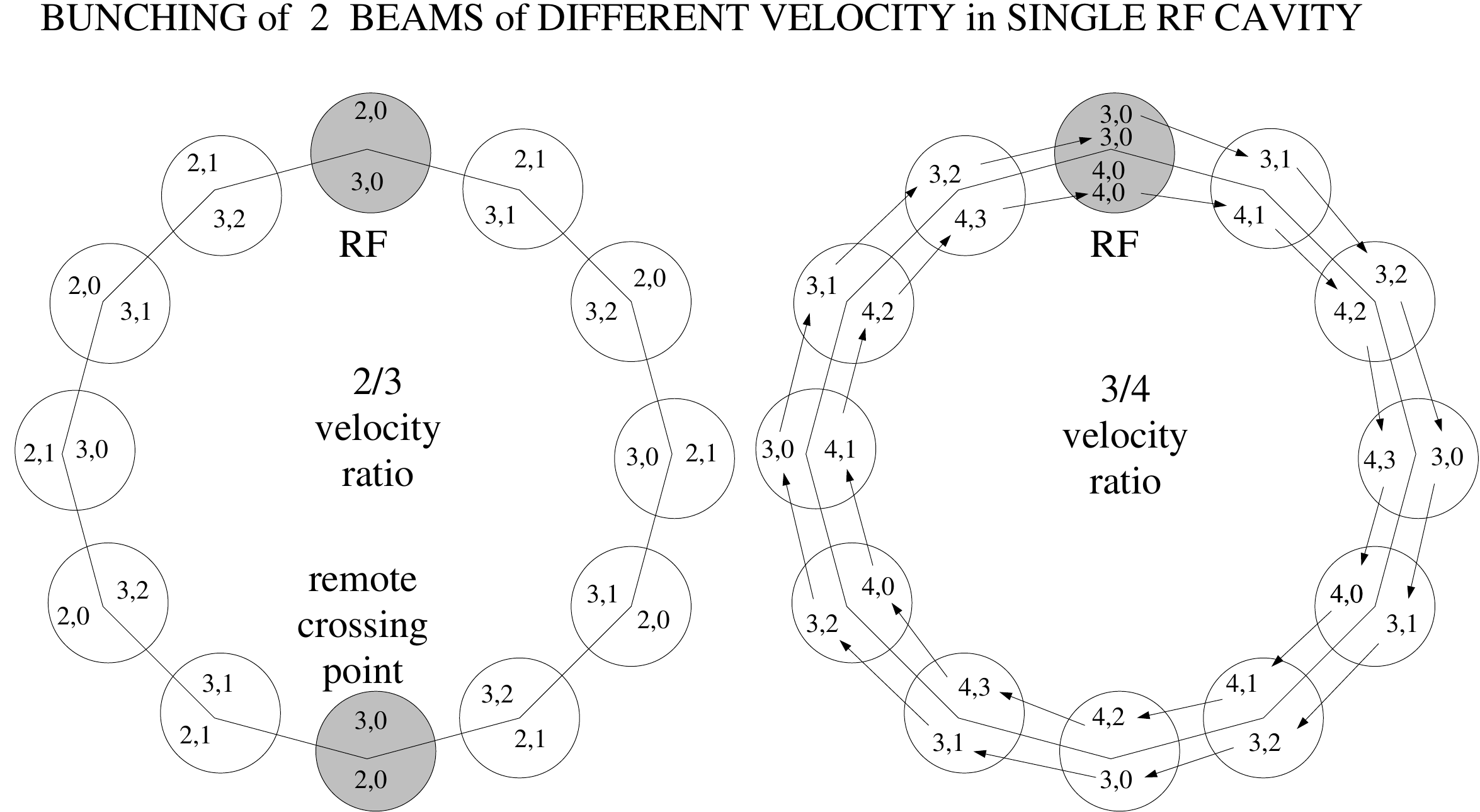}
\caption{
Stable RF buckets for beam velocity ratios of 2/3 or 3/4.  Shaded circles indicate locations at which bunch positions 
coincide.  The undesirable collisions at the RF location should be removable with symmetrically split RF locations.
\label{fig:THA1I3f3b}}
\end{figure*}

The condition for bunch collision points to occur at fixed ring locations is met by the beam 
velocities being in the ratio of integers; e.g. $\beta_1/\beta_2=8/7$ in Table~\ref{tbl:BendParams-h-d}.
\emph{Both circulating beams can be bunched by a single RF cavity in spite of their different velocities.}
For more nearly equal velocities the figure becomes more complicated. With 8/7 velocity ratio and $7\times8=56$, 
the RF frequency can be the 56th harmonic of a standard base frequency, $f_{\rm base}$, itself a harmonic number 
$h_n$ multiple $f_{\rm base} = h_nf_{\rm rev.}$ of the revolution frequency.
Stable buckets are labeled for simple cases in Fig.~\ref{fig:THA1I3f3b}. 
(Hint: when the second indices are both zero, the populated bunches superimpose.)  A ``remote'' bunch collision point 
appears on the left, but not on the right. 

\subsection*{Bend field stabilization and resettability}\mbox{}
\setlabel{Bend field stabilization and resettability}{sec:Set-Reset}

Magnetic bending with superimposed electric field is illustrated in Figures~\ref{fig:SectorPerspective} and 
\ref{fig:Helmut-iron-enhancedment}, the latter of which shows the possible, but probably unnecessary, inclusion
of soft iron flux return paths.  The designs have been performed by Helmut S\"oltner, as cited in the figure. 
Further details are contained in reference~\cite{IKP-EM-bend}

When contemplating the high precision measurement of nuclear parameters, especially their anomalous magnetic moments $G$,
one assumes that all intentional electric and magnetic field components are known with high precision and all unintentional
field components are known to vanish with high accuracy.  The degree to which this can be achieved in a ``small'' accelerator,
say of 100\,m circumference, needs to be established.  

Though it is possible to measure both magnetic and electric field components to high accuracy in many locations, it is not 
possible to make such measurements exactly along the storage ring design central orbit. In this respect, polarized beams can 
come to the rescue.  

As regards the orientation of the beam polarization, it is essential to distinguish between ``in-plane'' 
and ``out-of-plane'' orientations, where ``the plane'' refers to the ring beam plane, which is presumed to be horizontal.
\emph{In-plane precession}, induced by ideal magnetic fields acting on beam particle magnetic dipole
moments (MDMs) is routinely the dominant spin precession.  

Assuming the absence of non-zero electric dipole moments (EDMs) as is required by time reversal invariance, 
\emph{out-of-plane precession} can be induced only by electric or magnet field imperfection---radial, in-plane magnetic 
field components or vertical out-of-plane electric field components. In practice, the inevitable existence of unintentional 
fields acting on particle MDMs will induce out-of-plane precession.  The radial magnetic field average or the 
$\langle B_r\rangle$ and the vertical electric field $\langle E_y\rangle$ average are expected to be the dominant source of 
spurious MDM precession.   

The leading strategy for setting and resetting conditions will be to monitor the beam polarizations to feedback-stabilize the
beam polarizations.  Before this, however, this condition can be achieved by adjusting local beam deflection components;
 $\langle B_r\rangle$ can be canceled by canceling the  out-of-plane (vertical) orbit separation of (sequential)
counter-circulating beams. (Hysteresis in the possible soft iron cylinder mentioned previously would impair this
compensation significantly.)  We refer to this capability as  
``self-magnetometry''. The precision with which the orbits can be matched vertically depends on the precisions of the 
beam position monitors (BPMs) that measure the vertical beam positions, and on the ring lattice sensitivity to the 
magnetic field errors causing the orbits to be vertically imperfect.  Because of the weak vertical focusing this 
sensitivity is excellent.  

Assuming both beam spins are frozen, at least the ``primary'' beam-1 will, by convention, be globally frozen, 
with spin tune $Q_{s1}=0$.  The presence of magnetic bending guarantees that this condition can be satisfied.
Ideally both beams would have $Q_s=0$ but, with only a few exceptions, the ``secondary'' beam-2 can only be  
locally frozen; $Q_{s2}$ exactly equal to a rational fraction other than $0/1$.  

In this condition both beam polarizations 
can be phase-locked,  allowing both beam spin tunes to be set and re-set with frequency domain precision. This means that 
synchronism can be maintained for runs of arbitrary duration.  Since the RF frequency can also be restored to arbitrarily 
high precision, conditions can be set and  re-set repeatedly, without depending upon high precision measurement of the 
electric and magnetic bend fields.  

This also allows, for example, the magnetic bending field to be reversed with high precision, as would be required to interchange 
CW and CCW beams. This capability can be referred to as \emph{stabilizing all fields  by phase locking both revolution frequencies
and both beam polarizations, using their own MDMs as ``magnetometric gyroscopes''}.

\subsection{Proposed E\&m ring properties} \mbox{}

To represent a small part of the required bending force at radius $r_0$ being replaced by magnetic bending
while preserving the orbit curvature we define ``electrical and magnetic bending fractions'' 
$\eta_E$ and $\eta_M$ satisfying $$\eta_E + \eta_M=1, \hbox{\ where\ } |\eta_M/\eta_E| < 0.1 $$
This perturbation ``splits'' a unique velocity closed circular orbit solution into two slightly separated velocity 
circular solutions.  As a result there are periodic ``rear-end'' collisions between two particles co-moving 
with substantial, but different, velocities in the laboratory.  Their CM KEs can be in the several 100\,KeV range.
All incident and scattered particles then have convenient laboratory KEs, two orders of magnitude higher,
in the tens of MeV range.

Our proposed ``E\&m'' storage ring is ideal for investigating low energy nuclear processes.
With careful tuning of E and B, certain nucleon bunch pairs of different particle type, such as $p$
and $d$ or $d$ and $h$, can have appropriately different charge, mass, and velocity for their rigidities to be
identical.  Both beams can then co-circulate indefinitely, with different velocities.
For two beams \emph{of identical particle type}, higher velocity bunches will ``lap'' and pass through lower 
momentum bunches, thereby enabling ``rear-end'' elastic or inelastic nuclear collisions.
For nuclear beams \emph{of different particle type}, depending on the sign of magnetic field B, either 
lighter or heavier particle bunches will be faster, ``lapping'' the slower bunches periodically,
and enabling ``rear-end'' nuclear fusion events.

Only in such a storage ring can ``rear-end'' collisions occur with heavier particle bunches
passing through lighter particle bunches, or vice versa. 
From a relativistic perspective, treated as point particles, the two configurations 
just described would be indistinguishable.
But, as observed in the laboratory, to the extent the particles are composite, such collisions 
would classically be expected to be quite different or, at least, distinguishable.  

Pavsic, in a 1973 paper reproduced in 2001\cite{Pavsic}, develops a 
``mirror matter'' Hamiltonian formalism, distinguishing between ``external'' and ``internal'' symmetry.  
He points out, for example, that ``the existence of the anomalous proton or neutron magnetic moments indicates 
the asymmetric internal structure of two particles''; a comment that applies directly to the present paper.
Otherwise, Pavsic is agnostic, suggesting that his formalism provides only a parameterization for experiments
sensitive to internal structure, with possible implications concerning mirror matter.

\subsection{Tentative BNL site}\label{sec:BNL-site}\mbox{}

\begin{figure*}
\centering
\includegraphics[scale=0.08]{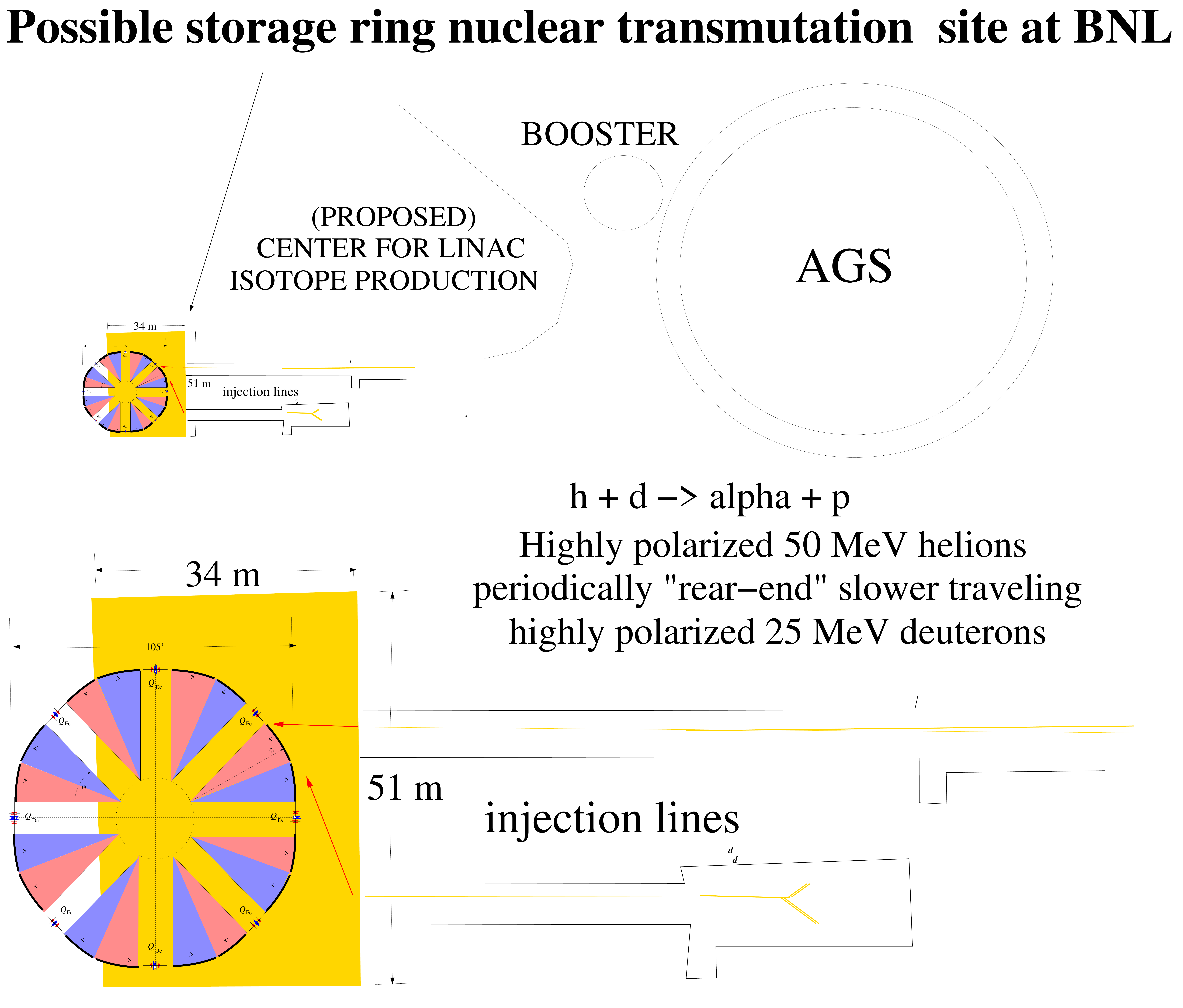}
\caption{\label{fig:Snowmass-PTR-extract-4-BNL}{\bf Above:\ }Tentative PTR location near the AGS at BNL, 
using existing, high current isotope sources. {\bf Below:\ } Magnified image insert of PTR complex.}
\end{figure*}

To indicate the size scale of the proposed PTR ring a tentative site location at Brookhaven National Laboratory (BNL) is 
shown in Figure~\ref{fig:Snowmass-PTR-extract-4-BNL}.  Most of the nuclear isotope beams mentioned in this paper can be made available
at suitable power levels in that vicinity\cite{Deepak}.  What makes this important is that the cost of producing
these beams from the ground up would significantly exceed the PTR fabrication costs shown on the right in Figure~\ref{fig:ElectrodeShappes_cost}
for the PTR ring illustrated in Figures~\ref{fig:RTR-figures} and \ref{fig:ElectrodeShappes_cost}.

\section{Nuclear physics investigation with E\&\lowercase{m} storage ring}\label{sec:Em-rings} \mbox{}

\subsection{``Rear end'' collisions: $h + d \rightarrow \alpha + p$} \mbox{}

``Rear-end'' collisions occurring during the passage of faster bunches through slower bunches can be
used to study spin dependence of nucleon, nucleon collisions in a semi-relativistic moving coordinate frame.
Such rear-end collisions allow the CM KEs to be in the several 100\,KeV range, while all incident and 
scattered particles have convenient laboratory KEs, two orders of magnitude higher, in the tens of MeV range.

This permits incident beams to be established in pure spin states and the polarizations of scattered particles 
to be measured with high analyzing power and high efficiency; Wilkin\cite{Wilkin} 
to Lenisa \cite{KolyaFrankPaulo}.  In this way the E\&m ring satisfies the condition that all nuclear collisions 
take place in a coordinate frame moving at convenient semi-relativistic speed in the laboratory, with CM KEs 
comparable with Coulomb barrier heights.

As a first example, this paper concentrates on $d$ and $h$ beams co-circulating concurrently in the same storage ring, 
with parameters arranged such that, in the process $d+h\rightarrow p+\alpha,$ rear-end collisions always occur 
in a detector at the intersection point (IP).  See Appendix~B.

(In a conventional (magnetic) contra-circulating colliding beam storage ring the energy would be 
above the pion production threshold, with production into this transmutation channel negligibly small.)

Consider $d$ and $h$ beams co-circulating concurrently in the same storage ring, with parameters arranged 
such that, in the process $d+h\rightarrow p+\alpha,$ rear-end collisions always occur in the detector at 
an intersection point (IP).  The center of mass kinetic energies (where their momenta are equal and opposite) 
have been adjusted to be close to the Coulomb barrier height for this nuclear scattering channel.
With judicious adjustment, all nuclear events will occur at the ring intersection point (IP) of a full acceptance 
interaction detector/polarimeter.  Such a device is illustrated schematically in Figure~\ref{fig:dodecahedron-1}\,.

\begin{figure}[h]
\centering
\includegraphics[scale=0.38]{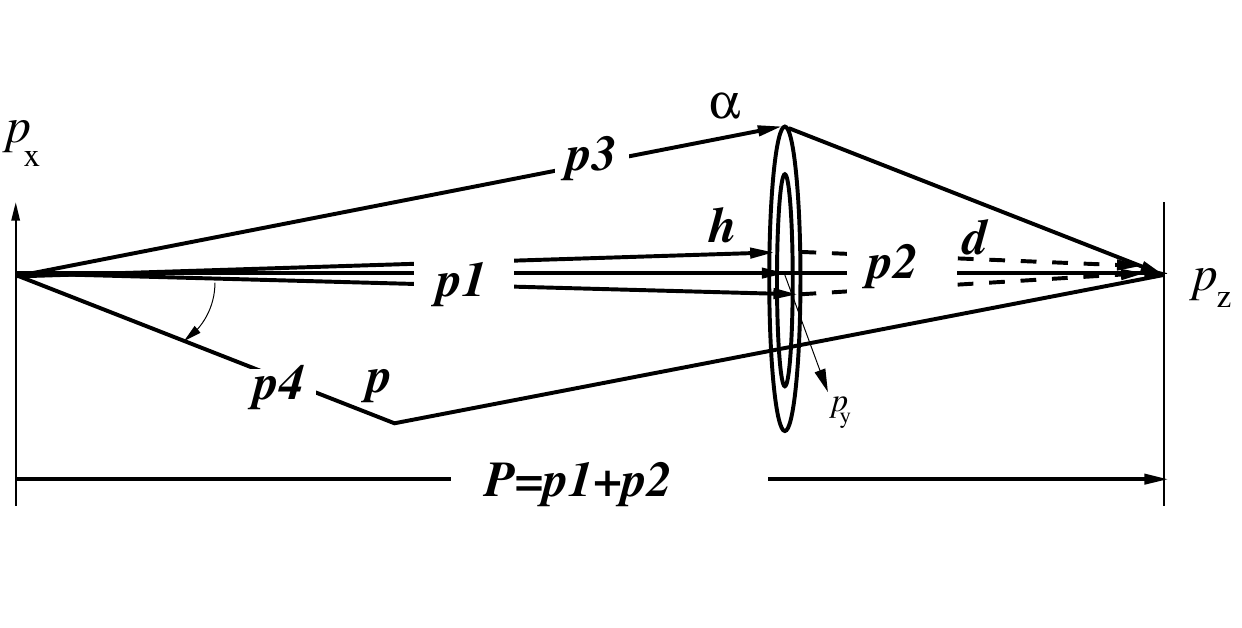}
\caption{\label{fig:h-d-alpha-p-kinematics}Laboratory frame momentum vector diagram.
The vector ${\bf P}={\bf p_1} + {\bf p_2}$ is the sum of the lab momenta of
one particle from beam~1($h$) and one from beam~2($d$).  Scattered alpha particle direction (3) is
shown above the beam axis; the scattered proton direction (4) would then be below, as displayed by parallelogram 
construction.}
\end{figure}

In this configuration the rest mass of the $h,d$ system will be fine-tunable on a KeV scale, for example 
barely exceeding the threshold of the $h+d\rightarrow \alpha+p$ channel, but below pion production and other
inelastic thresholds.  Tentatively neglecting spin dependence, the expected radiation pattern can be described as 
a ``rainbow'' circular ring (or rather cone) formed by the more massive ($\alpha$-particles) emerging from, and centered 
on, the common beam axis.  This ``view'' has not been observed previously in nuclear measurements since it requires 
a ``rear end'' collision.

Consider $d$ and $h$ beams co-circulating concurrently in the same storage ring, with parameters arranged 
such that, in the process $d+h\rightarrow p+\alpha,$ rear-end collisions always occur in the detector at 
an intersection point (IP).  The center of mass kinetic energies (where their momenta are equal and opposite) 
have been adjusted to be close to the Coulomb barrier height for this nuclear scattering channel.
With judicious adjustment, all nuclear events will occur at the ring intersection point (IP) of a full acceptance 
interaction detector/polarimeter.  Such a device is illustrated schematically in Figure~\ref{fig:dodecahedron-1}\,.

In this configuration the rest mass of the $h,d$ system will be fine-tunable on a KeV scale, for example 
barely exceeding the threshold of the $h+d\rightarrow \alpha+p$ channel, but below pion production and other
inelastic thresholds.  Tentatively neglecting spin dependence, the expected radiation pattern can be described as 
a ``rainbow'' circular ring (or rather cone) formed by the more massive ($\alpha$-particles) emerging from, and centered 
on, the common beam axis.  This ``view'' has not been observed previously in nuclear measurements since it requires 
a ``rear end'' collision.

\begin{figure}[hb]
\centering
\includegraphics[width=0.46\textwidth]{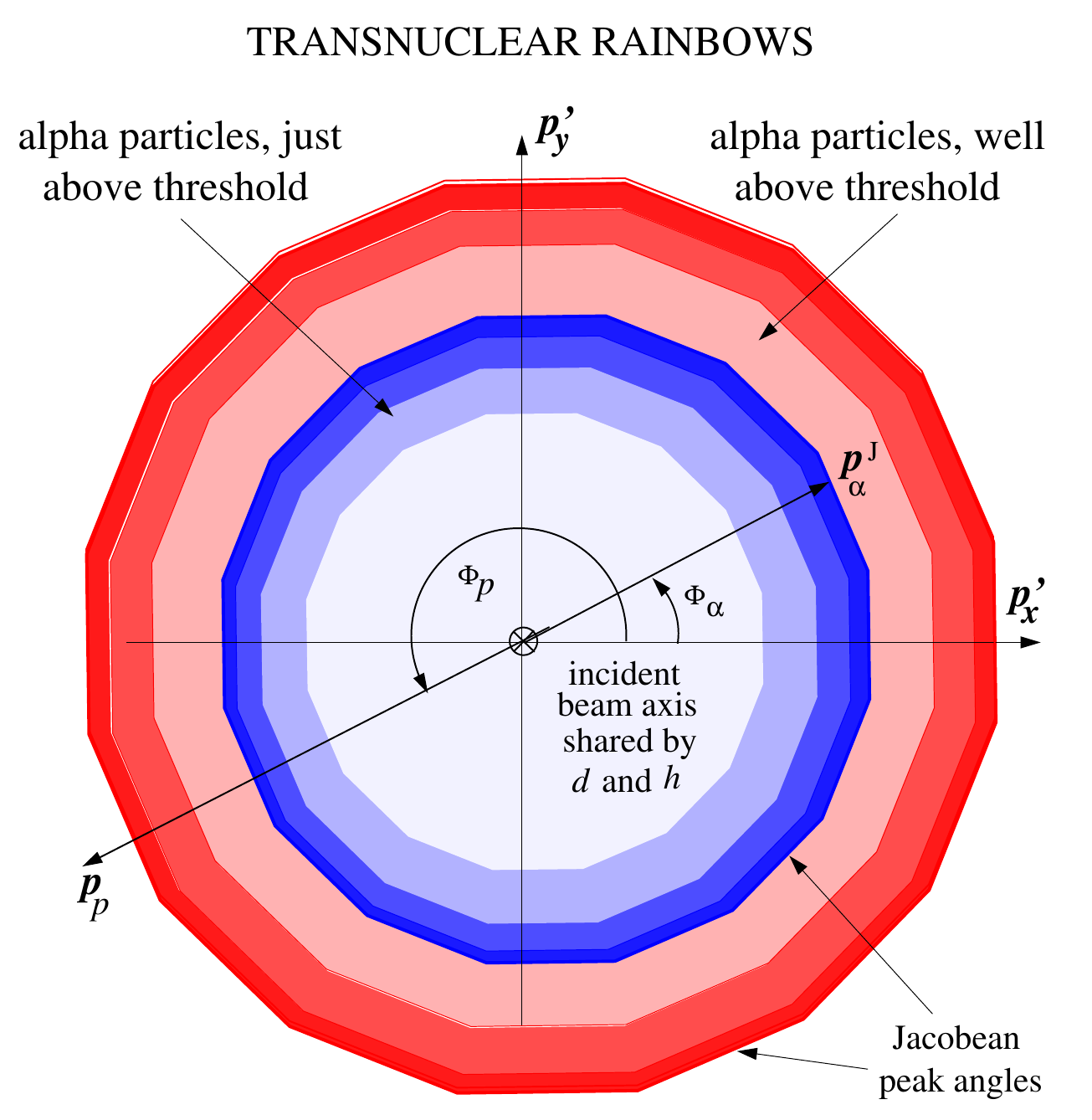}
\caption{
Transnuclear ``rainbows'' produced in the reaction $h+d\rightarrow\alpha +p$.
Shading represents scattered differential cross section.  Rainbow radii increase proportional to incident energy excess 
above threshold.  Superscript ``J'' labels the rainbow divergence edge caused by the vanishing Jacobean at the laboratory 
scattering angle maximum.
Jacobean peak ``rainbow'' production patterns for the nuclear transmutation process 
$h + d \rightarrow \alpha + p$. The $p + d \rightarrow h + \gamma$ channel would exhibit 
only a single, more acute angle $h$ rainbow.  For correlation with, say, terrestrial rainbows, 
refer to Van-de-Hulst\cite{Van-de-Hulst}.  For actual rainbows a further empirical numerical factor, $\pi(m-1)$ of order 1, 
multiplying the diameter, is required, where $m$ is ``index of refraction''.}
\label{fig:Rainbows}
\end{figure}

Consider $d$ and $h$ beams co-circulating concurrently in the same storage ring, with parameters arranged 
such that, in the process $d+h\rightarrow p+\alpha,$ rear-end collisions always occur in the detector at 
an intersection point (IP).
The center of mass kinetic energies (where their momenta are equal and opposite) are close to the 
Coulomb barrier height for this nuclear scattering channel.
With judicious adjustment, all nuclear events will occur at the ring intersection point (IP) of a full acceptance 
interaction detector/polarimeter.

Table-\ref{tbl:BendParams4-hd} provides kinematic parameters for the $h+d\rightarrow \alpha+p$ channel.  The first and last columns
identify incident beams $h$ and $d$ as beams 1 and 2.  Columns 2,3,4 contaim beam~1 parameters;  column 5 gives the electric field,
and column 6 gives the magnetic bending fraction $\eta_M$;  columns 7,8,9 contain beam~2 parameters; the remaining columns
give CM quantities, which are identified by asterisks ``*''. 

The columns labeled $Q_s$ are spin tunes.  In this paper nothing else is said about polarization, but \emph{support for scattering 
highly polarized beam particles with high quality final state polarimetry capability provides the main motivation for the proposed 
E\&m project.} 

The electric/magnetic field ratio produces perfect $\beta_h/\beta_d$=8/7 velocity ratio so that, for every 7 
deuteron turns, the helion makes 8 turns.  Notice, also, the approximate match of Q12=317\,KeV in this table, with Coulomb barrier energy, 
$V_{d,He3}$=313.1\,KeV.  This matches the incident kinetic energy to the value required to surmount the repulsive Coulomb barrier.

\begin{table*}[htb]\scriptsize
\caption{\label{tbl:BendParams4-hd}Fine-grain scan to center the collision point for co-traveling  KE1=49\,MeV helion energy and 24.9\,MeV deuteron energy.
With velocity ratio 8/7, multiplying by 7 produces the central entry in second last column. The bend radius is $r_0=11$\,m.} 
\centering
\begin{tabular}{c|ccc|cc|ccc|cccc|c|c} 
\toprule
  bm & $\beta_1$ & Qs1 &  KE1     &  E0  & $\eta_{M1}$ & $\beta_2$  &  Qs2 &  KE2 & $\beta*$ & $\gamma*$ & $M*$ & Q12 & $7\beta_1/\beta_2$  &   bm  \\ 
   1 &        &        &    MeV   &   MV/m   &            &        &         &    MeV &         &         &    GeV  &    KeV  &         &    2 \\ \midrule
   h & 0.1826 & -0.666 &  48.000  &  4.96139 &  -0.14662  &  0.1597 & -1.097  & 24.391  & 0.17343 & 1.01539 & 4.68432 & 311.21468   & 8.00083 &    d \\ 
   h & 0.1844 & -0.666 &  49.000  &  5.06742 &  -0.14742  &  0.1613 & -1.098  & 24.901  & 0.17519 & 1.01571 & 4.68432 & 317.54605   & 8.00015 &    d \\ 
   h & 0.1862 & -0.666 &  50.000  &  5.17355 &  -0.14822  &  0.1630 & -1.098  & 25.410  & 0.17693 & 1.01603 & 4.68433 & 323.87133   & 7.99947 &    d \\ 
\bottomrule
\end{tabular}
\end{table*}

Temporarily neglecting spin dependence, the expected radiation pattern can be described as 
a ``rainbow'' circular ring (or rather cone) formed by the more massive ($\alpha$-particles) emerging from, and centered 
on, the common beam axis.  This ``view'' has not been observed previously in nuclear measurements since it requires 
a ``rear end'' collision.  The $h + d \rightarrow \alpha + p$ nuclear transmutation channel,
are illustrated as ``rainbows'' in Fig.~\ref{fig:Rainbows}\,.

Table-\ref{tbl:BendParams-h-d} provides kinematic parameters for this $h+d\rightarrow \alpha+p$ channel.  
The columns labeled $Q_s$ are spin tunes.  In this paper little more is said about polarization, but 
\emph{support for scattering highly polarized beam particles with high quality final state polarimetry capability 
provides the main motivation for the proposed E\&m project.} 

The electric/magnetic field ratio produces perfect $\beta_h/\beta_d$=8/7 velocity ratio so that, for every 7 
deuteron turns, the helion makes 8 turns.  
Notice, also, the approximate match of Q12=317\,KeV in this table, with Coulomb barrier energy, 
$V_{d,He3}$=313.1\,KeV.  This matches the incident kinetic energy to the value required to surmount the 
repulsive Coulomb barrier.

\subsection{Rate calculation: $h + d \rightarrow \alpha + p$}\label{sec:Rate-calc}\mbox{}

In this case the $\beta$-ratio is 7/8.  Typical parameters include
\begin{align}
f_{\rm sr}            &= \hbox{ring proton revolution frequency} &= \SI{1e6}{Hz},       \notag\\
N_d, & N_h             = \hbox{numbers of stored particles}       &= 10^{11},      \notag\\
A_b                  &= \hbox{beam area} = \SI{0.1}{cm}\times \SI{0.1}{cm}       &= \SI{1e-2}{cm^2},  \notag\\
\sigma               &= \hbox{nuclear cross section}                                &= \SI{1e-24}{cm^2} \, . \notag
\end{align}

The (deuterium) ``target bunch nuclear opacity'' is
$$O_{\rm N} = N_d\sigma/A_b =  10^{11}\times10^{-24}/10^{-2} = 10^{-11} \, ,  \label{eq:params-1b}$$
which gives the fraction of particle passages that results in a nuclear event. 
The rate of particle passages is  
$$r_{\rm pass} = \frac{f_{\rm sr}}{7}\,N_h = \frac{10^6}{7}\times10^{11} = 0.142\times \SI{1e17}{s^{-1}}.\label{eq:params-1c}$$

The resulting nuclear event rate is
$$r_{\rm event} = O_{\rm N}\times r_{\rm pass} = 10^{-11}\times 0.142\times10^{17} = 1.42\times10^5\,{s^{-1}}\,.$$

Figure~\ref{fig:h-d-alpha-p-kinematics} contains a laboratory frame momentum diagram for the process. Rolled 
around the longitudinal axis, the figure is intended to show how azimuthal symmetry imposes the rainbow 
scattering pattern shown in Figure~\ref{fig:h-d-alpha-p-rainbow}, with cone angle increasing proportional 
to the incident energy excess over threshold energy.

\subsection{``Rainbow'',``rear-end''  $p + d \rightarrow p + d$ collisions}\label{sec:p-d-rainbow}

Here we consider $p + d \rightarrow p + d$ ``elastic'' (including weakly inelastic) scattering in 
the E\&m storage ring.  $p$ and $d$ beams co-circulate concurrently with different velocities in the same ring, 
such that ``rear-end'' collisions always occur at the same intersection point (IP).  {The CM kinetic energies are to 
be varied continuously, keV by keV, from below the several hundred keV Coulomb barrier height}, through the 
(previously inaccessible for spin control) range up to tens of MeV and beyond.
With the scattering occurring in a moving frame, initial and final state laboratory momenta are in the convenient
tens of MeV range.

All nuclear events occur within a full acceptance interaction detector/polarimeter. See Appendix~B.
Temporarily neglecting spin 
dependence, the CM angular distributions will be approximately isotropic~\cite{Fermi}\cite{Hagedorn}.
(Especially with heavier particles being faster) most final state particles end up traveling ``forward'' to 
produce ``rainbow'' circular rings (or rather cones) formed by the final state particles.
(In the absence of ``rear-end'' collisions) this ``view'' has yet to be self evident in nuclear scattering
experiments.

\begin{figure}[hbt]
\centering
\includegraphics[scale=0.40]{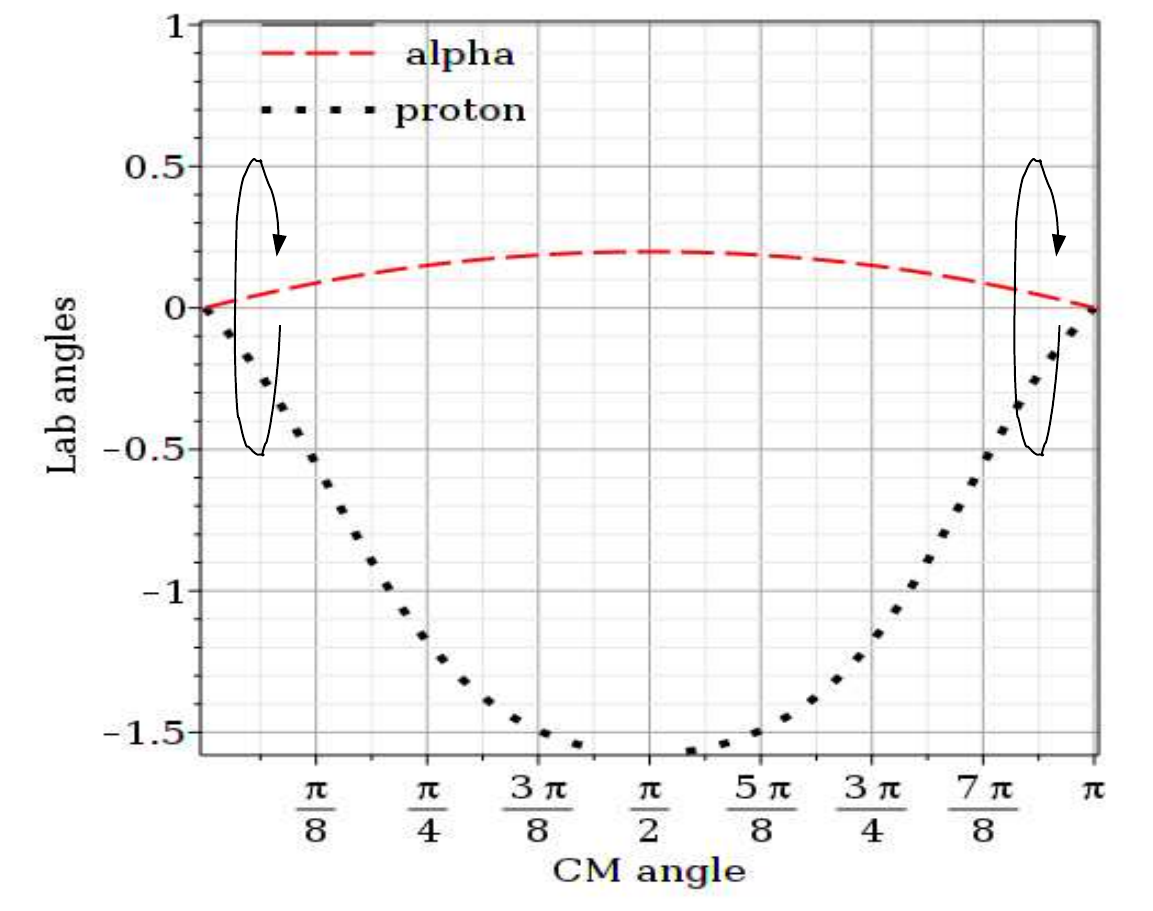}
\caption{\label{fig:CM-plot-CM-angle}Plot of lab angles vs c.m. angle for the $ h + d \rightarrow \alpha + p$ process.
Both incident beams are at (0,0) in this plot, which can be rolled azimuthally, as indicated..}
\end{figure}

\begin{figure}[hbt]
\centering
\includegraphics[scale=0.80]{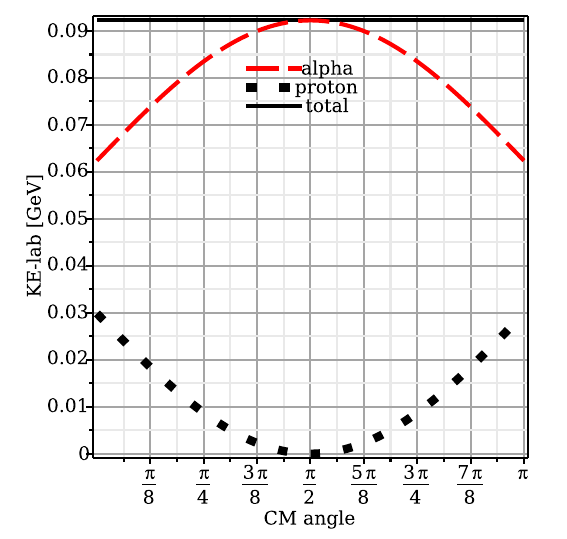}
\caption{\label{fig:CM-plot-CM-KE}Plot of lab KEs vs c.m. angle for the $ h + d \rightarrow \alpha + p$ process.}
\end{figure}

Kinematic parameters for a fine grained scan are shown in Table~\ref{tbl:BendParams4-pd2}\,. Center row, the electric/magnetic field ratio  
produces perfect $\beta_p/\beta_d$=3/2 velocity ratio such that, for every t=2 deuteron turns, the protons make 3 turns.  
The sum of CM kinetic energies in this case is $Q_{12}*=$\SI{3.5659}{MeV}.  
A coarse grained scan is shown in Table~\ref{tbl:p-d-energy-scan}. In the ``rear-end'' collision of (laboratory) \SI{10}{MeV} $p$ bunches 
passing through  \SI{8.419}{MeV} $d$ bunches, the (CM) KE is $Q_{12}*=$\SI{818.5}{keV}.   
With both beams polarized, spin tunes $Q_s$ are given in the tables.
\emph{Little more is said about polarization in this paper, but support for 
scattering highly-polarized beam particles with high-quality final state polarimetry capability provides the main 
motivation for the proposed E\&m storage ring project.}  
In recent years there have been many developments in beam polarization
control and in polarimetric spin orientation detection, many of which have been produced at the COSY storage ring in J\"ulich, 
Germany~\cite{Wilkin,Eversmann,Hempelmann,Rathmann-Kolya-Slim,Slim-Rathmann,RathmannWienFilter}. 
\setlength{\tabcolsep}{1pt}

\begin{table*}[htb]  \scriptsize
\caption{\label{tbl:BendParams-h-d}Fine-grain scan to center the collision point for co-traveling  KE1=49\,MeV helion energy and 24.9\,MeV deuteron energy.
The first and last columns identify incident beams $h$ and $d$ as beams 1 and 2.  Columns 2,3,4 contain beam~1 
parameters;  column 5 gives the electric field, and column 6 gives the magnetic bending fraction $\eta_M$;  
columns 7,8,9 contain beam~2 parameters; the remaining columns
give CM quantities, which are identified by asterisks ``*''. Q12* is the sum of kinetic energies in the CM frame.  In the next to last column,
with velocity ratio 8/7, multiplying by 7 produces the central entry in second last column. The bend radius is $r_0=11$\,m.} 
\centering
\begin{tabular}{c|ccc|cc|ccc|cccc|c|c} 
\toprule 
  bm & $\beta_1$ & Qs1 &  KE1     &  E0  & $\eta_{M1}$ & $\beta_2$  &  Qs2 &  KE2 & $\beta*$ & $\gamma*$ & $M*$ & Q12* & $7\beta_1/\beta_2$  &   bm  \\ 
   1 &        &        &    MeV   &   MV/m   &            &        &         &    MeV &         &         &    GeV  &    KeV  &         &    2 \\ \midrule
   h & 0.1826 & -0.666 &  48.000  &  4.96139 &  -0.14662  &  0.1597 & -1.097  & 24.391  & 0.17343 & 1.01539 & 4.68432 & 311.21468   & 8.00083 &    d \\ 
   h & 0.1844 & -0.666 &  49.000  &  5.06742 &  -0.14742  &  0.1613 & -1.098  & 24.901  & 0.17519 & 1.01571 & 4.68432 & 317.54605   & 8.00015 &    d \\ 
   h & 0.1862 & -0.666 &  50.000  &  5.17355 &  -0.14822  &  0.1630 & -1.098  & 25.410  & 0.17693 & 1.01603 & 4.68433 & 323.87133   & 7.99947 &    d \\ 
\bottomrule
\end{tabular}
\end{table*}

\begin{table*}[htb]\scriptsize
\caption{\label{tbl:BendParams4-pd2}Fine-grain scan to center the collision point for co-traveling beams: (lab) KE1=\SI{45.290}{MeV} protons 
and \SI{38.665}{MeV} deuterons. Beam~1(labeled in first column) parameters are shown on the left, followed by $E_0$ and (fractional) M value $\eta_M$,
then Beam~2 (labeled in final column) parameters and, finally, CM parameters indicated by asterisks $^\ast$. The columns labeled $Q_s$ are spin tunes. $Q^\ast_{12}$ is the sum of initial state kinetic energies in the CM system. In this fine-grained scan, the cryptic, comma-separated, $t, t^\ast \beta_1/\beta_2$ column heading, $t$ labels the entry in the top row (namely 2) which, multiplying a value (for $\beta_1/\beta_2$) close to 3/2 yields 3 as exactly as necessary to enable feedback stabilization of this condition. Entries in Table~\ref{tbl:p-d-energy-scan} have not been pre-tuned to the same degree.}
\setlength{\tabcolsep}{3pt}
\centering
\begin{tabular}{c|ccc|cc|ccc|cccc|c|c} 
\toprule 
  bm   & $\beta_1$ &    $Q_{s1}$ &    KE1   &     $E_0$   & $\eta_{M_1}$ & $\beta_2$ & $Q_{s2}$ & KE2 & $\beta^\ast$ & $\gamma^\ast$ &  M$^\ast$ & $Q^\ast_{12}$ & $t, t^\ast \beta_1/\beta_2$ & bm  \\
   1 &        &        &    MeV   &   MV/m   &            &        &          &    MeV &         &         &    GeV  &    keV      &      2  &    2 \\ \midrule
   p & 0.2996 &  0.294 &  45.190  &  4.77556 &  0.40511  &  0.1998 & -0.723  & 38.578  & 0.23366 & 1.02847 & 2.81744 & 3558.4   & 3.00019 &    d \\ 
   p & 0.3000 &  0.294 &  45.290  &  4.78686 &  0.40499  &  0.2000 & -0.724  & 38.665  & 0.23391 & 1.02853 & 2.81745 & 3565.9   & 3.00001 &    d \\ 
   p & 0.3003 &  0.294 &  45.390  &  4.79817 &  0.40487  &  0.2002 & -0.724  & 38.751  & 0.23416 & 1.02860 & 2.81746 & 3573.4   & 2.99983 &    d \\  
\bottomrule
\end{tabular}
\end{table*}
\begin{table*}[htb]\scriptsize
\caption{\label{tbl:p-d-energy-scan}Coarse-grained $p,d$ scattering energy scan. Entries in the second to last column have not been pre-tuned to the 
same precision as in Table~\ref{tbl:BendParams4-pd2}.}
\setlength{\tabcolsep}{3pt}
\centering
\begin{tabular}{c|ccc|cc|ccc|cccc|c|c} 
\toprule 
  bm   & $\beta_1$ &    $Q_{s1}$ &    KE1   &     $E_0$   & $\eta_{M_1}$ & $\beta_2$ & $Q_{s2}$ & KE2 & $\beta^\ast$ & $\gamma^\ast$ &  M$^\ast$ & $Q^\ast_{12}$ & $t, t^\ast \beta_1/\beta_2$ & bm  \\
   1 &        &        &    MeV   &   MV/m   &            &        &          &    MeV &         &         &    GeV  &    keV      &      2  &    2 \\ \midrule
   p & 0.1448 &  0.284 &  10.000  &  1.00030 &  0.44692  &  0.0944 & -0.702  &  8.419  & 0.11131 & 1.00625 & 2.81470 &  818.5   & 3.06776 &    d \\ 
   p & 0.2032 &  0.287 &  20.000  &  2.03242 &  0.43519  &  0.1334 & -0.708  & 16.906  & 0.15685 & 1.01253 & 2.81550 & 1618.9   & 3.04789 &    d \\ 
   p & 0.2470 &   0.29 &  30.000  &  3.09668 &  0.42334  &  0.1631 & -0.714  & 25.459  & 0.19142 & 1.01884 & 2.81629 & 2401.7   & 3.02856 &    d \\ 
   p & 0.2830 &  0.293 &  40.000  &  4.19343 &  0.41137  &  0.1881 & -0.720  & 34.079  & 0.22024 & 1.02517 & 2.81705 & 3167.5   & 3.00976 &    d \\ 
   p & 0.3140 &  0.296 &  50.000  &  5.32300 &  0.39927  &  0.2100 & -0.726  & 42.763  & 0.24535 & 1.03153 & 2.81780 & 3916.7   & 2.99145 &    d \\ 
   p & 0.3415 &  0.299 &  60.000  &  6.48572 &  0.38706  &  0.2297 & -0.732  & 51.510  & 0.26781 & 1.03791 & 2.81853 & 4649.7   & 2.97362 &    d \\  \bottomrule
\end{tabular}
\end{table*}

\section{Positron induced tritium two-body $\beta$-decay}\mbox{}
\setlabel{Positron induced helion two-body $\beta$-decay}{sec:Positron-induced}

This section might, just as logically, have been included as the final subsection of the preceding section.  
A separate section has been introduced in order to stress the fundamental importance and distinction of weak 
interaction investigation. This does not imply that the experimental methods already explained will need to
be greatly altered, though with significantly different particle detection/polarimetry.

A better reason for setting apart positron induced tritium two-body $\beta$-decay will be obvious from the following
``Natural tritium reference $\beta$-decay events'' section, which discusses the role to be played by ``background'' tritium reference 
decay events.

\begin{figure}
\centering
\includegraphics[scale=0.60]{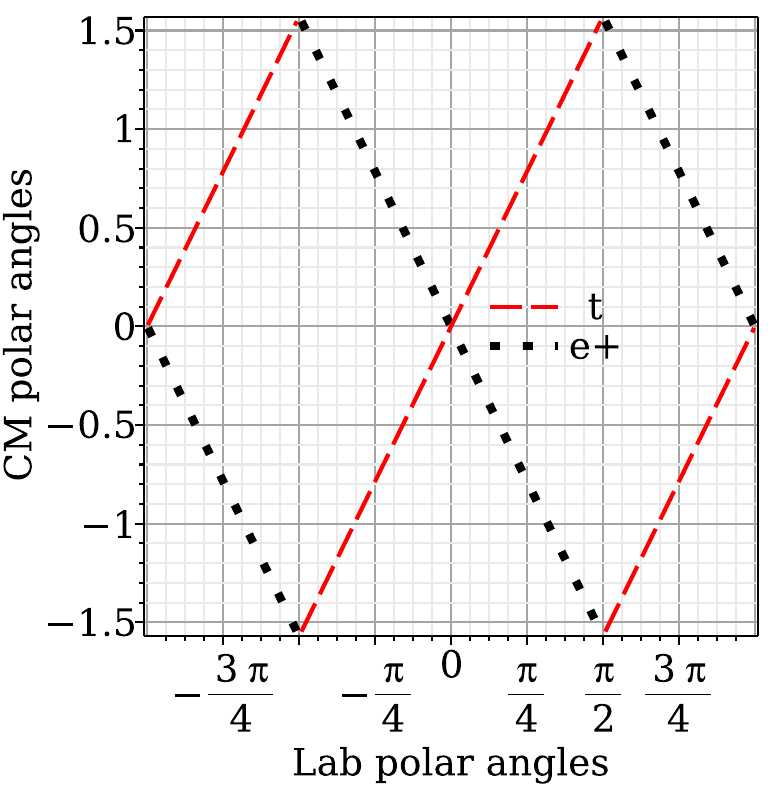}
\caption{\label{fig:Induced_beta_decay1}Plot of CM polar angles vs initial lab polar angles, for the $e^+ + {\rm triton}$ initial state channel.
Both incident beams are at (0,0) in this plot. This plot is annoying but essential.  Annoying because it is confusing and shows no physical data, 
essential because it provides the MAPLE representation of the transformation relating two different coordinate systems. This accounts for the 
curious appearance of figures such as Figure~\ref{fig:Induced-beta-decay2} .}
\end{figure}

\begin{figure}
\centering
\includegraphics[scale=0.60]{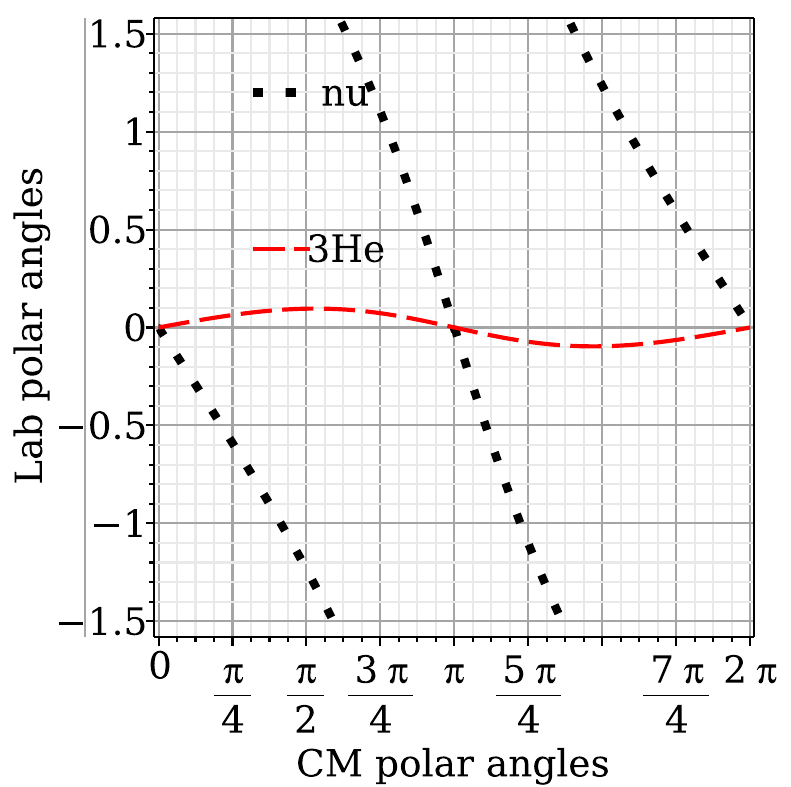}
\caption{\label{fig:Induced-beta-decay2}Plot of lab polar angles vs CM polar angles for the $e^+ + {\rm triton}$ final state ${\rm 3He} + \nu$ channel.  
Notice that, viewed in the laboratory, the produced helions are reasonably well collimated, while the (invisible) scattered neutrinos 
are more or less isotropic. In the legend ``nu'' stands for $\nu$.}
\end{figure}

\begin{figure}
\centering
\includegraphics[scale=0.60]{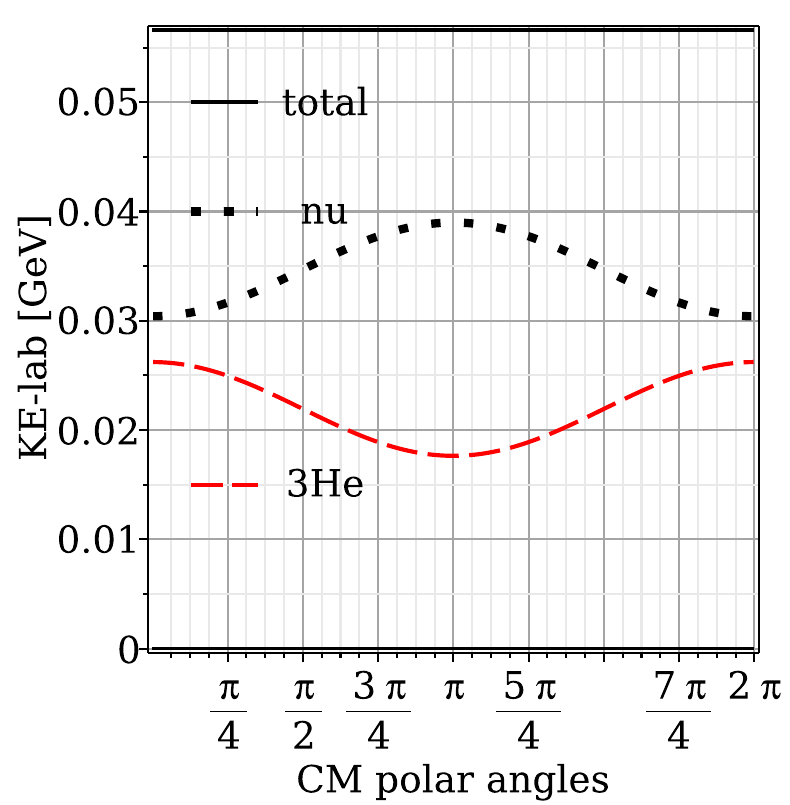}
\caption{\label{fig:Induced_beta_decay3}Plot of final state kinetic energies vs CM polar angles 
for the $e^+ + {\rm triton}$ final state ${\rm 3He} + \nu$ channel. In the legend ``nu'' stands for $\nu$. }
\end{figure}

Figures~\ref{fig:Induced_beta_decay1} and \ref{fig:Induced-beta-decay2} illustrate the kinematics of the 
$\lowercase{t}$ +\ $e^+\ \rightarrow$ $\lowercase{h}$ + $\nu$ channel in the form of graphs relating initial state CM polar angles to 
lab polar angles and final state lab polar angles to CM polar angles.  Figure~\ref{fig:Induced_beta_decay3} plots 
final state kinetic energies vs CM polar angles.  

Tables~\ref{tbl:h-t-co-and-contra} and \ref{tbl:h-t-energy-scan} provide lab parameters for helions and tritons following identical orbits.
with $h$ as master (beam 1) on the left and $t$ on the right, and shared field values in the middle.  Note, in this case, because of almost 
identical masses, but charges differing by a factor of two, that the magnetic bending is strong and destructive in both cases.

Because the process is exothermic, the sum of scattered kinetic energies, as well as being independent of scattering 
angle, exceeds the sum of initial state laboratory kinetic energies.  However, the helion and neutrino shares of lab kinetic energy 
are comparable, within a factor of two, as Figure~\ref{fig:Induced_beta_decay3} shows.

In reconstructing the two-body kinematics, the only significant theoretical unknown is the neutrino rest mass.  To 
the extent that helion direction and momentum are perfectly measured, the neutrino rest mass can be inferred.

An alternative way of assessing the impact of this statement, based on the already experimentally established upper limit 
of order 1\,eV for the neutrino mass, is that, to all intents and purposes other than for measuring the neutrino mass,
the kinematics satisfies, exactly, two-body kinematics with one of the final state particles mass-less.

For visualizing the angular distribution of scattered particles, one should roll both curves in Figure~\ref{fig:Induced-beta-decay2} 
around the longitudinal axis, anticipating cylindrical symmetry, at least to the extent that spin dependence can be neglected.

Unlike $p,p$ scattering, because the positron mass is so small, the Rutherford scattering should not be important,
even in near-forward directions. As a consequence most of the scattered helion foreground will be essentially free of background 
radiation.  On the other hand, in spite of the quite low positron kinetic energy, the determination of neutrino rest mass may
require radiate correction. 

Except at very small angles the signature of scattered helions should permit the foreground to be easily distinguishable from the 
background. Note though, that the largest helion laboratory angle will never exceed a quit small angle of order 15 degrees.  

\begin{table*}[htb]\scriptsize
\caption{\label{tbl:h-t-co-and-contra}Co- and contra-rotating closed orbit solutions, distinguished respectfully by the sign of beta2 in the two rows, 
with matching $h$ and $t$ beam parameters.  Notice, unusually in this table, that the sign of $\beta_2$ is opposite in the two rows. }
\centering
\begin{tabular}{c|cccccc|cc|ccccccc|c} 
\toprule 
     bm &     m1 &     G1 & q1 & $\beta_1$ &    Qs1     &    KE1   &     E0   &     etaM1 &     m2 &   G2   & q2 & $\beta_2$ &    KE2   &  bratio  &    Qs2     & bm \\ 
      1 &    GeV &        &    &        &            &    MeV   &   MV/m   &           &    GeV &        &    &          &    MeV   &          &            & 2  \\ \midrule
     h & 2.8084 & -4.1842 &  2 & 0.15196 & 4.233e-01 &  33.0000 &  4.4131  &  -0.4796  & 2.8089 & 7.9150 &  1 & -0.15020 &  32.2316 &  -0.9884 & 1.331e+00  & t  \\ \midrule
     h & 2.8084 & -4.1842 &  2 & 0.15196 & 4.233e-01 &  33.0000 &  4.4131  &  -0.4796  & 2.8089 & 7.9150 &  1 &  0.11399 &  18.4302 &  0.75013 & -3.736e+00 &  t \\
\bottomrule
\end{tabular}
\end{table*}
\begin{table*}[htb]\scriptsize
\caption{\label{tbl:h-t-energy-scan}Fine-grained $h$ and $t$ energy scans. As always, the bending radius is 11\,{\rm m}.}
\centering
\begin{tabular}{c|ccc|cc|ccc|cccc|c|c} 
\toprule 
  bm   &  beta1 &    Qs1 &    KE1   &     E0   &   etaM1 &    beta2  &    Qs2 &    KE2 &  beta*   & gamma*  &     M*  &   Q12*  &  t,t*bratio  &   bm  \\ 
   1 &        &        &    MeV   &   MV/m   &            &        &          &    MeV &         &         &    GeV  &    KeV      &    3     &    2 \\ \midrule
   h & 0.1443 &  0.423 &  29.700  &  3.96487 &  -0.47620  &  0.1082 & -3.728  & 16.582  & 0.12628 & 1.00807 & 5.61826 & 945.86939   & 4.00148 &    t \\ 
   h & 0.1467 &  0.423 &  30.700  &  4.10054 &  -0.47724  &  0.1100 & -3.730  & 17.142  & 0.12836 & 1.00834 & 5.61829 & 977.28706   & 4.00081 &    t \\ 
   h & 0.1490 &  0.423 &  31.700  &  4.23635 &  -0.47828  &  0.1117 & -3.733  & 17.702  & 0.13041 & 1.00861 & 5.61832 & 1008.67706   & 4.00015 &    t \\ 
   h & 0.1513 &  0.423 &  32.700  &  4.37230 &  -0.47932  &  0.1135 & -3.735  & 18.262  & 0.13243 & 1.00889 & 5.61835 & 1040.03943   & 3.99948 &    t \\ 
   h & 0.1535 &  0.423 &  33.700  &  4.50839 &  -0.48036  &  0.1152 & -3.737  & 18.822  & 0.13441 & 1.00916 & 5.61838 & 1071.37419   & 3.99882 &    t \\ 
\bottomrule
\end{tabular}
\end{table*}

\subsection{Natural tritium beta-decay reference events}\label{sec:Reference-decays}\mbox{}

Associated with a circulating tritium beam during positron-induced tritium $\beta$-decay data collection
there will be a naturally occurring ``background'' of ordinary tritium $\beta$-decays . Even though these decays
are distributed uniformly around the ring, they will provide valuable ``reference'' events at a conveniently
high data rate. 

The half-life of tritium is 12.33 years.  For a beam of, say, $N_T=10^{10}$ tritons, there will be about 18 decays
per second, uniformly distributed around the ring. The decay distributions in direction and energy of foreground and 
the natural decay events will not be very different, especially if all events emanate from the same origin which, for 
simplicity, we temporarily assume.

The difference between foreground and background distributions can be simplified by sequential, perturbatively justified
transformations, based first on the $0.511/2808\approx0.00018$ ratio of electron and triton (or helion) rest masses,
and next on the $10^{-6}/0.511$ ratio of present day approximate neutrino mass upper limit\cite{Otten-Weinheimer} and the electron mass. 
Though two body and three body distributions are very different for the light particles, they are not very different 
for the heavy final state helions.

The dominant randomly distributed variables are the azimuthal angles of the final state helions.  An analytically
exact roll transformation around the common beam axis of all background measurements, can eliminate the
difference of these azimuthal variables without biasing any dependence on the unknown neutrino mass that is statistically 
implied by the measurements under analysis.

One can artificially treat, as two-body, the three-body natural $\beta$-decay of the background events by treating the 
$e$ and $\nu$ particles as a single $e+\nu$ system which, along with the final state $h$ particle, are radiated 
isotropically in the CM system.  This too will not bias any dependence on the unknown neutrino mass that 
is statistically implied by the measurements under analysis.

Any implicit dependence on the neutrino mass can make its presence known only by its influence on the
final state distribution resulting from the decay of the $e+\nu$ system, also assumed to be isotropic.
This distribution is fixed by momentum conservation, which depends explicitly on the neutrino mass.
The two body kinematics in induced beta decays is not at all conducive to the experimental determination 
of the neutrino mass, since the only measurable final state kinematic parameters are carried by the high mass 
nuclear isotope.
 
\subsection{Polarimetry of natural $\beta$ decay produced helions}\label{sec:Helion-polarimetry}\mbox{}

During initial investigation of induced tritium $\beta$-decay it will not be possible to measure the
energy of the produced helions with the precision necessary for the precise determination of the neutrino mass.
It will, however, be possible to measure the spin dependence of this weak interaction channel.  As explained in
previous sections, the initial spin states can be pure and the final spin polarizations measured with high
efficiency and high analyzing power, as described in Appendix~B.        

\begin{table*}[htb]\scriptsize
\caption{\label{tbl:GOLDEN-1}Field strengths, kinematic data, and spin tunes for the initial state of
the reaction $e^+ + t \rightarrow h + \nu$. }
\setlength{\tabcolsep}{3pt}
\centering
\begin{tabular}{c|cccccc|cc|ccccccc|c} 
\toprule 
     bm &     m1 &     G1 & q1 &  beta1 &    Qs1     &    KE1   &     E0   &      etaM &     m2 &     G2   & q2 &    beta2 &    KE2 &  bratio   &    Qs2 &     bm \\
      1 &    GeV &        &    &        &            &    MeV   &   MV/m   &           &    GeV &          &    &          &    MeV &           &        \\ \midrule
     t & 2.8089 & 7.9150 &  1 & 0.11174 & -7.965e-01 &  17.7020 &  3.1730  &  0.3170  & 0.0005 & 0.0012 &  1 & 0.99991 &  37.8856 &   8.9485 & 7.502e-02 &    pos \\ 
\bottomrule
\end{tabular}
\end{table*}
\begin{table*}[htb]\scriptsize
\caption{\label{tbl:GOLDEN-3}Field strengths, kinematic data, and spin tunes for the reaction $e^+ + i \rightarrow h + \nu$. 
In the legend ``nu'' stands for $\nu$.  }
\setlength{\tabcolsep}{3pt}
\centering
\begin{tabular}{c|ccc|cc|ccc|cccc|c|c} 
\toprule 
  bm &  beta1 &    Qs1 &    KE1   &     E0   &   etaM   &  beta2 &  Qs2   &    KE2 &  beta*  & gamma*  &     M*  &   Q12*      & 9*bratio &   bm  \\ 
   1 &        &        &    MeV   &   MV/m   &          &        &        &    MeV &         &         &    MeV  &    KeV      &          &    2 \\ \midrule
   t & 0.1108 & -0.796 &  17.400  &  3.11842 &  0.01125 & 0.9999 &  0.074 & 37.321 & 0.12254 & 1.00759 & 2.84257 & 33136.76939 & 0.99722  &  pos \\ 
   t & 0.1111 & -0.796 &  17.500  &  3.13650 &  0.01118 & 0.9999 &  0.074 & 37.508 & 0.12290 & 1.00764 & 2.84272 & 33291.47709 & 1.00006  &  pos \\ 
   t & 0.1114 & -0.796 &  17.600  &  3.15460 &  0.01111 & 0.9999 &  0.075 & 37.695 & 0.12327 & 1.00769 & 2.84288 & 33445.98681 & 1.00288  &  pos \\ 
\bottomrule
\end{tabular}
\end{table*}
\begin{table*}[htb]\scriptsize
\caption{\label{tbl:GOLDEN-2}Field strengths, kinematic data, and spin tunes for the final state of the reaction $e^+ + t \rightarrow h + \nu$. }
\setlength{\tabcolsep}{3pt}
\centering
\begin{tabular}{c|ccc|ccc|cc|cccc|c} 
\toprule 
     bm &    m3 &   G3    & q3 & th3-90 & th4-90  &  KE3  &  E0    &  etaM  &   KE4   &  m4    &  G4    & q4 & bm   \\ 
     3 &    GeV &         &    &        &         &  MeV  & MV/m   &        &    GeV  &        &        &    &  4  \\ \midrule
     h & 2.8084 & -4.1834 &  2 &   8.64 & -111.14 & 21.93 & 3.1730 & 0.0110 & 34.6981 & 0.0000 & 0.0000 &  0 & nu  \\
\bottomrule
\end{tabular}
\end{table*}

 \section{Electron induced triton $\beta$-reincarnation}\mbox{}
\setlabel{Electron induced triton $\beta$-reincarnation}{sec:Electron-induced}

For various reasons, there is another permutation of the input and output weakly interacting particles that
seems more favorable experimentally; namely  $e^- + h \rightarrow t +\nu$.  This channel could be referred to as 
\emph{electron induced triton two-body $\beta$-reincarnation}. In traditional nuclear physics terminology the
process would be referred to as \emph{electron capture} (EC) or \emph{internal conversion}.  

As explained in Appendix~F, the (boxed) quartic Eq.~(\ref{eq:AbbrevFieldStrengths.3-rev}) is satisfied 
by the design orbit of every circular storage ring with arbitrarily superimposed E\&M bending.  

One might suppose that a predominantly magnetic ring, like an $e^+/e^-$ collider, would be needed to store 
positive and negative beams at the same time.  This is not correct in our case, however.  For one thing, to obtain
rear-end collisions both beams must travel in the same direction.  Also, though inconvenient, what makes it possible 
with predominantly electric bending, in our case, is that electrons are three orders of magnitude lighter than 
helions.   

For the experiment to work, a negative (electron) beam and a positive (He3) beam have to circulate in the same direction.  
This means the magnetic field bending has to be destructive in its effect on the He3 beam.  So, the electric field 
has to be stronger than it would be with no magnetic field.  Because of its charge being +2, the He3 has a two-fold 
advantage as regards the electric bending force, and it is also slow, not very responsive to the wrong sign magnetic 
field.

The magnetic field is constructive in its effect on the electron beam.  Fortunately, the electron velocities are 
four times greater than the helion velocities.  As a result, the (constructive) magnetic force on the electrons is 
two times greater than is the (destructive) magnetic force on the helions.
 
The net effect is that the bending is centripetal in both cases, even though it is destructive in both cases 
as regards the coherency of electric and magnetic contributions. 

The He3 bending remains, therefore, predominantly electric, but stronger than would be needed to store just the helions.
It may be, therefore, that the ring circumference needs to be increased from its 102.5\,m current value, but that the
current superposition design of Figure~\ref{fig:SectorPerspective} may remain unchanged.

In the context of the present paper, in contrast with the \emph{status quo,} the primary distinction is that 
the initial electron would be under direct experimental control, both in kinetic energy and spin orientation.
This causes initial state visualization to be more particle-like than wave-like. 

There are other considerations, mainly concerning event rate, that makes the $e^- + h \rightarrow t +\nu$ channel 
more attractive than the $e^+ + t \rightarrow h +\nu$ channel.  One is that intense polarized helion beams are 
already available---for example at BNL\cite{Zelenski-He3}. Another is that electron beams are easy, positron beams 
are difficult.

\subsection{The importance of unitarity}\mbox{}

Table~\ref{tbl:h-t-energy-scan} showed two examples of simultaneously-stored triton and helion beams.
For helion kinetic energies roughly twice as great as tritium energies these isotopes can co-circulate
compatibly---with velocities related by an integer ratio as desired.

Here we contemplate the possibility of electrons and helions co-circulating with velocities that are in integer 
ratio.  As before, the purpose is to reduce the $e^-$ and $h$ CM kinetic energies so as to enable rear-end
collisions in a moving frame of reference.

Now, however, since there is no Coulomb barrier to overcome, the motivation is different.  In this case the Coulomb force is 
attractive.  Presumably, this is helpful for the collision cross section.  It may also relax the ring space charge acceptance 
limitation.

A strong motivation for enabling rear-end collisions is to influence the detectable weak interaction rate by suppressing
competing inelastic channels other than electron capture (EC).  

This may seem to be counter-productive.  As well as being very small, neutrino cross sections
are proportional to the laboratory neutrino energy. Zuber's ``Neutrino Physics'' book,\cite{Zuber} provides total 
neutrino cross sections;
\begin{align}
\sigma(\nu N)     &= (0.677 \pm 0.014) \times 10^{-38}\,{\rm cm}^2 \times E_{\nu}/({\rm GeV})\notag\\
\sigma(\bar\nu N) &= (0.334 \pm 0.008) \times 10^{-38}\,{\rm cm}^2 \times E_{\nu}/({\rm GeV})
\label{eq:neutrino-total-cross-sections}
\end{align}
where $N$ stands for $p$ or $n$ or their average.  

Surely one wants the neutrino energies to be as large as possible, consistent with the maximum achievable electric
field (which is inversely proportional to the ring bending radius)?  

This reasoning is fallacious, however.  During the compound nucleus phase of nuclear collisions there is a competition
(to escape) amongst the various possible final state exit channels.  By unitarity, only one of the possible 
exit channels can be responsible for any given detectable event.  In this competition, low probability channels are 
strongly disadvantaged. 

The only situation in which (low probability) electron capture has an advantage is when most or all other (high probability)
inelastic channels are forbidden by energy conservation.  This condition can be met for weak interactions by enabling only 
rear-end collisions in the $e^- + h \rightarrow t +\nu$ channel, for which competing channels are forbidden by energy conservation. 

As with predominantly magnetic rings, head-on collisions would have large CM energies which would enable large cross section 
inelastic collisions which will suppress weak scattering elastic channels of lower energy.  In other words ``for CM energy, 
higher is not always better''. 

Nowadays one rarely hears accelerators referred to as ``atom smashers'', a term that was in common usage in spite 
of always being somewhat misleading.  The term ``nucleus smasher'' has never caught on, even though, at multi-GeV levels 
it would be quite apt.  What is being proposed here could then be called a ``nucleus tickler''.




%

\begin{figure}
\centering
\includegraphics[scale=0.60]{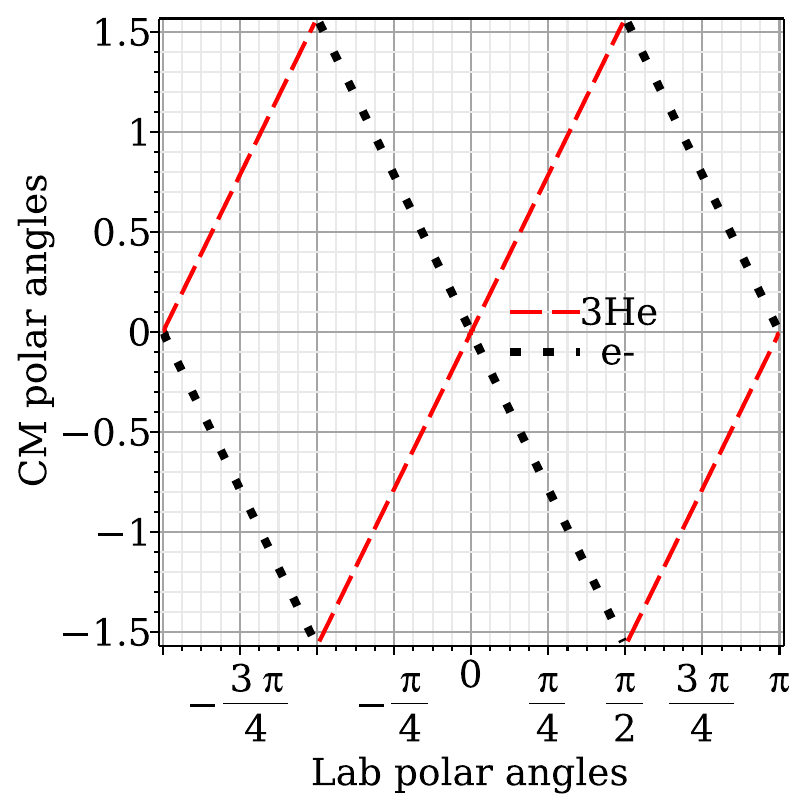}
\caption{\label{fig:3He-e_CM_angs_vs_Lab_angs}Plot of CM polar angles vs initial lab polar angles, for the initial state of
the reaction $e^- + {\rm 3He} \rightarrow {\rm triton} + \nu$. Both incident beams are at (0,0) in this plot. }
\end{figure}
\begin{figure}
\centering
\includegraphics[scale=0.60]{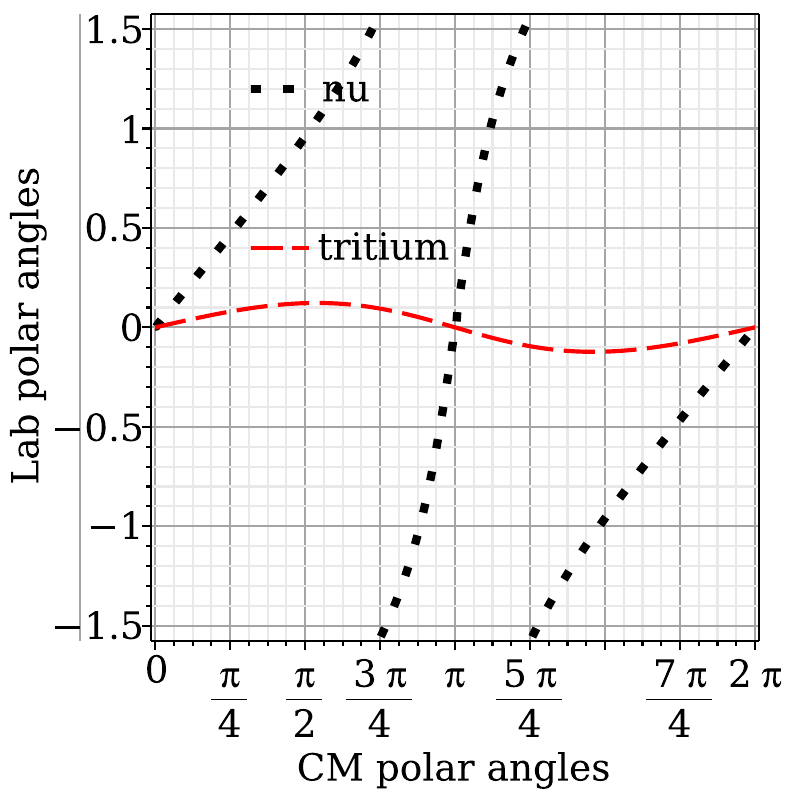  }
\caption{\label{fig:3He-e_Lab_Lab_angs_vs_CM_angle}Plot of lab polar angles vs CM polar angles, for the final state of
the reaction $e^- + {\rm 3He} \rightarrow {\rm triton} + \nu$.  Both incident beams are at (0,0) in this plot. 
Notice that, viewed in the laboratory, the produced helions are reasonably well collimated, while the (invisible) 
scattered neutrinos are more or less isotropic. In the legend ``nu'' stands for $\nu$.}
\end{figure}
\begin{figure}
\centering
\includegraphics[scale=0.60]{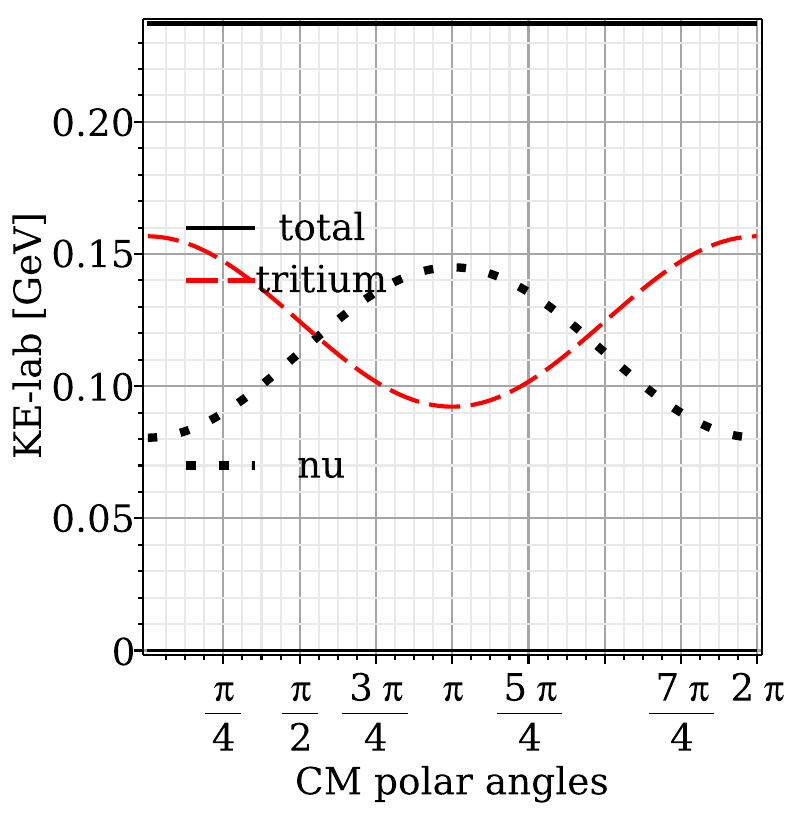}
\caption{\label{fig:3He-e_KE_lab_vs_CM_angle}Plot of final state kinetic energies vs CM polar angles, for the 
$e^- + {\rm 3He} \rightarrow {\rm triton} + \nu$. In the legend ``nu'' stands for $\nu$.}
\end{figure}
\begin{table*}[htb]\scriptsize
\caption{\label{tbl:GOLDEN-EC-1}Field strengths, kinematic data, and spin tunes for the initial state of the reaction $e^- + h \rightarrow t + \nu$.}
\setlength{\tabcolsep}{3pt}
\centering
\begin{tabular}{c|cccccc|cc|ccccccc|c} 
\toprule 
 bm &     m1 &      G1 & q1 &  beta1  &    Qs1     &    KE1   &     E0   &   etaM  &   m2   &    G2  & q2 &  beta2  &   KE2    &   Qs2   & bratio   & bm \\
  1 &    GeV &         &    &         &            &    MeV   &   MV/m   &         &  GeV   &        &    &         &   MeV    &         &          & 2  \\ \midrule
  h & 2.8084 & -4.1834 & 2  & 0.25000 & -8.641e-01 &  92.1000 &  9.2400  & -0.1214 & 0.0005 & 0.0012 & -1 & 0.99999 & 145.1413 & 0.32435 & 16.00013 & e  \\
\bottomrule
\end{tabular}
\end{table*}
\begin{table*}[htb]\scriptsize
\caption{\label{tbl:GOLDEN-EC-2}Field strengths, kinematic data, and spin tunes for the reaction $e^- + h \rightarrow t + \nu$. 
In the legend ``nu'' stands for $\nu$.   }
\setlength{\tabcolsep}{3pt}
\centering
\begin{tabular}{c|ccc|cc|ccc|cccc|c|c} 
\toprule 
  bm &  beta1 &    Qs1 &   KE1  &   E0   &   etaM   & beta2 &  Qs2  &  KE2   & beta* & gamma* &   M*  &   Q12*  & 4*bratio & bm  \\ 
   1 &        &        &   MeV  &   MV/m &          &       &       &  MeV   &       &        &  GeV  &    MeV  &          & 2 \\ \midrule
   h & 0.2500 & -0.864 &  92.09 &  9.238 & -0.12138 & 1.00  & 0.324 & 145.12 & 0.285 & 1.043  & 2.919 & 110.120 & 0.99994  & e \\ 
   h & 0.2500 & -0.864 &  92.10 &  9.240 & -0.12139 & 1.00  & 0.324 & 145.14 & 0.285 & 1.043  & 2.919 & 110.131 & 0.99999  & e \\ 
   h & 0.2500 & -0.864 &  92.11 &  9.241 & -0.12140 & 1.00  & 0.324 & 145.15 & 0.285 & 1.043  & 2.919 & 110.142 & 1.00004  & e \\ 
\bottomrule
\end{tabular}
\end{table*}
\begin{table*}[htb]\scriptsize
\caption{\label{tbl:GOLDEN-EC-3}Field strengths, kinematic data, and spin tunes for the final state of the reaction $e^- + h \rightarrow t + \nu$. }
\setlength{\tabcolsep}{3pt}
\centering
\begin{tabular}{c|ccc|ccc|cc|cccc|c} 
\toprule 
 bm &    m3  &   G3   & q3 & th3-90 & th4-90 &   KE3  &   E0   &   etaM  &   KE4    &   m4   &    G4  & q4 & bm  \\ 
  3 &   GeV  &        &    &        &        &   MeV  &  MV/m  &         &    MeV   &        &        &    &  4  \\  \midrule
  t & 2.8089 & 7.9150 & 1  &  10.97 &  86.32 & 124.48 & 9.2400 & -0.1214 & 112.7400 & 0.0000 & 0.0000 &  0 & nu  \\ 
\bottomrule
\end{tabular}
\end{table*}

\subsection{Unambiguous event reconstruction}\label{sec:EventReconstruction}\mbox{}

\begin{figure*}
\centering
\includegraphics[scale=0.24]{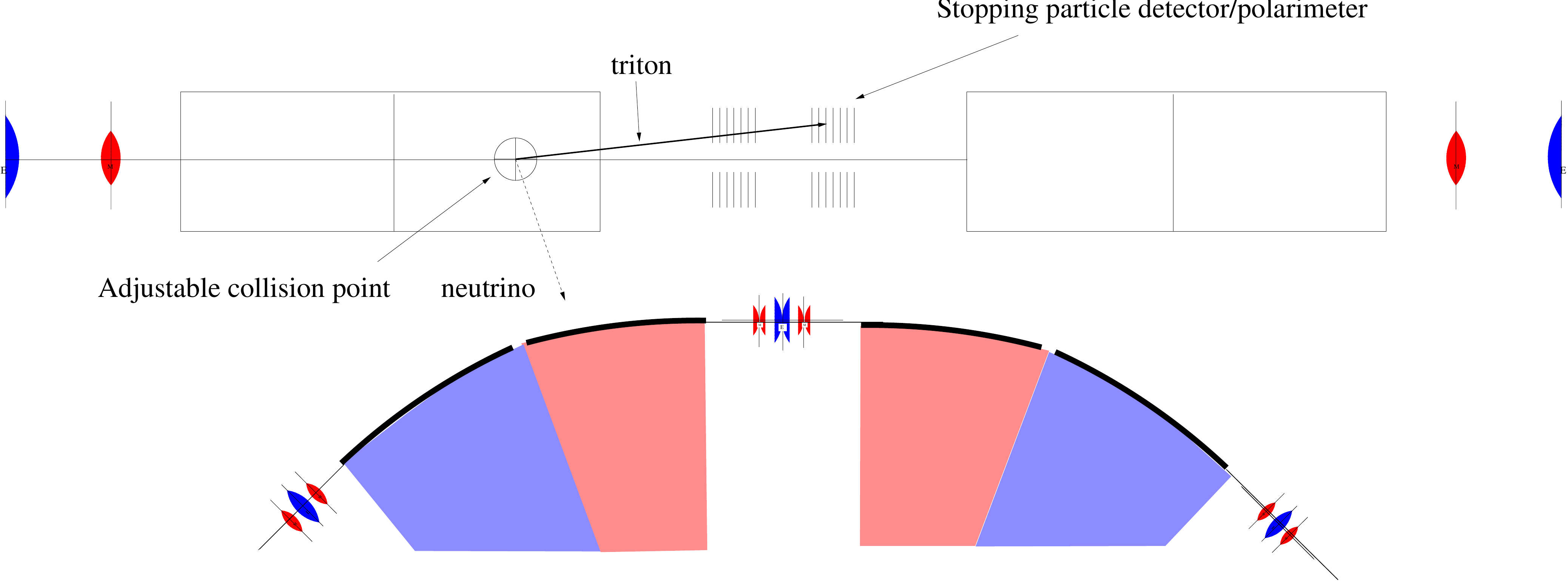}
\caption{\label{fig:PTR-layout-Toroidal8_102p2-tracking-only}. Figure showing an expected $h + e^- \rightarrow t + \nu$} event,
with the triton stopping in the tracking chamber.  With half the quadrupoles removed, for injection and extraction, the ring 
super-periodicity has been reduced from 8 to 4.
\end{figure*}

Consider the channel $e^- + h \rightarrow t +\nu$, concentrating in particular on the unambiguous reconstruction of a foreground event such as 
illustrated in Figure~\ref{fig:triton-extraction}, which is a blow-up of the central part of 
Figure~\ref{fig:PTR-layout-Toroidal8_102p2-tracking-only}.
With the quadrupoles removed from alternating straight sections, one of the
four free straight sections can be instrumented with annular tracking/polarimiter chambers with graphite film plates. A candidate triton 
trajectory begins, on axis at point~(1), somewhere along the bunch overlap region. Its subsequent history can be reconstructed sequentially.
\begin{enumerate}
\item
The initial scattering point is known to be a longitudinally centered point close to a fixed, but adjustable, intersection point (IP).
\item
With the chamber azimuthally symmetric, the entry point (2) can be measured with high precision. 
\item 
Within one of the graphite chamber (actually construction grade graphene) a triton-carbon nuclear scatters elastically at point (3)
(with efficiency, i.e. probability, of say 1/500, from a carbon nucleus in one of the chamber plates.  See Appendix-B.  Such a scatter measures the 
triton polarization with high analyzing power, of say 0.4.  
\item
The triton ranges out at point~4. The full 2,3,4 range is necessarily bounded (precisely, from above) by the stopping power of tritons in graphite.   
\end{enumerate}

This sequence of events provides unambiguous evidence for the sequence of events just described. Such events can be expected to be azimuthally 
symmetric around the longitudinal axis. Five hundred times this many events, similarly azimuthally symmetric, will be reasonably be ascribed to the
process $e^- + h \rightarrow t +\nu$; but no polarization information is available for these events, especially those that range out with 
almost maximum range, under the assumption they are tritons.

\subsection{Triton extraction, tracking, and polarimetry}\label{sec:Extraction}\mbox{}

\begin{figure*}
\centering
\includegraphics[scale=0.35]{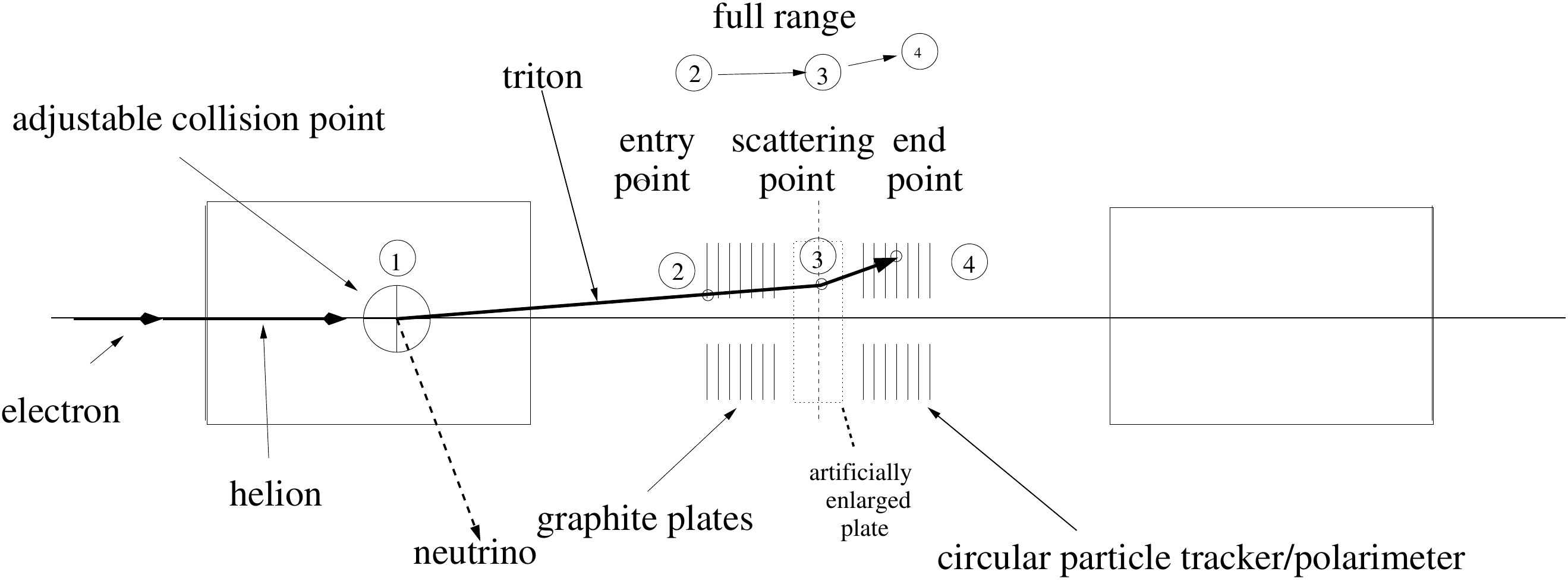}
\caption{\label{fig:triton-extraction}Figure showing an expected $h + e^- \rightarrow t + \nu$} event,
with the triton stopping in the tracking chamber. The helion kinetic energy has been selected arbitrarily,
though possibly higher in energy than could be handled in a ring with 11\,m bending radius. 
Experimental considerations concerning practical experimental optimization have not been formulated.
\end{figure*}

Figure~\ref{fig:triton-extraction} shows an expected $h + e^- \rightarrow t + \nu$ event,
with the triton stopping in a tracking chamber. The angles of the scattered particles are taken from Table~\ref{tbl:GOLDEN-EC-3}.
where they are listed as ``th3-90'' and ``th4-90'', representing particle ``3'' (t) and particle ``4'' ($\nu$) radiated at 
$90^{\circ}$ in the CM system. These are maximum angles in the laboratory, at the Jacobean peak of the rainbow pattern.
The corresponding kinetic energies are given in the same table.

Far more significant than the scattered triton direction is their factor of two reduced charge, compared to the helions they
replace.  Much like electron stripping injection, one has charge reduction extraction, as shown in Figure~\ref{fig:triton-extraction}. 

At the cost of reducing the ring periodicity, the quadrupoles can be removed from half of the straight sections,
in order to make room for both injection and extraction equipment.  Injection is covered in some detail in reference~\cite{CYR}. 

An azimuthally limited outer half-cone could then be passively extracted, at a cost of reduction of extraction efficiency by
at least a factor of two.  To take best advantage of the Jacobean peeking the beam could then be passed through a C-shaped
collimator. which would make the extracted beam more nearly monochromatic, without changing the energy of any single scattered 
nucleon.  Most of the cross section is near the limb of the rainbow pattern.  This is beneficial for the efficient collection and 
re-injection of the emitted tritons.  The most promising route may be to choose a more massive nuclear process in order to reduce the 
cone angle of the extracted beam.  

\section{Artificial production of any stable nuclear isotope}\label{sec:Artificial}\mbox{}

\begin{figure}
\centering
\includegraphics[scale=0.5, angle=89.6]{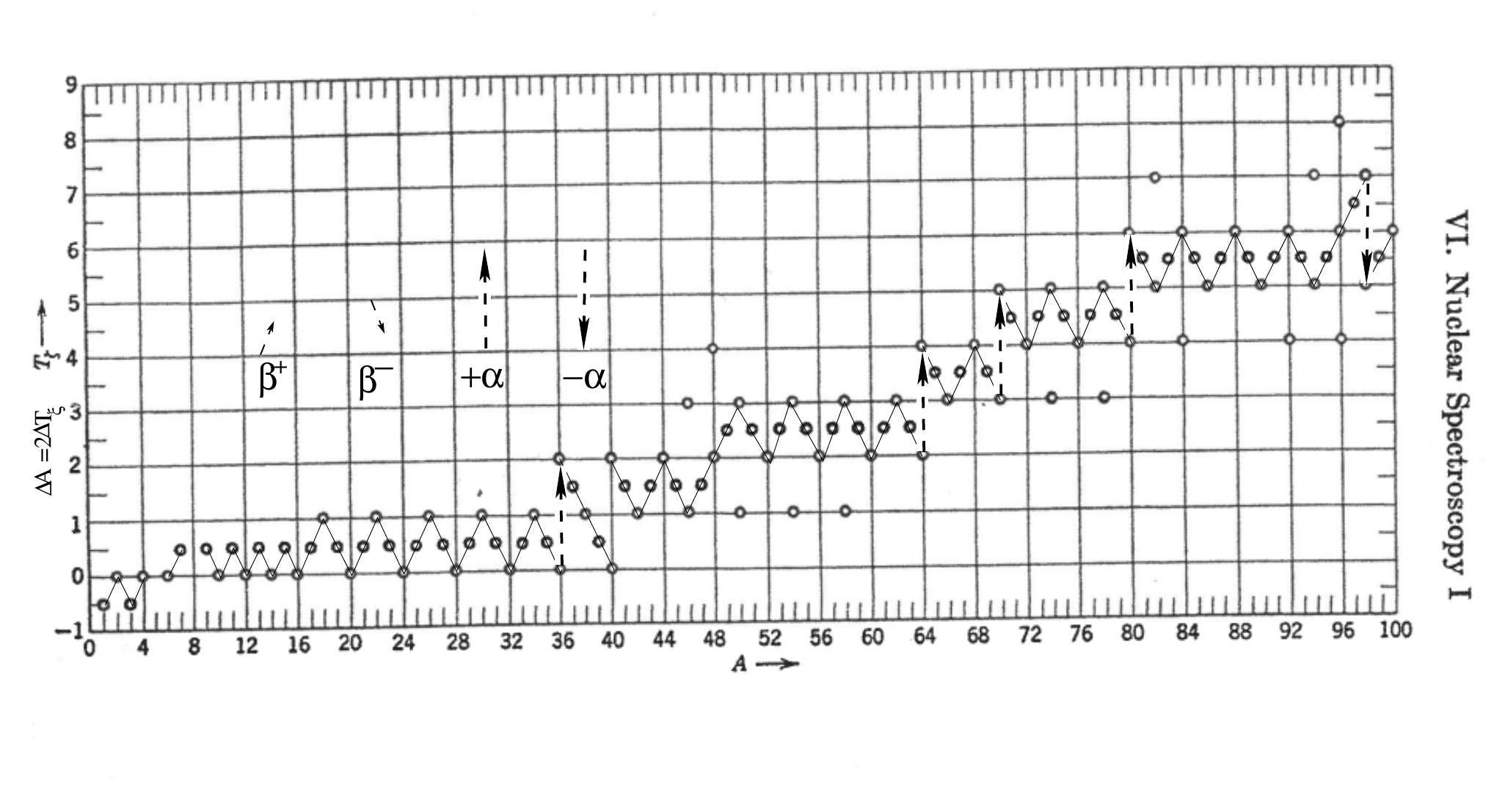}
\caption{\label{fig:Restore-stable-nuclei}Table of all stable nuclear isotopes copied from Blatt and Weisskopf\cite{BlattWeisskopf}
with the original caption being, ``Abscissa: mass number $A$; ordinate neutron excess $T_{\xi}$; each circle represents a stable nuclear species.''
The stable species have been connected by diagonal lines or vertical arrows produced by one of the four inset operations 
$\beta^-$, $\beta^+$, $\pm\alpha^+$, which represent artificial transmutations that could, in principle, be performed in an E\&m storage ring.  
The remaining dozen or so, of unconnected stable isotopes, can also be produced by $\pm\alpha^+$ operations; these are not shown connected, 
in order to make the figure less confusing.  Long sequences connected by zig-zag lines are promising experimentally, since they allow the
comparison of many redundant measurable parameter determinations.  Straight line connections, joining as many as five stable isotopes
are the most promising, since they can be performed with only minor readjustments of the same storage ring process.}
\end{figure}

Figure~\ref{fig:Restore-stable-nuclei}, based on a figure from Blatt and Weisskopf,\cite{BlattWeisskopf}, shows a plot of all stable 
nuclear isotopes.  As described in the caption, the plot represents transmutations that could, in principle, be performed 
in an E\&m storage ring to produce artificially any stable (or unstable) nuclear isotope.
\section{Recapitulation and conclusions}\label{sec:Physics-goals}\mbox{}

From the perspective of the present authors this paper represents the culmination of one sub branch of an experimental program
to study time reversal invariance initiated by Norman Ramsey and Robert Purcell in 1957, with measurement of the neutron EDM 
being their first target.  This program has continued ever since, most recently reducing the current neutron EDM upper limit to 
\cite{PSI-neutron};    $d_n = (0.0\pm1.1_{\rm stat}\pm0.2_{\rm sys}\times10^{-26})\,e$\,cm.  

In a 2004 paper by Farley et al.\cite{Farley},
it was suggested that the muon EDM could be measured by freezing the muon spins (i.e. Q\_s=0) of counter-circulating muons.
By 2011 a plan for measuring the proton EDM in the all-electric 232\,MeV storage ring (referred to previously in this paper as 
the ``Nominal All-Electric Proton EDM ring'') had been developed.  

Shortly after this, a CPEDM Group was formed, led, initially, by BNL and by the COSY lab in Juelich, Germany, with participation 
more or less balanced between European (including Russian) and North American physicists, with co-leaders from the lead laboratories.

By 2018, however, for reasons best forgotten, having to do with program priorities, the CPEDM group had split, largely, 
but not entirely, into an American and a European proton EDM group.  Nevertheless the work of the CPEDM group (including 
significant previous contributions from the American contingent) but authored by the European contingent, was documented
in the CERN Yellow Report\cite{CYR} referenced frequently in this paper.  

The combination of COVID-19 and the CPEDM group schism naturally slowed progress on the proton EDM project.  But it also led to 
the development by the ``sub-branch'' mentioned in the first sentence of this section---especially in the form of
reference\cite{RT-Compromise}, in which the concept of small and inexpensive predominantly electric E\&m 
storage rings were proposed in which ``doubly magic'' protons and helions (i.e. both spin tunes zero) could counter-circulate 
simultaneously, thereby enabling measurement of the difference of proton and helion EDM's 
(which are expected to vanish individually, by time reversal) could be measured.  
Since the dominant systematic error cancels in this difference, this test of T-violation is
more sensitive than either EDM measurement could be individually.  

The same paper also showed that both spin tunes in the nominal all-electric proton EDM ring could not be globally frozen 
without the intentional inclusion of some magnetic bending.  Furthermore, with minimally strong magnetic field (to freeze 
both beam spin tunes) the spin tune difference would be inconveniently small.  Overcoming this problem would lead
more nearly to the configuration described in the present paper. 

Significant as it was, this particular proton minus helion EDM measurement development cannot be said to make the EDM 
measurement easy. \emph{What follows in the present paper is easy}, at least by comparison. 

It soon became apparent that, as well as enabling counter-circulation, E\&m bending would also enable co-circulation of 
various particle types.  This led to the rear-end collisions and the capability of controlling the initial-state spins and
measuring the final state spins which constitutes the main body of the present paper.  Unlike the EDM measurement,
these low energy nuclear measurements seem to be not very challenging to perform experimentally.

Subsequently, an especially promising weak interaction $\beta$-decay channel, namely  $e^- + h \rightarrow t +\nu$,
suggested itself, as described in ``section \ref{sec:Electron-induced}''.

Based on the numerous low energy nuclear channels analyzed in this paper, \emph{what is being proposed is an
inexpensive, yet powerful, low energy nuclear physics program, rather than any individual project}.
The program can be expected to include investigation of the spin dependence of nuclear scattering and transmutation,
including weak nuclear interactions.

The goals therefore are to provide experimental data sufficient to refine our understanding of the nuclear force
(to the extent it can be disentangled from the electromagnetic force) and nuclear physics.  

Pure incident spin states, high analyzing power final state polarization measurement, and high data rates should
initiate a qualitatively and quantitatively new level of experimental observation of nuclear reactions.

Especially important is the investigation of wave particle duality and spin dependence of ``elastic'' 
$p,d$ scattering below the pion production threshold.  Precision comparison of ``light on heavy'' and 
``heavy on light'' collisions (which would be identical for point particles, but necessarily for compound
particles) can also probe the internal nuclear structure; perhaps distinguishing experimentally between ``prompt'' 
and ``compound nucleus'' scattering.  This promises to provide a more instructive visualization of internal 
structure than can be produced by the parton picture obtained by ultra-high momentum transfer inelastic electron 
scattering.  

This paper has described an E\&m storage ring capable of the room temperature laboratory 
spin control of two particle nuclear scattering or fusion events.
The novel equipment making this possible is a storage ring with superimposed electrical and magnetic bending.
Rings like this were introduced by Koop but have not yet been built.

Serving as a demonstration of nuclear to electrical energy conversion, such apparatus can perform measurements needed to 
refine our understanding of thermonuclear power generation and cosmological nuclear physics.
It is the novel capability of such rings to induce ``rear-end'' nuclear collisions that makes this possible.     

Emphasizing the measurement of spin dependence in low-energy nuclear physics, the goal is to provide experimental 
data to refine our understanding of nucleon composition along with the nuclear force and its influence on 
elementary-particle physics.  The better understanding of low energy nuclear processes that can be obtained from 
the proposed improvement of experimental measurement methods seems certain to enhance cosmological nuclear physics. 

Ironically, this improvement will be at least partly produced by the use of storage rings to investigate nuclear processes
at the \emph{reduced kinetic energies of cosmological nuclear physics}, compared to presently available
fixed target measurements.

\section{\bf APPENDICES}

\appendix

\subsection{\bf A. Compromise E\&m quadrupole design}\label{sec:Compromise}\mbox{}

Ideally, the discrete lattice quadrupoles would have electric and magnetic fields superimposed in the
same ratio $\eta_E/\eta_M$ as in the bend regions.  In frozen-spin proton PTR application the magnetic 
bending is hardly ``perturbatively small'' compared to the electric bending; $\eta_M/\eta_E\approx1/3$.
This presents a serious design problem. 

Clearly the coils producing the magnetic field need to be outside the electrodes producing the electric field.
Any scheme with (insulated) current carrying conductors situated between electrodes
is doomed to fail, since the electric and magnetic quads would be skew relative to one another.
Yet, the electric and magnetic quadrupole field reference radii would preferably be equal, for the field ratio 
to be constant over the full ring aperture. 

\begin{figure}[t]
\centering
\includegraphics[scale=0.45]{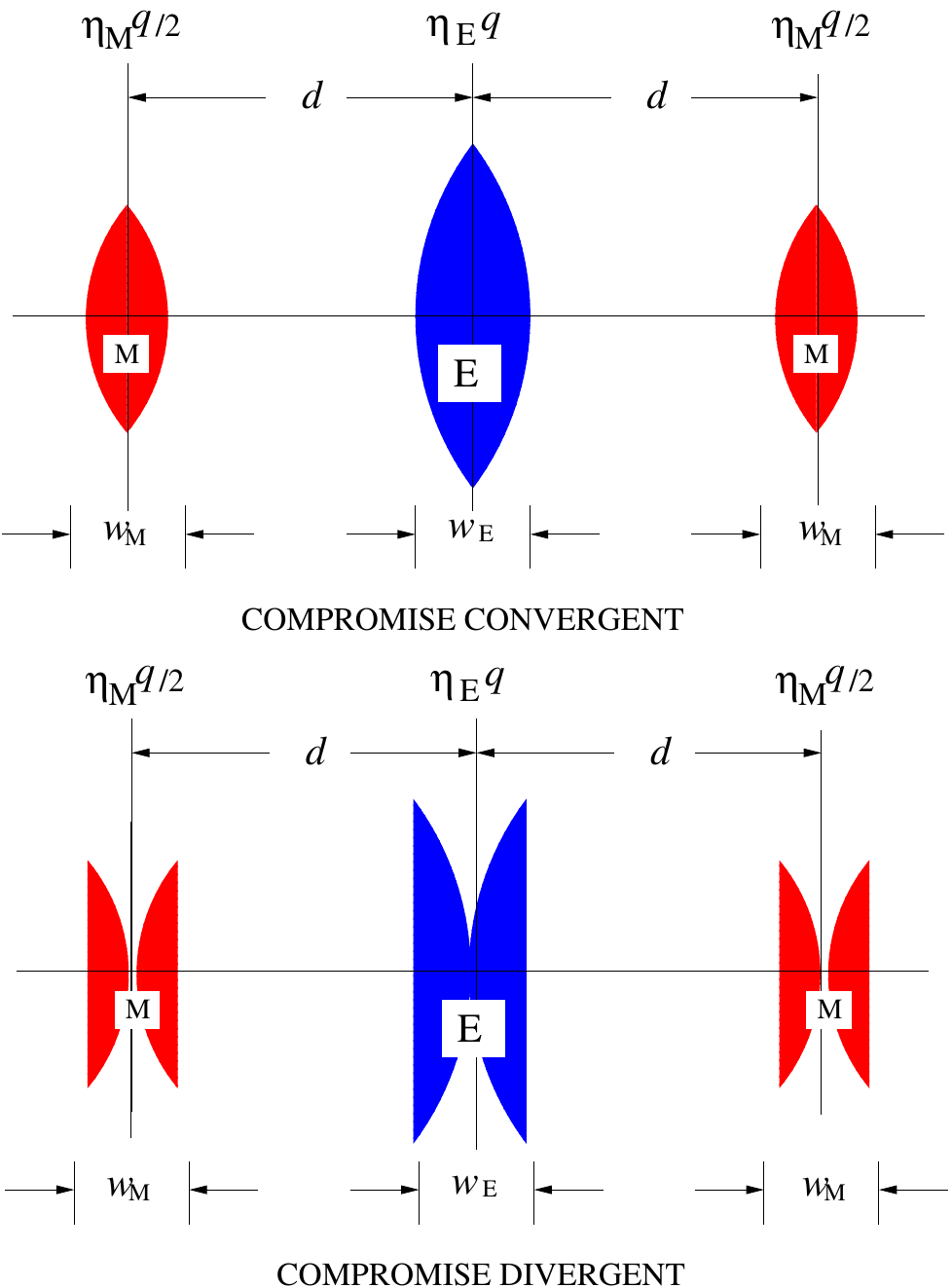}
\caption{\label{fig:BF-ReversibleQuad}''Compromise triplet'' representation of an electric quadrupole with (weak)
magnetic quadrupole superimposed. Blue quads are electric, red magnetic. For the primary beam all quads combine 
constructively; for the secondary (backward-traveling) beam the red signs reverse.  In thin element 
approximation, with the quad strengths shown, the triplet approximates a thin lens of strength $q=1/f$, with
(perturbatively small but) unequal forward and backward focal lengths $f$.  The purpose for this complication is 
to make backward and forward $\beta$-functions as nearly identical as possible.  For sufficiently small values of 
both $d$ and $\eta_M/\eta_E$ the triplet acts like a singlet quadrupole with weak magnetic field superimposed 
on strong electric field.}
\end{figure}

As a compromise an approximate match can be obtained by replacing 
each quadrupole by the symmetric quadrupole triplet arrangement shown in Figure~\ref{fig:BF-ReversibleQuad}.  
This provides the required match in thin-element approximation, though not in higher order approximation.
This provides the primary motivation for reducing lumped quadrupole strengths to the extent possible.

With each of the three quadrupoles forming the ``triplet'' on the top of Figure~\ref{fig:BF-ReversibleQuad}
being ``focusing'', treated as a single quadrupole, its symbol would be ``F''.  Similarly, the ``triplet''
on the bottom, treated as a single quad,  symbolized as ``D''.

Because the thin element approximation is valid only perturbatively, and the $\eta_M/\eta_E\approx1/3$  ratio is not
negligible, the forward and backward optical functions can never be exactly equal.  This does not seriously impair the EDM 
determination, provided that both forward and backward optics are stable.  With quads situated at crossing points,
as in the design shown in Figure~\ref{fig:RTR-figures}, there should be comfortably broad bands 
of joint stability.

\subsection{\bf B. Stopping-proton polarimetry}\label{sec:StoppingProtons}\mbox{}

\begin{figure}[h]
\centering
\includegraphics[scale=0.30]{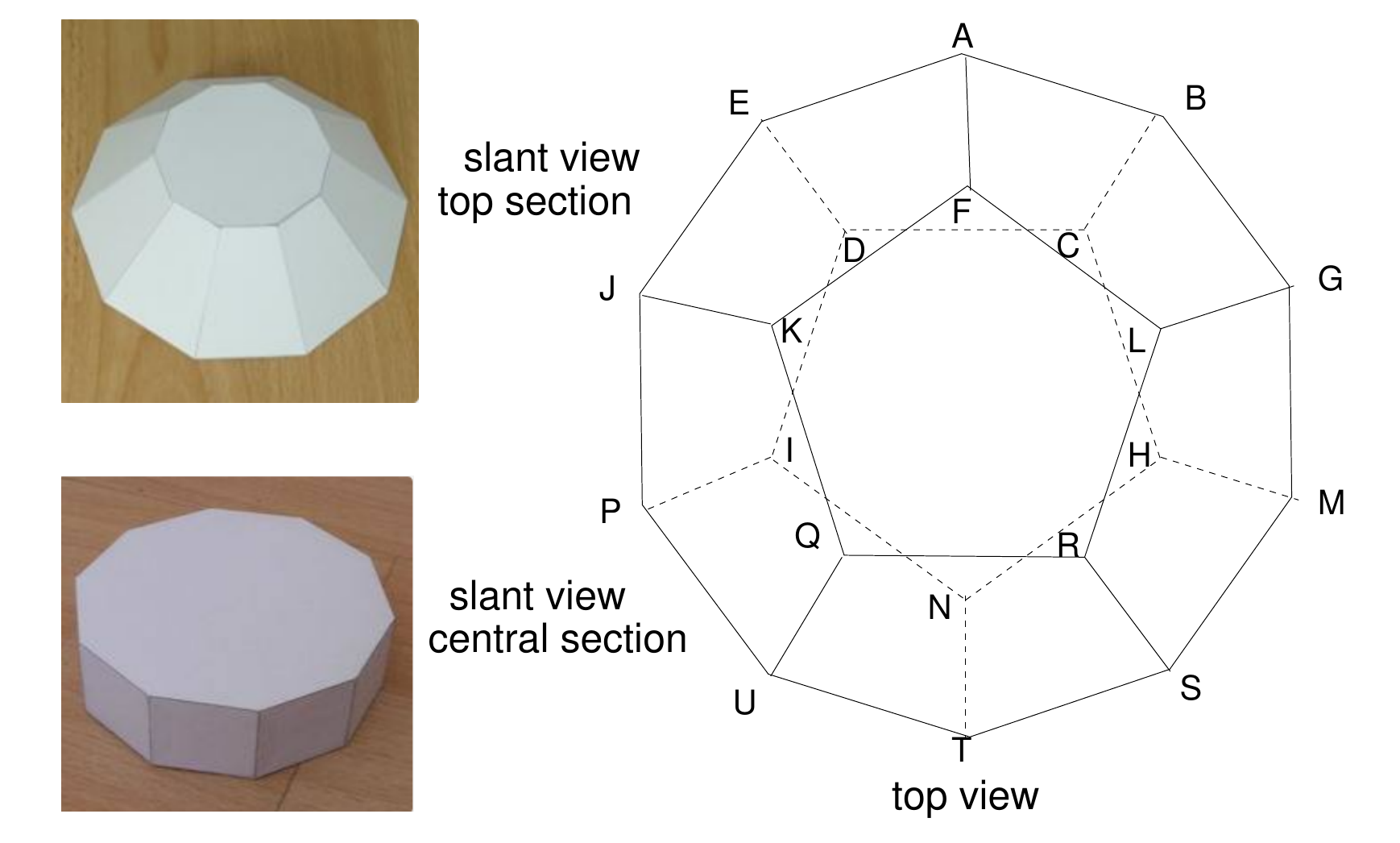}
\caption{\label{fig:dodecahedron-1}On the right is an artist's conception top view of an almost full-acceptance 
tracking/stopping/polarimeter at the storage ring intersection point IP. Their dodecahedral faces subtend roughly 
equal solid angles. The figures shown on the left are slant views of horizontal slices.  To accommodate passage 
of the colliding beams there is little useful particle detection in the up-down central section.}
\end{figure}

In every nuclear scattering or transmutation channel described in this paper, there is a slow proton or other 
stable (or semi-) charged nuclear isotope. Apparatus appropriate for stopping these particles in graphite is
shown in Figure~\ref{fig:dodecahedron-1}.  This provides unambiguous channel identification, percent level
energy measurement, and high efficiency, high analyzing power, polarimetry. 

Table~\ref{tbl:graphite-stopping} shows stopping powers and ranges for low energy protons stopping in graphite. 
With graphite density of 1.7\,g/cm$^2$, all final state protons will stop in a graphite plate chamber, with
measured range producing accurate energies for most final state protons. Precision energy determination (for 
example to exclude inelastic scatters) depends on the full stopping range.  But, since the $p,C$ polarization 
analyzing power falls with decreasing proton energy, polarimetric analyzing power is provided mainly by the 
left/right asymmetry of $p,C$ elastic scatters in the first half of their ranges, while their energies 
remain high.
\begin{table}[h]\scriptsize
\medskip
\centering
\begin{tabular}{|c|ccc|c|c|}             \hline
K.E. & Stopping   & Power,   & $dK/d(\rho_0l)$ & range, $\rho_0l$ & 20\,col3/col4\\
     &          & MeV cm$^2$/gm &           & gm/cm$^2$ &          \\
MeV  & electronic(e) & nuclear(n) & total(t) &          & n-prob. \\  \hline
20   & 2.331E+01  & 1.006E-02  & 2.332E+01  & 4.756E-01 & 0.00862  \\ 
40   & 1.331E+01  & 5.221E-03  & 1.331E+01  & 1.662E+00 & 0.00784  \\ 
60   & 9.642E+00  & 3.553E-03  & 9.645E+00  & 3.453E+00 & 0.00736 \\ 
80   & 7.714E+00  & 2.703E-03  & 7.717E+00  & 5.786E+00 & 0.00700 \\ 
100  & 6.518E+00  & 2.186E-03  & 6.520E+00  & 8.616E+00 & 0.00670 \\ 
120  & 5.701E+00  & 1.838E-03  & 5.703E+00  & 1.190E+01 & 0.00644 \\ 
140  & 5.107E+00  & 1.587E-03  & 5.109E+00  & 1.561E+01 & 0.00621 \\ 
160  & 4.655E+00  & 1.398E-03  & 4.656E+00  & 1.971E+01 & 0.00600 \\ 
180  & 4.299E+00  & 1.250E-03  & 4.301E+00  & 2.418E+01 & 0.00581 \\ 
200  & 4.013E+00  & 1.130E-03  & 4.014E+00  & 2.900E+01 & 0.00563 \\ \hline
sum  &            &            &            &           & 0.06761 \\
\hline
\end{tabular}
\caption{\label{tbl:graphite-stopping}Stopping power for protons stopping in graphite,
density 1.70\,gm/cm$^2$, NIST\cite{NIST-PSTAR}.  The final column (col6) gives the probability of nuclear scatter
in the approximation that nuclear energy loss (in this energy range) is always negligible compared 
to electric energy loss.  The binning error associated with wide kinetic energy bins
is quite small because the probabilities vary slowly.}
\end{table}

\subsection{{\bf C. Transverse beam dynamics}}\label{sec:Transverse-dynamics}\mbox{}

This appendix expands upon Chapter 7: ``The EDM Prototype Ring (PTR)''.  
of CERN Yellow Report (CYR) ``Storage Ring to Search for Electric Dipole 
Moments of Charged Particles: Feasibility Study''\cite{CYR} That report
referenced the ``nominal all-electric'', 232\,MeV proton storage ring needed to
``freeze'' the spins of transversely polarized protons, as required to measure the
proton electric dipole moment(EDM).  For realistically achievable electric field value 
the ring circumference for the nominal all-electric ring needed to exceed 600\,m or so.  
The CYR report proposed to build a prototype ring ``PTR'' ring as
a precursor.  This ring is referred to here as PTR(2019).

The proposed role for PTR(2019) was to attack numerous design uncertainties.  Included in 
that proposal was the superposition of magnetic bending that would enable the ring to be 
predominantly electric, weakly magnetic (i.e. E\&m).  This would enable the spins of 
protons of much lower energy, such as 45\,MeV, to be frozen.  Rings of various sizes 
were investigated, eventually setting the PTR bending radius to 11\,m with ring 
circumference 102.5\,{\rm m}, which remains the same as the ring PTR(2019) proposed
in the present paper.

This appendix reviews the transverse optics of rings scaled down
from the nominal all-electric 232\,{\rm MeV} ring by factors ranging from 1 to 6.
Special emphasis was placed on optimizing the PTR circumference for the injection of 
polarized beams from COSY, having $\mathcal{C}_{\rm INJ}=182\,$m circumference.
Considerable effort went into the PTR(2019) design described in the CYR.
By now, in 2023, as already mentioned, this design has been superseded by a 
substantially smaller, 102.5\,m PTR(2023) design. This included using a
cannibalized COSY as injector, which amounted to nearly eliminating the COSY straight 
sections, but retaining the arcs to reduce the COSY circumference to the 102.5\,m value.

Though predominantly electric, storage rings can be expected to perform much
like magnetic rings, but many of the standard formulas describing magnetic rings
need to be re-derived for electric rings.  The present appendix therefore contains
many familiar-looking formulas as modified for electric bending, as well as
more challenging formulas specific to superimposed E\&m bending.

Appendix~D describes longitudinal performance in that PTR(2023) storage ring.  
But, since the transverse optical design methodology is unchanged, the PTR(2019) 
parameters have been retained for the present appendix.  To simplify single bunch 
train transfer injection, the main constraint is that INJ and PTR ring circumferences 
should be related by ``easy'' rational fractions, such as 1, 1/2 or 1/3.  Other significant 
issues are availability of free straight section length for  needed apparatus, required 
electric field maximum, and overall PTR cost. 

\subsubsection{Down-scaling the 232\,MeV proton EDM ring}\mbox{}

Following release of the CERN Yellow Report (CYR)\cite{CYR}, with its scaled down
PTR(2019) prototype ring design,  the sufficiency of total free drift space for required diagnostic 
and beam handling equipment became an issue.  One approach considered was to migrate 
from rounded-square ring shape to racetrack-shape, to make more space available.  
This idea has been rejected for two reasons. 

\begin{figure}[htbp]
\centering
\includegraphics[scale=0.50]{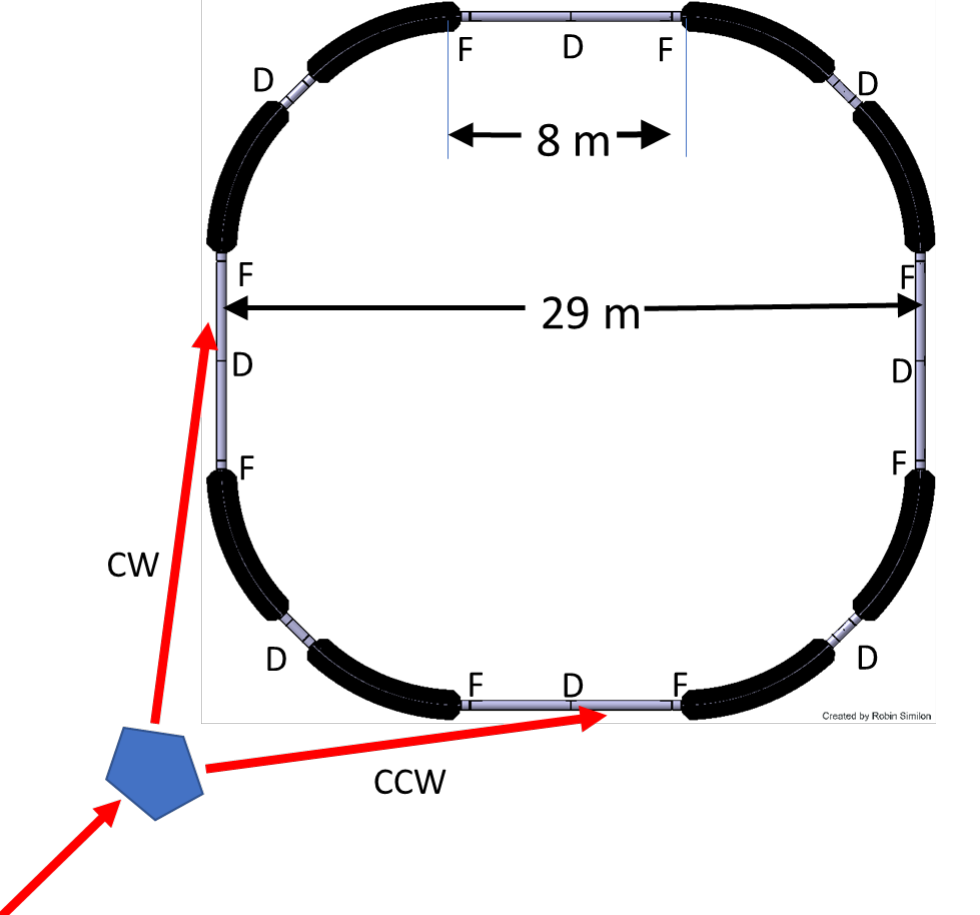}
\caption{\label{fig:PTR-corner-injection}(Copied from CYR) the basic layout of the prototype ring 
consists of 8 dual, superimposed electric and magnetic bends; 2 families of 
quadrupoles---focusing~(F) and de-focusing (D); 
with an optional skew quadrupole family at mid-points of the four 8\,m long straight sections. 
The total circumference, as described in the CYR, was about 100\,m. Three lattices described in
the present appendix have nominal circumferences of 90.8\,m, 136.5\,m, and 183\,m, intentionally 
adjusted close to 1/2, 2/3, and 1/1 times the COSY ring circumference.  Otherwise the ring designs 
are little changed from the CYR.} 
\end{figure}

Injection at a corner, as shown in Figure~\ref{fig:PTR-corner-injection}, seems economical 
and ideally symmetric for injecting counter-circulating beams. This symmetry would be lost
in a race-track configuration, complicating the task of exactly matching beam profiles.
A more serious problem is that the focusing implied by long straight sections
precludes the possibility of tuning the vertical tune $Q_y$ down close to zero. 
This penalizes self-magnetometry sensitivity, which scales as $1/Q_y$.

We consider only rounded-square lattice shapes. It is 
convenient that the required ring scaling was already faced in the far greater
range down to PTR scale from nominal all-electric scale in reference\ \cite{CYR}.  
We tentatively also assume the quadrupoles labeled ``D'' in 
Figure~\ref{fig:PTR-corner-injection}, located at the mid points of
long straight sections, are not required, at least for preliminary design.
The lattice name identifications were {\tt-0p5-COSY}, {\tt-0p75-COSY}, and {\tt-1p0-COSY}. 

Furthermore the (difficult) low $Q_y$ (high $\beta_y$) region was emphasized.
As shown in Chapter~7 of the CYR, the quite numerous QF and QD quadrupoles 
provide ample tuning range for both  tunes, $Q_x$ and $Q_y$. The three rings 
remained (at least nominally) capable of being tuned down arbitrarily close to the $Q_x=Q_y=0$) limit.  
This capability was provided by alternating gradient, weakly focusing $m \neq 0$ electrode shaping.  
But, for most ring commissioning, the focusing provided by QF and QD quads is dominant,
making the electrode shapes almost cylindrical (i.e. $m\approx\pm0$).

Focusing on the technically-difficult, low $Q_y$, limit makes the cases addressed
in this appendix hypersensitive.  Strengthening the QF/QD focusing from this limit
can be performed easily.  Technical discussion of these points has already been
given in CYR Chapter~7.

All parameter determinations in this appendix are based on coordinating two 
entirely different ring design programs. One of these, referred to as ``MAPLE'', is based on 
Wollnik linear transfer matrices, and is essentially equivalent to Valeri Lebedev's similar
program\cite{Lebedev-EDM}---traditionally there has been quite good agreement between 
these explicitly linearized Courant-Snyder codes. 

The other ring design program is referred to as 
``ETEAPOT''\cite{BNL-Electic-Analogue}\cite{UAL-ETEAPOT}. Patterned after
TEAPOT\cite{TEAPOT}\cite{Ztrack}\cite{UAL}\cite{UAL-ETEAPOT}, developed originally for the SSC
by Lindsay Schachinger and R. Talman and  others, ETEAPOT,  developed by J. and R. Talman,
is based entirely on particle tracking.  In this approach, ring transfer matrices are calculated 
as outputs rather than being provided analytically as inputs. Traditional lattice parameters, 
such as tunes, Twiss functions, and dispersions, are obtained from the derived transfer maps 
(including higher than linear order, if necessary).

For rings with electric bending
the figures in this appendix show that agreement between MAPLE and ETEAPOT models
is only fair, especially for ultra-weak vertical focusing.  The discrepancies are presumably due 
to an approximation built into ETEAPOT (which is nominally exact only for $m=1$ spherical electrode 
shapes, but needs to be extrapolated to more nearly cylindrical design electrode shapes. This 
extrapolation range is large enough for its perturbative approximation to degrade significantly. 

This discrepancy remains to be investigated.  For now
a single empirically chosen parameter, the same for all three rings, adjusts
the maximum vertical beta function values obtained by MAPLE and ETEAPOT to
best agree.  (Compare Figure~\ref{fig:E-PTR2-ETEAPOT-betax} and \ref{fig:E-PTR2-ETEAPOT-betay}
and accompanying tables, \ref{tbl:0p5}, \ref{tbl:0p75}, and \ref{tbl:1p0},
where the free parameter is indicated as an argument -0.032, of ETEAPOT[-0.032], for example.)
As a result the vertical beta functions obtained by ETEAPOT differ by as much as
ten percent from those calculated by MAPLE (in the hypersensitive high $\beta_y$ region).
This level of disagreement was accepted as tolerable at that time.  As already mentioned, 
this comparison has been made in the ``difficult fine-tuning limit'' for which the vertical 
betatron tune is close to zero.

\subsubsection{Twiss functions from transfer matrices}\mbox{}

Like the magnetic accelerator simulation code TEAPOT, the electric accelerator
code ETEAPOT, does not base particle tracking on theoretically-derived
transfer matrices. Rather, a standard set of small amplitude particles are 
tracked (i.e. to satisfy Newton's force law). Then, by differencing the outputs,
transfer matrices are extracted---in what amounts to numerical differentiation.
Then, from the transfer matrices, the lattice Twiss functions are derived. Because
this is an unconventional approach, the following sections explain the approach
in more detail.

In this section $x$ and $y$ and $ct$ subscripts are suppressed,
and only transverse evolution is to be discussed. 
The most general
transfer matrix is a six-by-six matrix ${\bf M}(s_i,s_j)$, which propagates a
phase space vector ${\bf x}(s_i)$ at $s_i$, to its value ${\bf x}(s_j)$ at $s_j$;
\begin{equation}
{\bf x}(s_j) 
 =
{\bf M}(s_i,s_j)\,
{\bf x}(s_i). 
\label{TM.1}
\end{equation}
ETEAPOT starts by assigning coordinates at the origin, $s_0=0$,
to the particles in a standard bunch (as described earlier), and then
evolves the standard bunch and records the coordinates ${\bf x}(s_i)$ 
at all points $s_i$.  From these results the transfer matrices 
${\bf M}(0,s_i)$ can be calculated (also as described earlier). By definition
\begin{equation}
{\bf M}(0,0)
 =
{\bf I}, 
\label{TM.2}
\end{equation}
where ${\bf I}$ is the $6\times6$ identity matrix. 

To extract Twiss lattice functions one needs periodic, ``once-around'' transfer 
matrices, distinguished by overhead tildes, and defined by
\begin{equation}
{\widetilde{\bf M}}(s_i)
 \equiv
{\bf M}(s_i,s_i+\mathcal{C}_0)
 =
{\bf M}(0,s_i)\,
{\bf M}(s_i,\mathcal{C}_0).
\label{TM.3}
\end{equation}
where $\mathcal{C}_0$ is the circumference of the design
orbit; the final step has been taken because knowledge of
$s_i+\mathcal{C}_0$ requires tracking for more than one
complete turn, but we are assuming that tracking has been
done only for exactly one complete turn. Propagation from
$s=\mathcal{C}_0$ to $\mathcal{C}_0+s_i$ is the same as
propagation from $s=0$ to $s_i$. The Twiss parameterization of 
(one partitioned diagonal
$2\times2$ block of) such a once-around, symplectic 
transfer matrix is
\begin{equation}
{\widetilde{\bf M}}(s_i)
 =
\begin{pmatrix}
\cos\mu + \alpha\sin\mu          & \beta\sin\mu \\
-\frac{1+\alpha^2}{\beta}\sin\mu & \cos\mu - \alpha\sin\mu 
\end{pmatrix}.
\label{TM.4}
\end{equation}
Extraction of the $\alpha_0$ and $\beta_0$, the Twiss parameters 
at the origin, can start from
\begin{equation}
\cos\mu
 = 
\frac{1}{2}\,\big({\widetilde{\bf M}}_{11}(0) + {\widetilde{\bf M}}_{22}(0)\big),
\label{TM.5}
\end{equation}
which fixes $\cos\mu$. Because of sign ambiguity, this determines only
$|\sin\mu|$. One also has the relations
\begin{equation}
\beta_0
 = 
\bigg|
\frac{\widetilde{\bf M}_{12}(0)}{\sin\mu}
\bigg|.
\label{TM.6}
\end{equation}
and
\begin{equation}
\alpha_0
 = 
\frac{1}{2\sin\mu}\,
\big({\widetilde{\bf M}}_{11}(0) - {\widetilde{\bf M}}_{22}(0)\big).
\label{TM.7}
\end{equation}
With $\beta$ being positive by convention, 
from the 1,2 element, sign($\sin\mu$) can be seen to be
the same as sign(${\widetilde{\bf M}}_{12}$). With $\cos\mu$ known,
this fixes $\sin\mu$.
Together, these relations fix $\sin\mu$, $\cos\mu$, $\alpha_0$, and $\beta_0$.

Conventionally one also introduces a third Twiss parameter
\begin{equation}
\gamma_0
 = 
\frac{1+\alpha_0^2}{\beta_0},
\label{TM.8}
\end{equation}
which can be obtained once $\beta_0$ and $\alpha_0$
have been determined. 

Because of the multiple-valued nature of inverse trig functions,
these relations do not determine a unique value for $\mu$. They do,
however, determine the quadrant in phase space in which the angle $\mu$
resides. For sign($\sin\mu$)$>$0 the angle $\mu$ resides in the first
or second quadrant, in which case the fractional tune is less than 1/2;
otherwise the fractional tune is greater than 1/2.
For sign($\cos\mu$)$>$0 the angle $\mu$ resides in the first
or fourth quadrant, in which case the fractional tune is below 1/4
or above 3/4. These considerations fix the fractional parts of
the tunes.

An ``aliasing'' or ``integer-tune'' ambiguity remains, however,
which cannot, even in principle, be obtained from the once-around 
matrix. Only if both transverse tunes are less than 1 (which is 
hardly ever the
case) would the tunes be equal to the fractional tunes that have
been determined. In general, to obtain the integer tunes, it is 
necessary to analyze the turn by turn data at sufficiently closely-space 
intermediate points in the lattice.

\subsubsection{Evolving the Twiss functions around the ring}\mbox{}

To find the Twiss parameters at an arbitrary location $s_i$ in the
ring requires the once-around transfer matrix ${\widetilde{\bf M}}(s_i)$. 
This can obtained most compactly by multiplying the equation
\begin{equation}
{\bf M}(0,s_j)
 =
{\bf M}(s_i,s_j)\,{\bf M}(0,s_i)
\label{TM.9}
\end{equation}
on the right by ${\bf M}^{-1}(0,s_i)$ to produce
\begin{equation}
{\bf M}(s_i,s_j)
 =
{\bf M}(0,s_j)\,{\bf M}^{-1}(0,s_i).
\label{TM.10}
\end{equation}
Substituting this with $s_j=\mathcal{C}_0$ into Eq.~(\ref{TM.3}) 
produces
\begin{equation}
\widetilde{\bf M}(s_i)
 =
{\bf M}(0,s_i)\,\widetilde{\bf M}(0)\,{\bf M}^{-1}(0,s_i).
\label{TM.11}
\end{equation}
Having obtained ${\widetilde{\bf M}}(s_i)$,
the procedure described in the previous subsection can
then be used to obtain $\alpha(s_i)$ and $\beta(s_i)$.
But the integer tune ambiguity can, again, not be
resolved. To resolve this ambiguity both $x$ and $y$ phases
have to be tracked continuously through the lattice,
requiring that they advance continuously and monotonically.
(Later, at least in principle, the same ambiguity will 
have to be faced for longitudinal motion. But the
integer longitudinal tune is almost always zero, so the
problem is usually absent in the longitudinal case.)

ETEAPOT requires the lattice description to be in the form
of an {\tt .sxf} file.  To be ``legal'' the granularity of such
a file has to be fine enough that no phase can advance by more
than a quarter integer through any element in the file. Before
working out the $\alpha$ and $\beta$ function evolution, 
ETEAPOT first works out the total phase advances from
the origin to every node specified by the {\tt .sxf} file
(or, if some elements are sliced more finely, by every node
after slicing). 

There is an alternative way of finding the betatron phase
advances. It starts with a Twiss parameterization
of ${\bf M}(0,s)$ from the origin to an arbitrary position $s$
in the lattice;
\begin{equation}
\frac{{\bf M}(0, s)}{\sqrt{\beta_0\beta(s)}}
 =
\begin{pmatrix}
\frac{\cos\psi(s)+\alpha_0\sin\psi(s)}{\beta_0} & \sin\psi(s) \\
   \cdot & \frac{\cos\psi(s)-\alpha(s)\sin\psi(s)}{\beta(s)}.         
\end{pmatrix}
\label{TM.15}
\end{equation}
The 2,1 element is quite complicated; it is not shown here since it will not be needed
for the following analysis; in any case its numerical value can be determined from the 
unit determinant requirement.
 
Dividing the 1,1 element by the 1,2 element produces
\begin{equation}
\frac{{\bf M}_{1,1}(0, s)}{{\bf M}_{1,2}(0, s)}
 =
\frac{\sqrt{\frac{\beta(s)}{\beta_0}}\Big(\cos\psi(s) + \alpha_0\sin\psi(s)\Big)}
     {\sqrt{\beta_0\beta(s)}\,\sin\psi(s)}
 =
\frac{\cot\psi(s) + \alpha_0}{\beta_0}.
\label{TM.16}
\end{equation}
Rearranging this equation produces
\begin{equation}
\psi(s)
 = 
\tan^{-1}\,
\frac{{\bf M}_{1,2}(0, s)}
     {\beta_0{\bf M}_{1,1}(0, s) - \alpha_0{\bf M}_{1,2}(0, s)}.
\label{TM.17}
\end{equation}
Like all inverse trigonometric formulas, this equation has multiple solutions.
But, with the {\tt .sxf} granularity being required to be fine enough, one
can (in principle) sequentially obtain unique phases. With
$\psi(s)$ starting from $\psi(0)=0$, as $s$ increases
from $s_i$ to $s_{i+1}$ there is a unique solution of Eq.~(\ref{TM.17}),
$\psi(s_i)\ge\psi(s_{i-1})$ such that the function $\psi(s)$ increases 
monotonically, as required.
Because the interval from $s_{i=1}$ to $s_i$ is non-zero, $\psi$ will,
superficially, advance discontinuously; the correct solution is the least
discontinuous. The same calculation has to be done for both the $x$ 
and $y$ betatron sectors. 
Unfortunately, even though, theoretically, the phase advances 
monotonically, numerical errors can cause Eq.~(\ref{TM.17}) to
give local phase decrease.  This can cause 
Eq.~(\ref{TM.17}) to give occasionally erratic results.

\subsubsection{ETEAPOT and MAPLE lattice functions\ }\mbox{}

This section, first written more than five years ago, concerns the 
up-scaling or down-scaling of Electric storage rings.  Most of the rest of this 
paper concentrates on a PTR ring of 102.5\,{\rm m} circumference, which is
perhaps the smallest practical size.  The appendix is now more likely to
provide parameters for the consideration of a more conservative, larger, ring.

\begin{figure*}[htbp]
\begin{minipage}[b]{0.49\linewidth}
\centering
\includegraphics[scale=0.7]{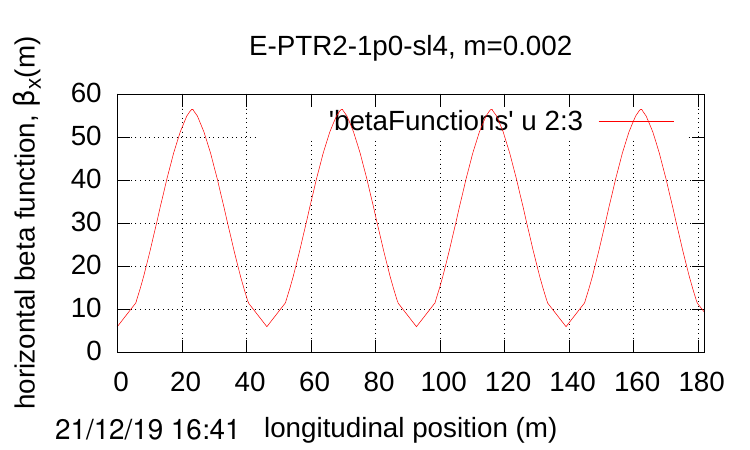}\includegraphics[scale=0.42]{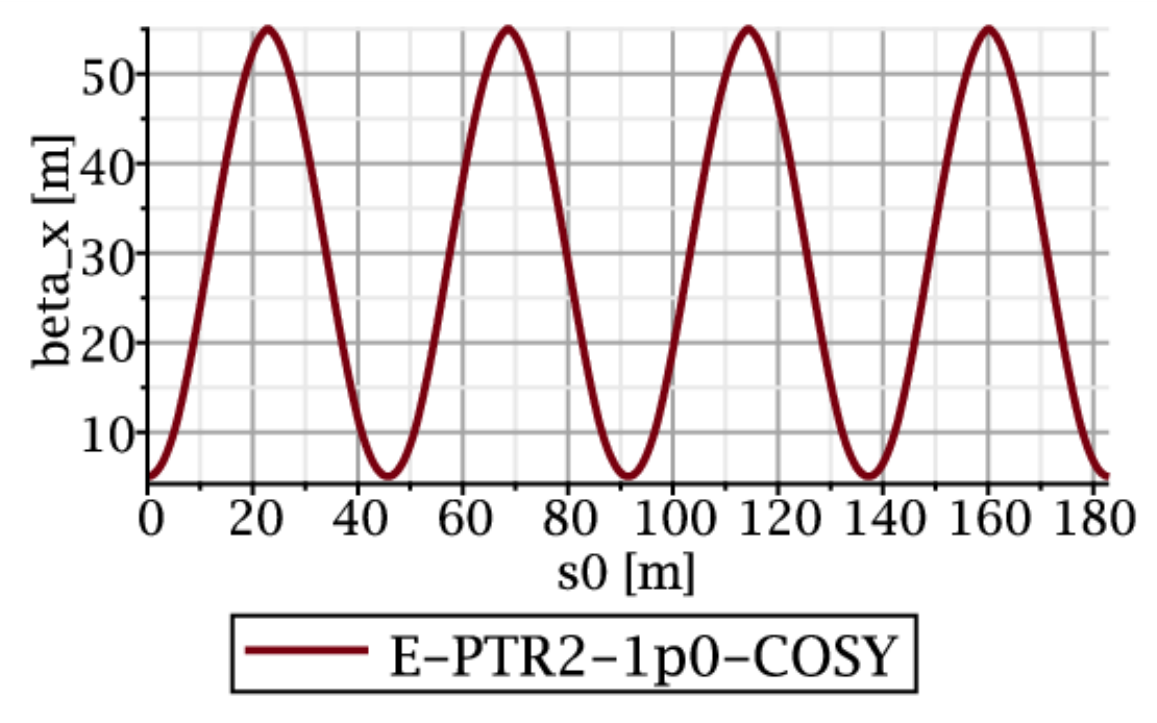}
\includegraphics[scale=0.7]{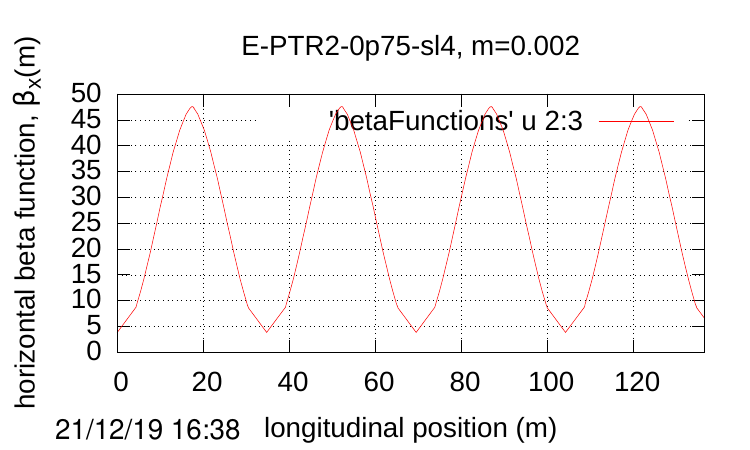}\includegraphics[scale=0.42]{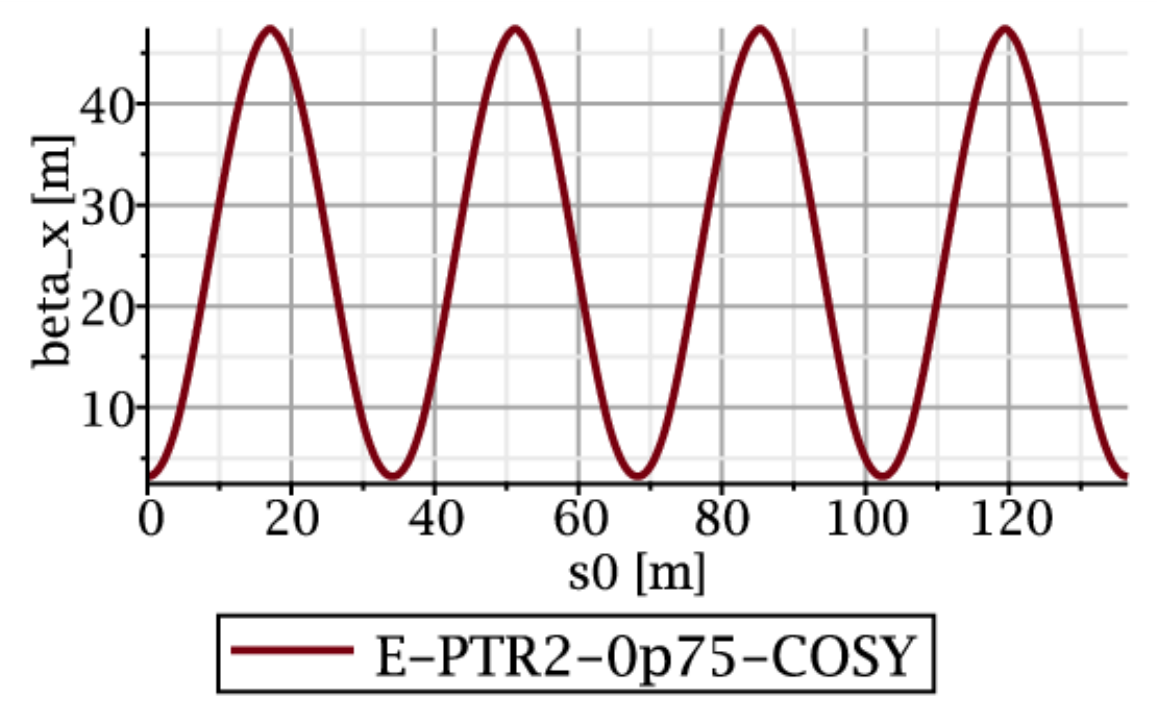}
\includegraphics[scale=0.7]{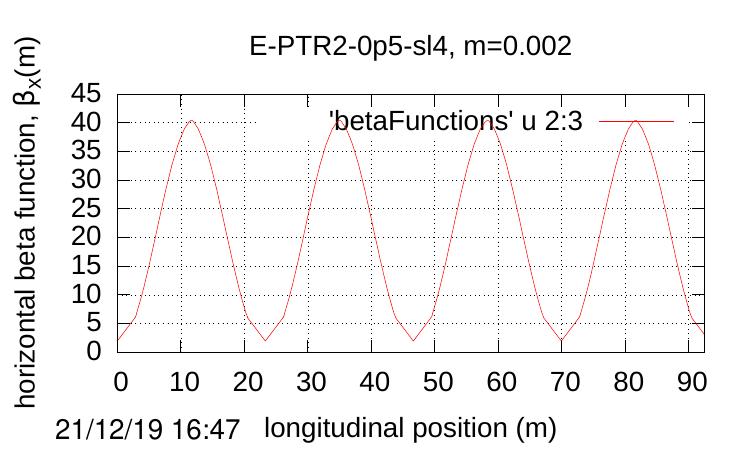}\includegraphics[scale=0.42]{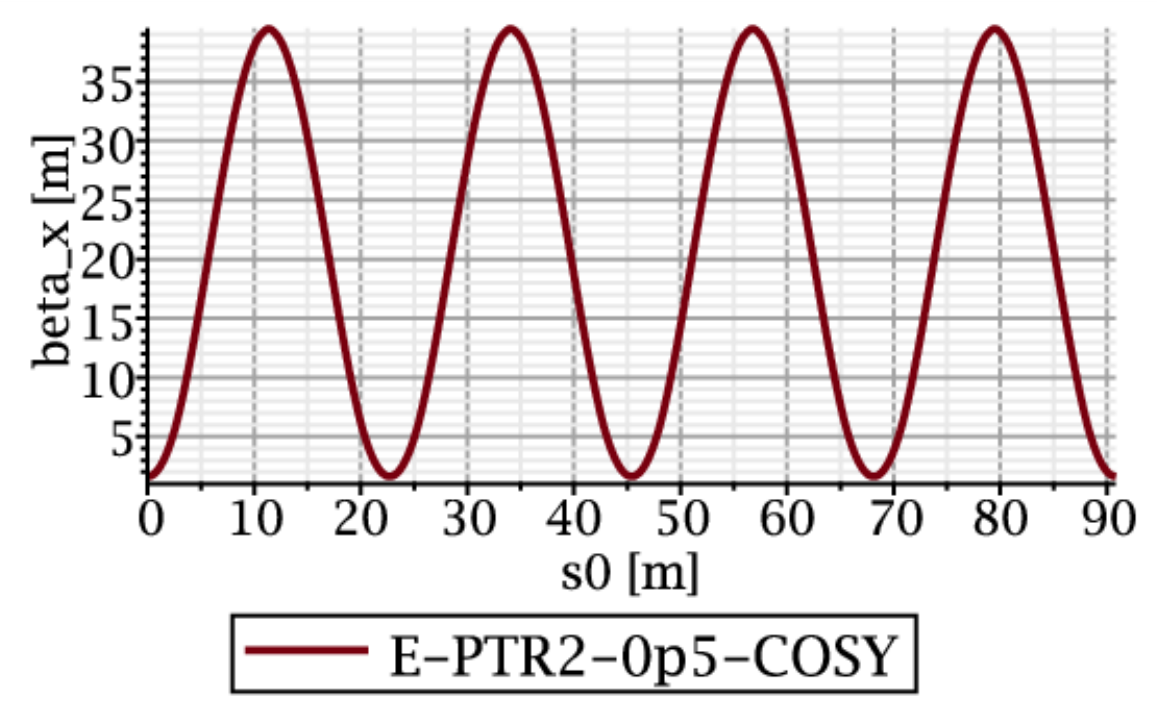}
\caption{$\beta_x(s)$ calculated by ETEAPOT for the {\tt E-PTR2} rings is shown on the left. $\beta_x(s)$ calculated by MAPLE linearized 
theory for the {\tt E-PTR2} ring is shown on the right.}
\label{fig:E-PTR2-ETEAPOT-betax}
\end{minipage}
\end{figure*}

Figures~\ref{fig:E-PTR2-ETEAPOT-betax}
compares $\beta_x$ evaluations for three candidate lattices.
Of the rings in this appendix, {\tt E-PTR2-0p5} is the 
one closest to the prototype ring described in the CERN Yellow Report. 
The alternating gradient pole shape
focusing, though ultra-weak, by itself, barely produces vertical stability.
``Geometric'' focusing, though also not very strong, is capable of
providing much stronger horizontal focusing, with horizontal tune $Q_x$
greater than 1.0. The
QF and QD quads, arranged in FODO pattern, can potentially provide stronger focusing
in both planes. As described in CYR, by strengthening QF and QD, there is a large 
tuning range over which $Q_y$ can smoothly increase from barely above 0 to just below 1,
and $Q_x$ can smoothly decrease from barely below 2 to just above 1.   
The figures on the left are evaluated by ETEAPOT, the ones on the right by 
MAPLE, linearized (Wollnik) formulas.

\begin{figure*}[htbp]
\hspace{-0.6cm}
\begin{minipage}[b]{0.45\linewidth}
\centering
\includegraphics[scale=0.72]{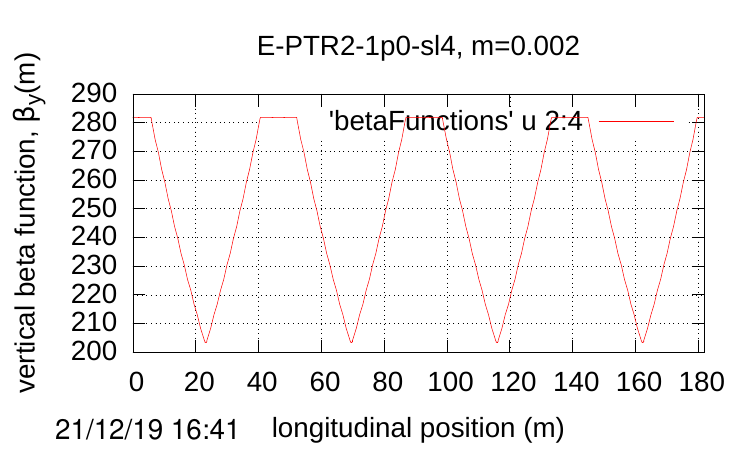}\includegraphics[scale=0.43]{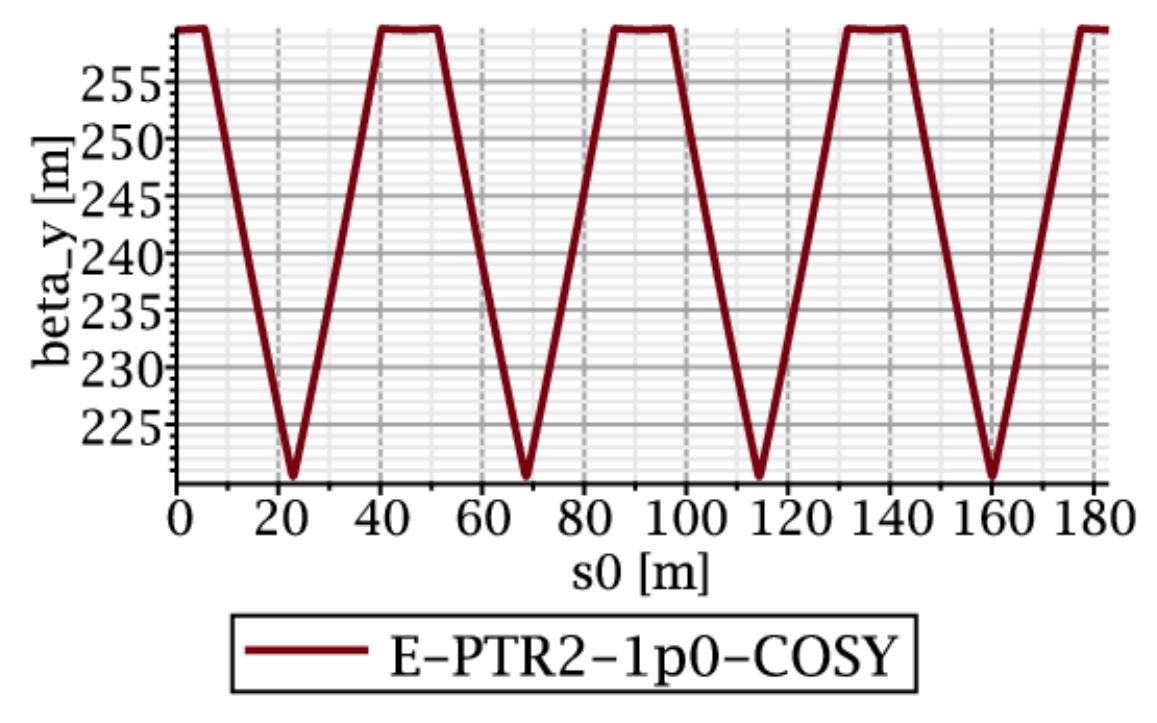}
\includegraphics[scale=0.72]{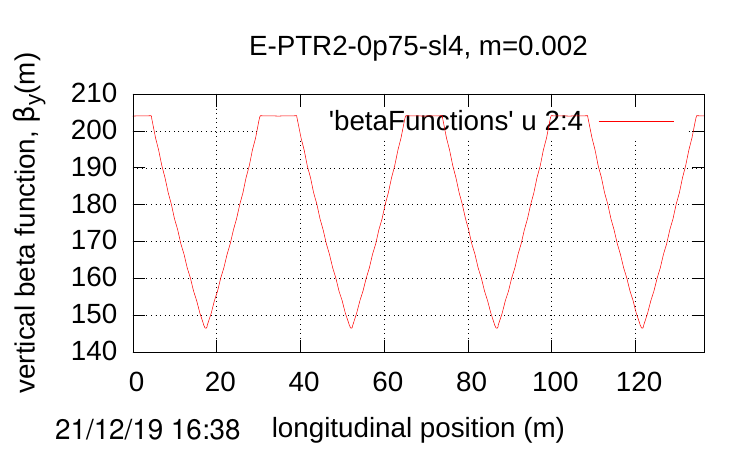}\includegraphics[scale=0.43]{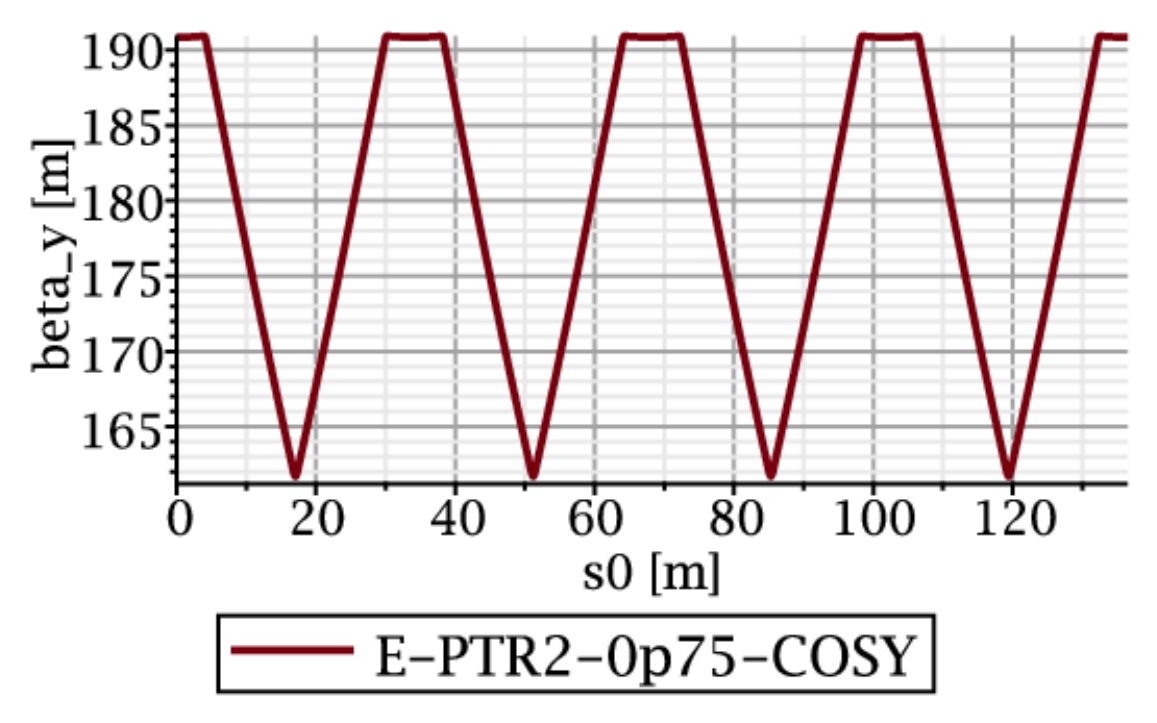}
\includegraphics[scale=0.72]{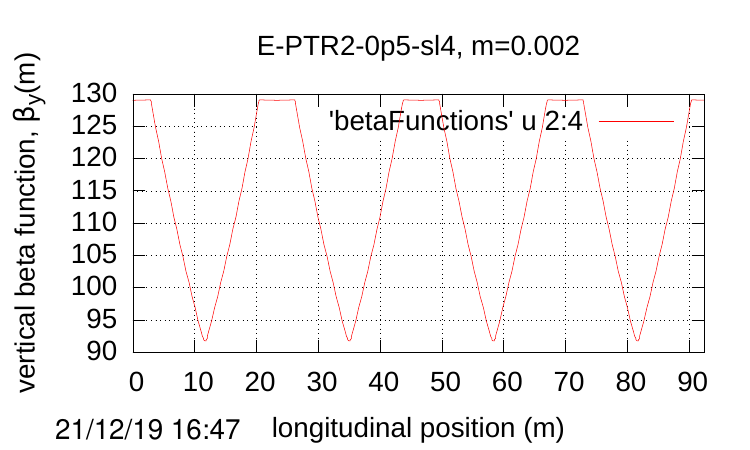}\includegraphics[scale=0.43]{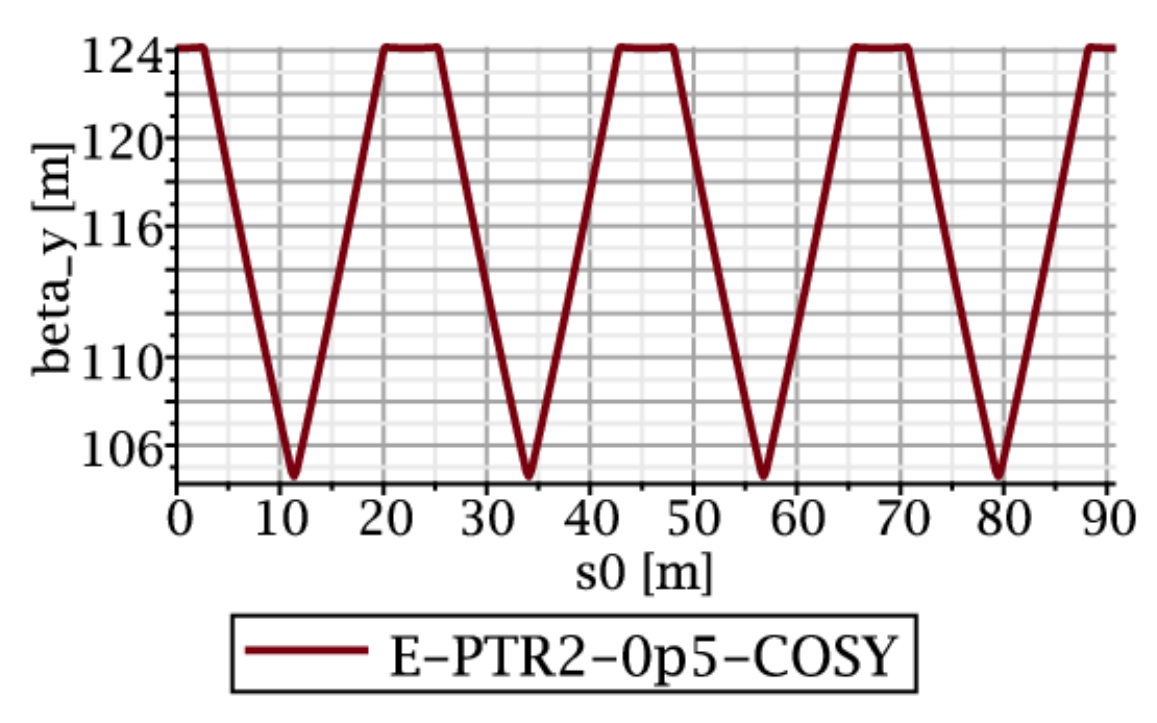}
\caption{$\beta_y(s)$ calculated by ETEAPOT the {\tt E-PTR2} rings is shown on the left. $\beta_y(s)$ calculated by MAPLE linearized  theory for 
the {\tt E-PTR2} rings is shown on the right.}
\label{fig:E-PTR2-ETEAPOT-betay}
\end{minipage}
\end{figure*}

\begin{figure*}[htbp]
\begin{minipage}[b]{0.45\linewidth}
\centering
\includegraphics[scale=0.45]{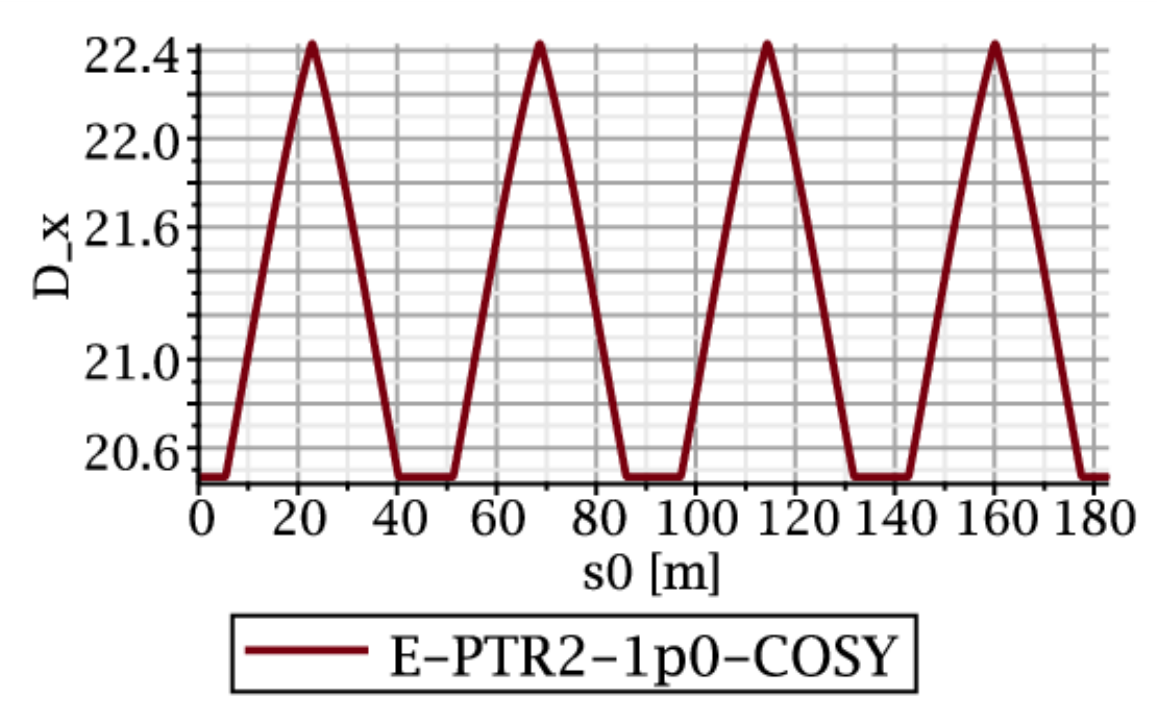}\includegraphics[scale=0.45]{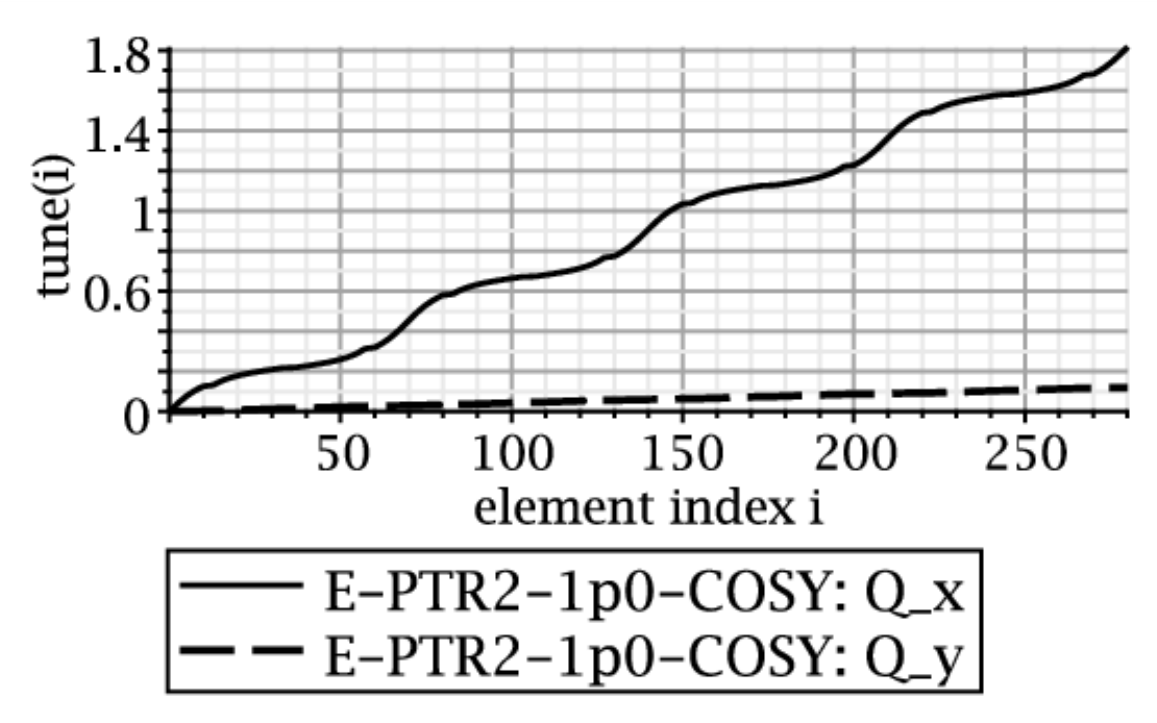}
\includegraphics[scale=0.45]{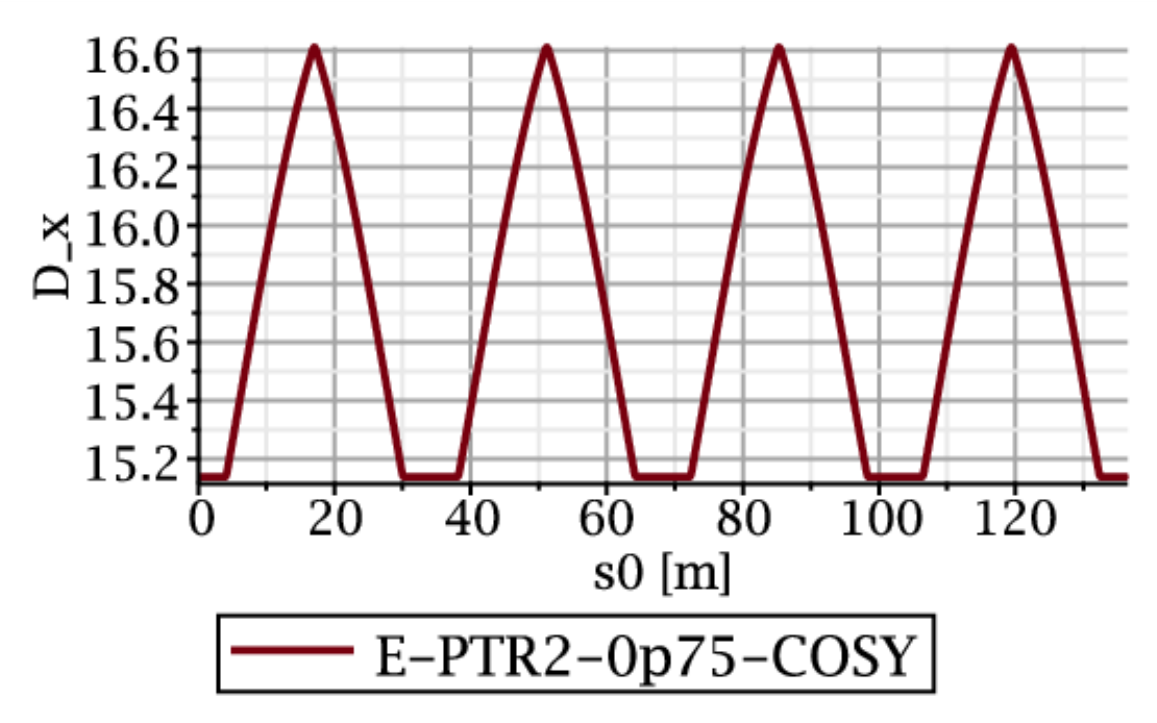}\includegraphics[scale=0.45]{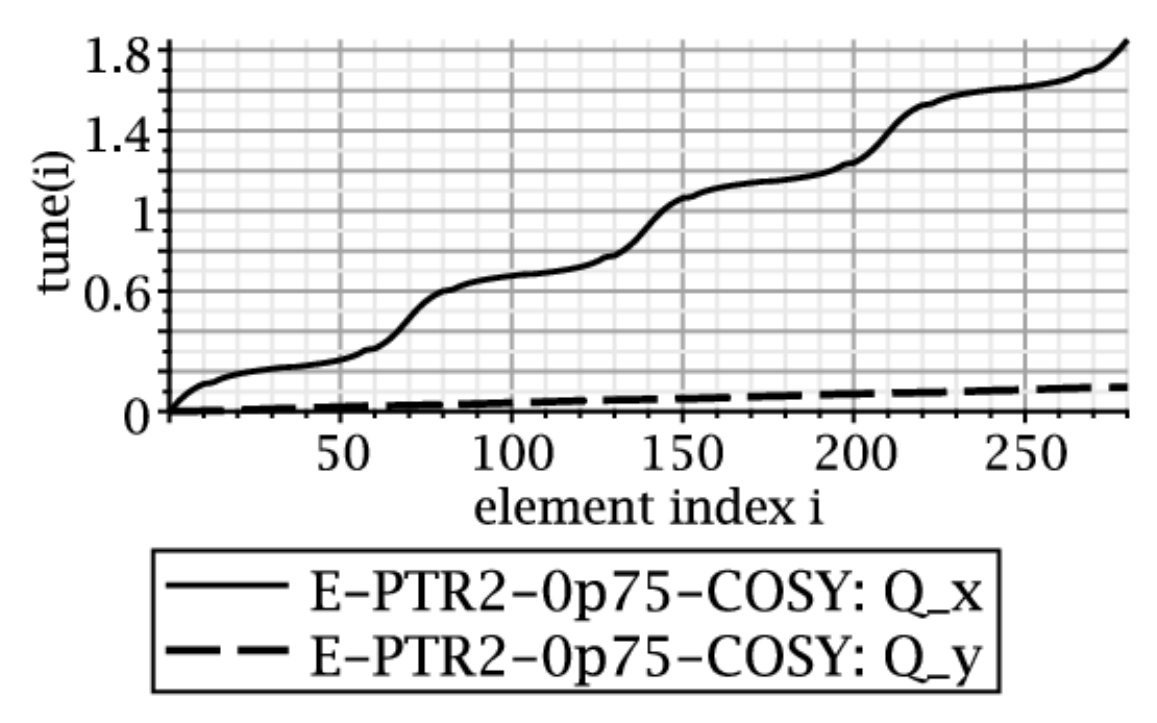}
\includegraphics[scale=0.45]{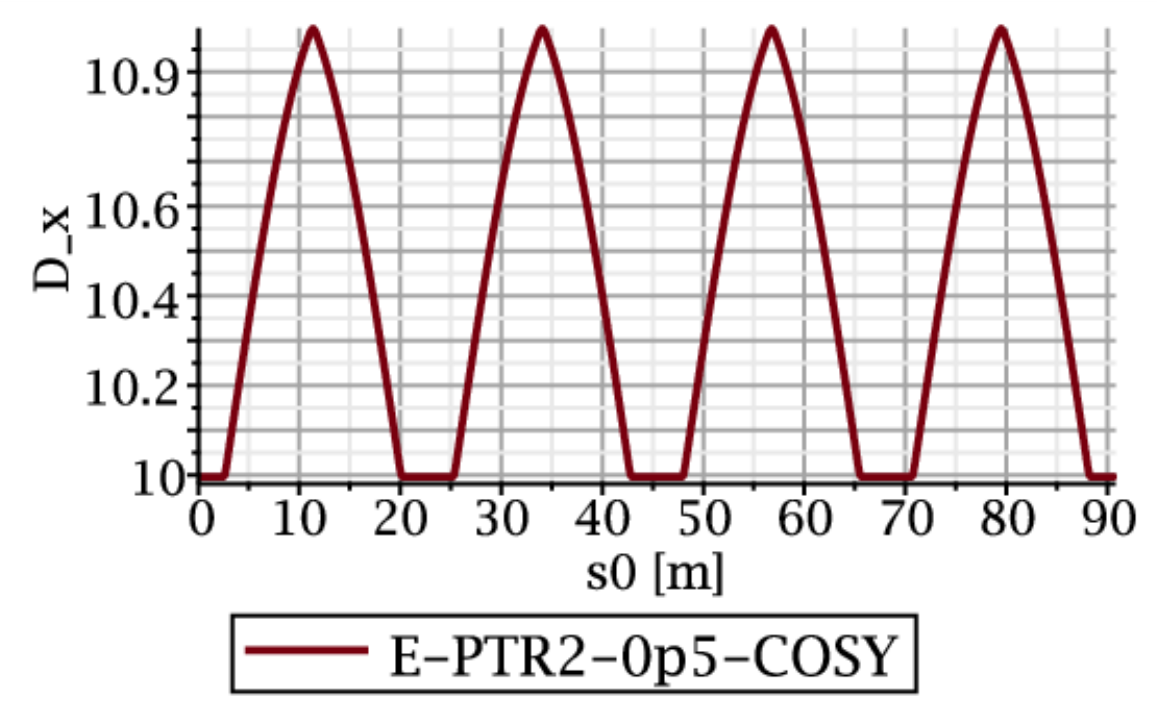}\includegraphics[scale=0.45]{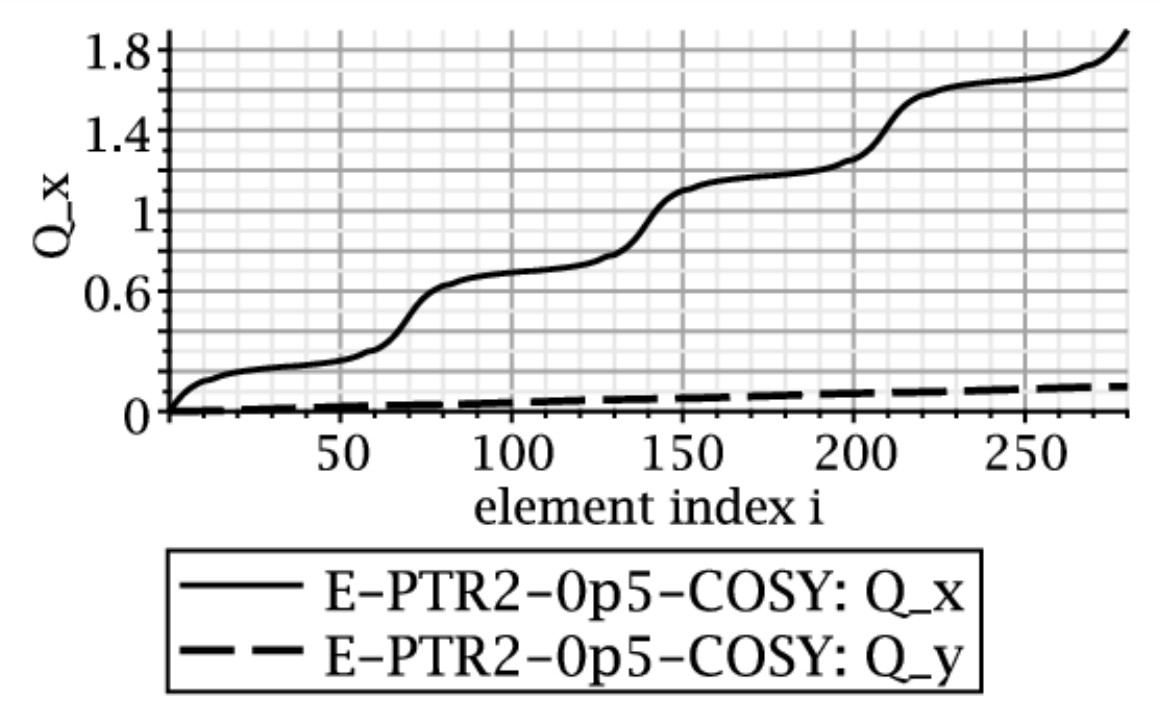}
\caption{\label{fig:E-PTR2-MAPLE-dispersion}$D(s)$ calculated by MAPLE linearized  theory for the {\tt E-PTR2} rings is shown on the left.
tune advance calculated  by MAPLE linearized  theory for the {\tt E-PTR2} rings is shown on the right.}
\end{minipage}
\end{figure*}

There is good qualitative agreement between MAPLE and ETEAPOT.
Quantitative agreement is not perfect though. This is at least partly
due to a significantly large ``perturbative'' correction from the 
$m=1$ electrode-shape focusing ``Mu\~noz-Pavic'' limit in which ETEAPOT is analytically
exact, to the $m\approx0$ ``cylindrical'' electrode case, applicable in this,
and all other rings in this appendix. Side-by-side horizontal beta function comparisons 
are exhibited in Figures~\ref{fig:E-PTR2-ETEAPOT-betax}. 
Vertical beta function comparisons are exhibited in Figures~\ref{fig:E-PTR2-ETEAPOT-betay} 
Horizontal dispersion functions are shown in Figure~\ref{fig:E-PTR2-MAPLE-dispersion}, along with 
betatron tune accumulation plots shown on the right..

Along with other parameters, much the same lattice function information is shown in 
compressed tabular form in the tables shown right after the figures; 
Table~\ref{tbl:1p0}, ~\ref{tbl:0p75}, and ~\ref{tbl:0p5}.

\begin{table*}[hbp]\footnotesize
\caption{\label{tbl:1p0}MAPLE/ETEAPOT comparisons for lattice {\tt E-PTR2-1p0-COSY}} 
\medskip
\centering
\begin{tabular}{|c|c|c|c|c|}           \hline
parameter            & parameter name         & unit & ETEAPOT [-0.032] &    MAPLE      \\ \hline
super-periodicity    &  $N_{\rm S}$             &     &      4       &       4       \\
bend radius          &  {\tt r0}              &  m   &   21.0      &      21.0      \\
long straight length &  {\tt llstot}          &  m   &  11.760     &  11.760        \\
half bend per cell   &  $\Theta=\pi/N_{\rm S}$ &  r    &  0.785398   & 0.785398      \\
half bend length     & {\tt leh}              &  m   &  8.247      &  8.247        \\
circumference        & $\mathcal{C}$          &  m   &   183.0     &  183.0        \\
quad strengths       & $qF,qD$                & 1/m  & 0.00476, -0.00476 & 0.00476, -0.00476 \\ 
(alternating) field index &  $m$              &      &  $\pm0.002$ & $\pm0.002$    \\  \hline
electric field, 30\,MeV p & $E$                  & MV/m &  2.813      &  2.813              \\
electrode voltages   & $V$                    & KV   & $\pm$98.45  & $\pm$98.45    \\
electrode separation & $gap$                  & cm   &     7       &      7        \\  \hline
horz beta, min/max   & $\beta_x$              &  m   & 5.898, 5.642 & 5.061, 55.08   \\ 
vert beta, min/max   & $\beta_y$              &  m   & 203.6, 281.8 &  220.7, 259.5  \\ 
dispersion, max,min  & $D_x$                  &  m   &             &  20.5, 22.4   \\
horizontal tune      &  {\tt Qx}              &      &  1.7390     &    1.8163    \\ 
vertical tune        &  {\tt Qy}              &      &  0.1193     &    0.1196    \\
\hline
\end{tabular}
\end{table*}
\begin{table*}[hbp] \footnotesize
\caption{\label{tbl:0p75}MAPLE/ETEAPOT comparisons for lattice {\tt E-PTR2-0p75-COSY}} 
\medskip
\centering
\begin{tabular}{|c|c|c|c|c|}           \hline
parameter            & parameter name         & unit & ETEAPOT [-0.032] &    MAPLE      \\ \hline
super-periodicity    &  $N_{\rm S}$            &      &      4        &       4       \\
bend radius          &  {\tt r0}              &  m   &    15.4      &    15.4       \\
long straight length &  {\tt llstot}          &  m   &   8.932      &  8.932        \\
half bend per cell   &  $\Theta=\pi/N_{\rm S}$ &  r    &  0.785398    & 0.785398      \\
half bend length     & {\tt leh}              &  m   &   6.047      &  6.047        \\
circumference        & $\mathcal{C}$          &  m   & 136.49       & 136.49         \\
quad strengths       & $qF,qD$                & 1/m  & 0.0065, -0.0065 & 0.0065, -0.0065  \\ 
(alternating) field index &  $m$              &      &  $\pm0.002$  & $\pm0.002$    \\  \hline
electric field, 30\,MeV p& $E$                  & MV/m &  3.836       &  3.836       \\
electrode voltages   & $V$                    & KV   & $\pm$134.2   & $\pm$134.2  \\
electrode separation & $gap$                  & cm   &     7        &      7        \\  \hline
horz beta, min/max   & $\beta_x$              &  m   & 3.809, 46.5  & 3.183, 47.43  \\ 
vert beta, min/max   & $\beta_y$              &  m   &   149, 204.0 & 161.7, 190.8  \\ 
dispersion, max,min  & $D_x$                  &  m   &              &  15.2, 16.6    \\
horizontal tune      &  {\tt Qx}              &      &  1.776       &  1.851        \\ 
vertical tune        &  {\tt Qy}              &      &  0.1237      & 0.1214        \\
\hline
\end{tabular}
\end{table*}
\begin{table*}[hbp]\footnotesize
\caption{\label{tbl:0p5}MAPLE/ETEAPOT comparisons for lattice {\tt E-PTR2-0p5-COSY}} 
\medskip
\centering
\begin{tabular}{|c|c|c|c|c|} \hline
parameter            & parameter name         & unit & ETEAPOT [-0.032] &    MAPLE      \\ \hline
super-periodicity    &  $N_{\rm S}$             &     &      4       &       4       \\
bend radius          &  {\tt r0}              &  m   &    10.0     &     10.0      \\
long straight length &  {\tt llstot}          &  m   &     6.0     &      6.0      \\
half bend per cell   &  $\Theta=\pi/N_{\rm S}$ &  r    &  0.785398   & 0.785398      \\
half bend length     & {\tt leh}              &  m   &   3.92699   &  3.92699      \\
circumference        & $\mathcal{C}$          &  m   &   90.832    &  90.832       \\
quad strengths       & $qF,qD$                & 1/m  & 0.01, -0.01 & 0.01, -0.01   \\ 
(alternating) field index &  $m$              &      &  $\pm0.002$ & $\pm0.002$    \\  \hline
electric field, 30\,MeV p & $E$                    & MV/m &   5.91      &   5.91        \\
electrode voltages   & $V$                    & KV   & $\pm$206.7  & $\pm$206.7     \\
electrode separation & $gap$                  & cm   &     7       &      7        \\  \hline
horz beta, min/max   & $\beta_x$              &  m   & 1.919, 40.0  & 1.04, 39.5   \\ 
vert beta, min/max   & $\beta_y$              &  m   & 90.2, 129.0  & 104.6, 124.9 \\ 
dispersion, max,min  & $D_x$                  &  m   &              & 10.00, 10.99 \\
horizontal tune      &  {\tt Qx}              &      &   1.833      &    1.898     \\ 
vertical tune        &  {\tt Qy}              &      &   0.1320     &    0.1247    \\
\hline
\end{tabular}
\end{table*}
\begin{figure*}[h]
\centering
\includegraphics[scale=1.2]{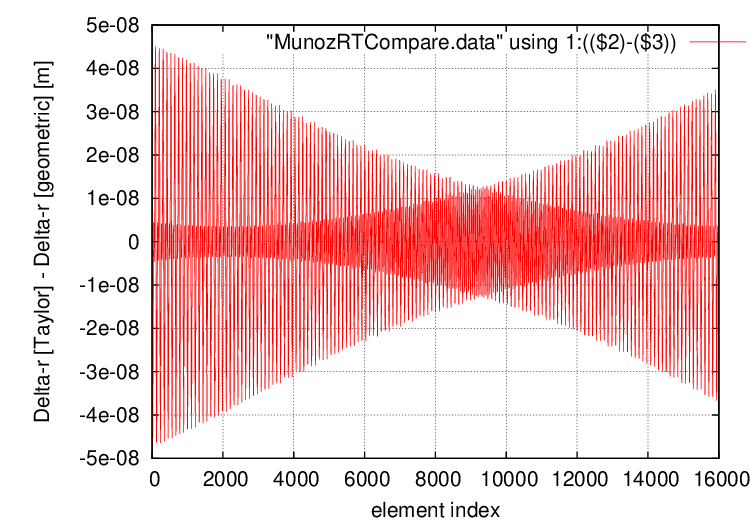}
\caption{\label{fig:MunozRT-compare}Deviation between revolutions of
transverse coordinate $x$ as calculated by (accelerator-style, periodicity-enforced) 
Taylor series formulation and the (geometric, or Mu\~noz, or 
``astronomical'') formulation.  Though the agreement is splendid over this
many revolution, one sees trouble ahead, in the form of exponential deviation 
with time.  Furthermore, close inspection of the plot, already shows indication of
the onset of chaotic motion---it is evident from irregularities visible in the overap regions.  
This shows that, in spite of its elegant analytic form,
the astronomical formulation is inappropriate for accelerator physics.}
\end{figure*}
Figure~\ref{fig:MunozRT-compare}  compares transverse evolution
as calculated by the ``geometric'' or ``Mu\~noz'' analytic (cosmological scale) 
approach and by the (more nearly conventional accelerator) truncated 
Taylor series approach.  

For inverse square law ($m$=1) electrode shapes, evolution through thick elements 
in ETEAPOT (as in TEAPOT) is analytically exact. This includes also spin evolution.
This exact treatment is only possible for certain electric or magnetic 
force fields---uniform magnetic fields for magnets, inverse square law 
(a.k.a. ``Kepler'' or ``spherical'') for electric fields. In neither case 
will the as-built, realistic accelerator force fields be ideal. 
In ETEAPOT (as in TEAPOT) the deviations from ideal are represented by thin
(and hence symplectic) elements. This approach proved valid in TEAPOT,
for both electrons and protons, for example in calculating dynamic 
apertures of SSC, LHC, CESR, etc.  We regard this approach to be essential
for faithful spin tracking.

The beauty of point sources in mechanics is that the force fields
are central. As a result, their 3D orbits are, in fact, restricted to two 
dimensional planes.  For non-relativistic Kepler planetary orbits the 
orbits are perfect ellipses. The protons in proposed E\&m storage rings 
are sufficiently relativistic that the orbits are not exactly elliptical,
but the deviations from ideal ellipse show up only in ``advances of
perihelion'' typically of the  order of tens of degrees per revolution.  
Through short bend elements the orbits look very much like 
ellipses---in fact they even look very much like circular arcs. The exact 
analytic expressions can be represented by ``ellipses'' with ``varying constant'' 
major axis and eccentricity. This evolution is well known and is known to be valid,
even at cosmological scale, as long as general relativity effects remains negligible.

The orientation of an (arbitrarily relativistic) Kepler orbit can be specified by
the orientation of its major and minor axes, and the particle position can be 
specified by a simple analytic formula (first obtained by Newton) depending 
only on an angle $\theta-\theta_0$ which is the angle between the particle 
angle $\theta$ and the perihelion angle $\theta_0$.
  
The description of accelerator orbits as (relativistic) Kepler orbits
has one inelegant feature; namely, on the central design orbit, the orbit is
circular and, in this limit, there is no unambiguous perihelion angle. 
In this sense, \emph{the central design orbit is singular}.  Superficially this
might seem not to be a problem, as the code can define the perihelion angles
arbitrarily for each particle and then keep track of their evolutions.
This includes the requirement, when the orbit switches from oblate 
to prolate, for $\theta_0$ to change discontinuously. 
\emph{Regrettably, this is easier said than done.}

Transverse evolution through the same 16,000 elements as
calculated two independent ways is compared in 
Figure~\ref{fig:MunozRT-compare}. The difference between
radial position calculations never exceeds 
$4.5\times10^{-8}\,$m and, on the average, the deviation
is less than $10^{-9}\,$m.  Even so, evidence of chaotic deviations,
presumably because of aphelion/perihelion confusion in the 
purely analytic treatment, remain visible. 

Because of the singularity of the central orbit, this chaotic motion 
causes chaotic uncertainty not unlike the Heisenberg uncertainty
implicit in wave-particle duality.

\subsection{{\bf D. Longitudinal beam dynamics}}\label{sec:longitudinal-dynamics}\mbox{}

\subsubsection{Fractional momentum and energy offsets.\ }\mbox{}

Different communities employ different definitions of the fractional
momentum or energy offset.  Wollnik (Section 4.1.1.2) defines the 
``rigidity offset'' $\Delta$ of an off-energy particle by  
\begin{equation}
\delta_{\rm Wollnik} \equiv \Delta = \frac{r - r_0}{r_0},
\end{equation}
where $r_0$ is the nominal bend radius, and $r=r_0(1 + \Delta)$ is the radius of
the concentric circular orbit followed by the off-momentum particle.
This is the fractional offset coordinate applicable to the linearized transfer 
matrix benchmark comparison results in this report.  For ``sector bends'' in 
which all orbits enter and exit more or less normal to the edges of the
bend element there is hardly any need to distinguish between radial
offsets outside and inside bend elements.
ETEAPOT (like MAD) defines a fractional energy offset,
\begin{equation}
\delta_{\rm UAL} 
 \equiv 
\frac{{\mathcal E}^O-{\mathcal E_0}^O}{p_0^Oc}.
\label{eq:Offset.2}
\end{equation}
The superscript $O$ is attached to ${\mathcal E}^O$ and $p^O$ here to specify 
that these quantities are restricted to regions ``outside'' bend regions;
i.e. where the potential energy vanishes. (For ${\mathcal E}$ this notation
is redundant since ${\mathcal E}$ is preserved except in RF cavities. As it
happens it is also redundant for $p_0$ since, by definition, the potential
energy vanishes everywhere on the design orbit.)
\begin{itemize}
\item
It is shown in the ETEAPOT manual that the Wollnik
$\Delta$ parameter is related to the MAD/UAL momentum deviation 
factor $\delta_{\rm UAL}$ by
\begin{equation}
\Delta 
 =
\Big(1 + \frac{1}{\gamma_0^2}\Big)\,\frac{1}{\beta_0}\,\delta_{\rm UAL}
\big(
 = 
2.744\,\delta_{\rm UAL}
\big)
\label{eq:chromatic.6q}
\end{equation}
for the nominal full scale all-electric proton EDM experiment. 

\item
A notation commonly employed in the electron and proton accelerator 
worlds (with jargon ``delta p over p'') is
\begin{equation}
\delta_p
 \equiv 
\frac{p^O-p_0}{p_0}.
\label{eq:Offset.4}
\end{equation}
(We intentionally refrain from introducing a
momentum-inside variable $\delta_{p^I}$; defined in terms of this variable the
dispersion $D_{\delta_{p^I}}$ would be singular for field index $m=0$.)
Relations among fractional \emph{outside} momentum or energy coordinates,
since they are unaffected by potential energy, can be
obtained using standard (for magnetic lattices) kinematic
formulas, starting from
\begin{equation}
{\mathcal{E}}^2 = p^2c^2 + m_p^2c^4.
\label{eq:Offset.9}
\end{equation}
For example,
\begin{equation}
\frac{dp}{p_0}
 =
\frac{1}{\beta_0}\,
\frac{d\mathcal{E}^O}{p_0c},
\quad\hbox{or}\quad
\delta_p = \frac{1}{\beta_0}\,\delta_{\rm UAL}.
\label{eq:Offset.10}
\end{equation}
To obtain $\delta_p$ (also known as ``delta p over p'') from 
$\delta_{\rm UAL}$ one must therefore divide $\delta_{\rm UAL}$ by 0.6.
\item
Should one want the absolute change in $\gamma$ corresponding to $\delta_{\rm UAL}$
one has to multiply $\delta_{\rm UAL}$ by $p_0c/(m_pc^2)=0.7/0.938=0.746$.
This relation can also be expressed as 
$d\gamma=\beta_0\gamma_0\delta_{\rm UAL}$.
\end{itemize}

Having established the principle behind the $O$-superscript notation,
from here on quantities without superscripts are to be interpreted
as \emph{outside} quantities. This simplifies the discussion of synchrotron
oscillations where it is assumed that the bend field 
electric potential vanishes throughout RF cavities.

\subsubsection{Analytic longitudinal parameters.\ }\mbox{}

This section discusses the analytical treatment used to estimate
parameters for checking the numerical ETEAPOT results. It is the
only section in which it is necessary to deal with the confusing
subject of kinematic quantities in the interior of bend elements.
The (weak) quadrupoles present in the lattice to trim the tunes are 
neglected for this treatment. This causes an uncontrolled, but 
hopefully small, uncertainty in the comparison values. 

For circular orbits in an electric bend element 
with radial electric field $-E_0$, the radius of 
the central orbit is given by
\begin{equation}
r_0 
 =
\frac{m_pc^2/e}{E_0}\,\gamma_0\beta_0^2.
\label{eq:Offset.5}
\end{equation}
For field index $m$ the radius of curvature of an off-momentum
circular orbit is given by
\begin{equation}
 r
 = 
r_0\,
\Big(
\frac{\gamma_0^2-1}{\gamma_0}\,
\frac{\gamma^I}{{\gamma^I}^2-1}
\Big)^{1/m},
\label{eq:Offset.6}
\end{equation}
where $\gamma_0$ is the value on the central orbit. This relationship
is singular for field index $m=0$ (which is also known as the 
``cylindrical'' case because the electrodes giving this radial
dependence are cylindrical). $m=0$ is
also the unique case in which the dependence of electric potential on
radius $r$ is logarithmic and cannot be expressed by a power law.

For values of $m$ near 1 the orbit velocities of off-energy circular 
orbits within bend elements are approximately independent of $r$.
This causes the time of flight of off-energy
closed orbits through arcs (which are perfect circles) 
to be dominated by the arc length, which
is the bend angle multiplied by $r$. Since the speed depends only
weakly on momentum this is ``above transition'' behavior.
On the other hand, off-energy closed orbit path lengths through drifts 
are independent of $\delta_{\rm UAL}$, which causes time of flight of 
off-energy closed orbits through drifts to be dominated by the 
particle speed; this is ``below transition'' behavior. Assuming
the drift lengths are fairly short, the net behavior is 
``above-transition'' like.

For negative $m$ values the radius of the off-energy closed orbit 
increases with increasing energy offset. The dispersion $D$ is 
therefore positive for $m<0$; furthermore the magnitude $|D|$ 
\emph{increases} as $m$ approaches zero. 
For positive $m$ values (such as the $m=1$ ``spherical case'')
the radius of off-energy closed orbits decrease with increasing
energy offset. The dispersion $D$ is therefore negative for $m>0$;
but the magnitude $|D|$ again \emph{increases} as $m$ approaches zero. 

Realistic lattices also have quadrupoles which affect the dispersion. 
But, since the quadrupoles are very weak in the three benchmark 
lattices, their influence on lattice dispersion can be expected to
be not very important. (For the minimized-dispersion, strong horizontal
focusing, combined function lattices to be recommended as a 
response to results in the present report,
this will no longer be even approximately valid.)  

Combining all these statements, the dispersion function $D$ has to
have a singularity in the vicinity of $m=0$. It is the longitudinal
dynamics (rather than the transverse) that is sensitive to lattice
dispersion. To make them as simple as possible, the $m$-values of the 
three benchmark lattices were chosen close to zero.  Inadvertently this
choice has caused the longitudinal dynamics of the benchmark lattices
to be hyper-sensitive.

An example of this hypersensitivity comes about in comparing computer
simulation and analytic results. The analytic results depend on the
dispersion. But the dispersion is not used, and is not available, during 
particle tracking. This makes it a priori unknown what RF phase will give 
stable longitudinal motion in simulation. We resolve this ambiguity by 
trying two RF phases differing by $\pi$ and choosing the phase that 
gives stable motion.

Another subtle complication of operation near $m=0$ comes
about because a particle having positive velocity offset outside
bends can have either positive or negative velocity offset inside 
bends, depending on the sign of $m$. This dependence is transparent
to the computer simulation but complicates the analytic calculation
of the ``slip factor'' needed to calculate the synchrotron oscillation 
frequency. 
 
In a truly uniform, weak-focusing, electric
ring the off-momentum closed orbits are true circles
with radius $r$, in which case Eq.~(\ref{eq:Offset.6}) 
provides an analytic formula valid for all amplitudes.
For our purposes this is somewhat academic however, as
explicit quadrupoles make the benchmark lattices 
slightly ``separated-function''. Deflections in the 
quadrupoles cause the off-momentum trajectories to be 
somewhat non-circular in the bends. 

\subsubsection{Dispersion of the benchmark lattices.\ }\mbox{}

Of the benchmark lattices, the closest to the ideal
case is the {\tt E\_BM\_Z.RF} lattice for which the
quadrupoles are very weak. We have found however,
after the quadrupoles have been adjusted to give
identical tunes for the three benchmark cases, that
the longitudinal behaviors of the three lattices are
essentially identical. For example we 
will use Eq.~(\ref{eq:Offset.6}) 
to estimate the dispersion for the {\tt E\_BM\_M1.0} lattice.

In Eq.~(\ref{eq:Offset.6}) the value $m=0$ is singular; this corresponds to
the fact that, for $m=0$, $\gamma^I$ is equal to $\gamma_0$, 
independent of $r$. Exploiting this independence, and neglecting
a small $m$-dependent term that comes from treating the radial
electric field as being independent of $r$, the dependence of
$\gamma$ (which is $\gamma^O$ by definition) and $r$ is
\begin{equation}
\mathcal{E}^O
 =
 m_pc^2\gamma
 \approx 
m_pc^2\gamma_0 + eE_0(r-r_0).
\label{eq:Offset.7}
\end{equation}
Differentiating with respect to $r$ gives
\begin{equation}
d\mathcal{E}^O = eE_0dr.
\label{eq:Offset.8}
\end{equation}
Using Eq.~(\ref{eq:Offset.2}), this can be re-expressed as  
\begin{equation}
dr = \frac{p_0c/e}{E_0}\,\delta_{\rm UAL}.
\label{eq:Offset.8p}
\end{equation}
The UAL-units dispersion is therefore given by
\begin{equation}
r_{\rm co}(\delta_{\rm UAL})
 =
D_{\rm UAL}\delta_{\rm UAL},
\label{eq:Offset.8qm}
\end{equation}
where
\begin{equation}
D_{\rm UAL} = \frac{p_0c/e}{E_0} = \frac{0.701}{10.48\times10^{-3}} = 66.9\,{\rm m}.
\label{eq:Offset.8q}
\end{equation}
UAL/ETEAPOT tracking, agree well with this estimate.  Figure~\ref{fig:QsVSsqrtVRF}, for which horizontal 
betatron motion has been largely eliminated, the straight line fit using Eq.~(\ref{eq:Offset.8q}) 
agrees well with the tracking results.

\subsubsection{Slip factor and synchrotron tune.\ }\mbox{}

The revolution period of the central particle is given by
\begin{equation}
T_{\rm rev}(\gamma_0)
 =
\frac{2\pi r_0 + D_{\rm tot}}{\beta_0c},
\label{eq:SlipFac.1}
\end{equation}
where $D_{\rm tot}=331.3274-2\pi\times40=80.0$\,m is the accumulated 
straight section length. 
Neglecting the effect of the quadrupoles,
for an off-momentum closed orbit, the
revolution period is
\begin{equation}
T_{\rm rev}(\gamma)
 =
\frac{2\pi r}{\beta_0c} + \frac{D_{\rm tot}}{\beta^Oc}.
\label{eq:SlipFac.2}
\end{equation}
The constancy of $\beta$ within the bend region in the $m=0$ 
case is exploited in the first term, but the actual $\beta^O$
value has to be used in the second term. (Because $D_{\rm tot}$
is so short for the benchmark lattices, this term is small for
the benchmark lattices.  But for long straight sections, as in 
the FNAL option, time of flight through straight regions 
strongly influences the longitudinal motion.)

From $\beta^2=1-1/\gamma^2$ we have, evaluated at $\beta=\beta_0$,
$d\beta=d\gamma/(\beta_0\gamma_0^3)$. Using this in 
Eq.~(\ref{eq:Offset.7}), and applying Eq.~(\ref{eq:Offset.8q}),
the off-momentum, outside, velocity is
\begin{equation}
\beta^O
 \approx
\beta_0
\Big(
1 + \frac{1}{\beta^2_0\gamma_0^3}\,
\frac{E_0}{m_pc^2/e}\,
D_{\rm UAL}\delta_{\rm UAL}
\Big)
 =
\beta_0
\Big(
1 + \frac{D_{\rm UAL}\delta_{\rm UAL}}{r_0\gamma_0^2}.
\Big)
\label{eq:SlipFac.3}
\end{equation}
Substituting $r-r_0\approx D_{\rm UAL}\delta_{\rm UAL}$ 
also into Eq.~(\ref{eq:SlipFac.2}) yields
\begin{equation}
\frac{T_{\rm rev}(\gamma) - T_{\rm rev}(\gamma_0)}{T_{\rm rev}(\gamma_0)}
  =
\frac{1}{\mathcal{C}}\,
\Big(
2\pi
 \mp
\frac{1}{\gamma_0^2}\,\frac{D_{\rm tot}}{r_0}
\Big)
D_{\rm UAL}\delta_{\rm UAL},
\label{eq:SlipFac.4p}
\end{equation}
where $\mathcal{C}$ is the circumference of the design orbit. The $\mp$
factor in the second term allows for the fact that $D_{\rm UAL}$
can have either sign, while the phase slip in the straight section
is necessarily negative.
Edwards and Syphers\cite{EdwSyph} define the ``slip factor'' $\eta_{\rm RF}$
by
\begin{equation}
\frac{T_{\rm rev}(\gamma+\Delta\gamma)-T_{\rm rev}(\gamma_0)}{T_{\rm rev}(\gamma_0)}
 =
\frac{\eta_{\rm RF}}{\beta_0^2}\,
\frac{\Delta\gamma}{\gamma_0},
\label{eq:SlipFac.5}
\end{equation}
and we obtain
\begin{equation}
\eta_{\rm RF}
 =
\beta_0\,
\Big(
2\pi
 \mp
\frac{1}{\gamma_0^2}\,\frac{D_{\rm tot}}{r_0}
\Big)\,
\frac{D_{\rm UAL}}{\mathcal{C}}
 =
-0.92.
\label{eq:SlipFac.6}
\end{equation}
As mentioned above, the negative sign is accounted for in the
tracking simulation by shifting the RF phase appropriately
to produce stable motion.

Edwards and Syphers give the synchrotron tune in terms of the slip
factor:
\begin{equation}
Q_s 
 =
\sqrt{
\frac{1}{2\pi}
\frac{h_{\rm RF}\eta_{\rm RF}\cos(2\pi{\rm lag})}{\beta_0^2\gamma_0}\,
\frac{V_{\rm RF}}{m_pc^2/e}
}
 =
0.0061.
\label{eq:SlipFac.7m}
\end{equation}
The $\eta_{\rm RF}\cos(2\pi{\rm lag})$ product needs to be positive for 
the synchrotron oscillations to be stable. The ${\rm lag}$
factor, which establishes the RF phase, is read in from the SXF
lattice description file. (This is discussed further in a 
concluding section.)

\subsubsection{Dispersion from transfer matrices.\ }\mbox{}

Orbit evolution from the origin at $s=0$ to
a general position $s$ can be expressed by
a transfer matrix ${\bf M}(s,0)$:
\begin{equation}
\begin{pmatrix}
x(s) \\ x'(s) \\ \delta
\end{pmatrix}
  =
\begin{pmatrix}
  M_{11}(s,0)  &  M_{12}(s,0)  &   M_{13}(s,0) \\
  M_{21}(s,0)  &  M_{22}(s,0)  &   M_{23}(s,0) \\
   0      &    0     &    1       
\end{pmatrix}\,
\begin{pmatrix}
x(0) \\ x'(0) \\ \delta
\end{pmatrix}.
\label{eq:Dispersion.2p} 
\end{equation}
Note that the third component is conserved except at RF cavities.
Also some of th matrix elements depend on the definition of $\delta$.
After associating a transfer matrix to each of the elements in the ring, 
${\bf M}(s,0)$ is found by ``concatenating'' (i.e. multiplying) these
matrices.

For fully-relativistic magnetic lattices the third component $\delta$ is customarily
referred to as $\delta p/p$, and this coordinate
can be identified almost exactly with $\delta\gamma/\gamma$ or with fractional
energy offset (from the design orbit). 
But for an only weakly-relativistic EDM lattice which, furthermore, has
electric bending, it is necessary to be more careful. For now we leave
the definition of $\delta$ open, planning to replace it by one or the
other of the fractional offset coordinates introduced in the previous
section. Of course the dispersion function $D_{\delta}(s)$ becomes definite only 
when $\delta$ is defined unambiguously.
 
The ``off-momentum closed orbit'' $x_{\rm c.o.}(\delta,s)$ is defined
to be the unique orbit which, for longitudinal phase space displacement 
$\delta$, closes on itself after a complete turn around the ring.
The dispersion is then defined, in linearized approximation, by
\begin{equation}
x_{\rm c.o.}(\delta,s)
 =
D_{\delta}(s)\,\delta.
\label{eq:Dispersion.1}
\end{equation}
Since the transfer matrix includes a description of the 
influence of energy offset, it can also be used
to find the dispersion function $D_{\delta}(s)$. First the
``fixed point'' at the origin, 
$(D_{\delta}(0), D'_{\delta}(0))$ has to be found.
If the lattice
has mirror symmetry, which is true for the benchmark lattices, 
the origin can be chosen on the axis of symmetry, and $D(s)$ is 
also mirror-symmetric; in this case $D'(s)|_{s=0}=0$. 

Evolution once around
the ring, starting from the origin, of an 
off-energy particle on the closed orbit corresponding to 
its energy is described by the ``once-around'' transfer
matrix ${\bf M}$:
\begin{equation}
\begin{pmatrix}
D(0)\,\delta \\ 0 \\ \delta
\end{pmatrix}
  =
\begin{pmatrix}
  M_{11}  &  M_{12}  &   M_{13} \\
  M_{21}  &  M_{22}  &   M_{23} \\
   0      &    0     &    1       
\end{pmatrix}\,
\begin{pmatrix}
D(0)\,\delta \\ 0 \\ \delta
\end{pmatrix}.
\label{eq:Dispersion.2} 
\end{equation}
This provides two formulas,
\begin{equation}
D(0) 
 =
\frac{M_{13}}{1-M_{11}}
 =
-\frac{M_{23}}{M_{21}},
\label{eq:Dispersion.3} 
\end{equation}
one of which can be used as a consistency check.
Evolution around the ring of $D(s)$ and $D'(s)$ is then
given by  
\begin{equation}
\begin{pmatrix}
D(s) \\ D'(s) \\ 1
\end{pmatrix}
  =
\begin{pmatrix}
  M_{11}(s)  &  M_{12}(s)  &   M_{13}(s) \\
  M_{21}(s)  &  M_{22}(s)  &   M_{23}(s) \\
   0      &    0     &    1       
\end{pmatrix}\,
\begin{pmatrix}
D(0) \\ 0 \\ 1
\end{pmatrix}.
\label{eq:Dispersion.4} 
\end{equation}
Spelled out explicitly,
\begin{align}
D(s)
 &=
D(0)\,M_{11}(s)
+M_{13}(s),       \notag \\
D'(s)
 &=
D(0)\,M_{21}(s)
+M_{23}(s).
\label{eq:Dispersion.5} 
\end{align}

\subsubsection{ETEAPOT longitudinal plots.\ }\mbox{}

\begin{figure*}[hbp]
\includegraphics[scale=0.60]{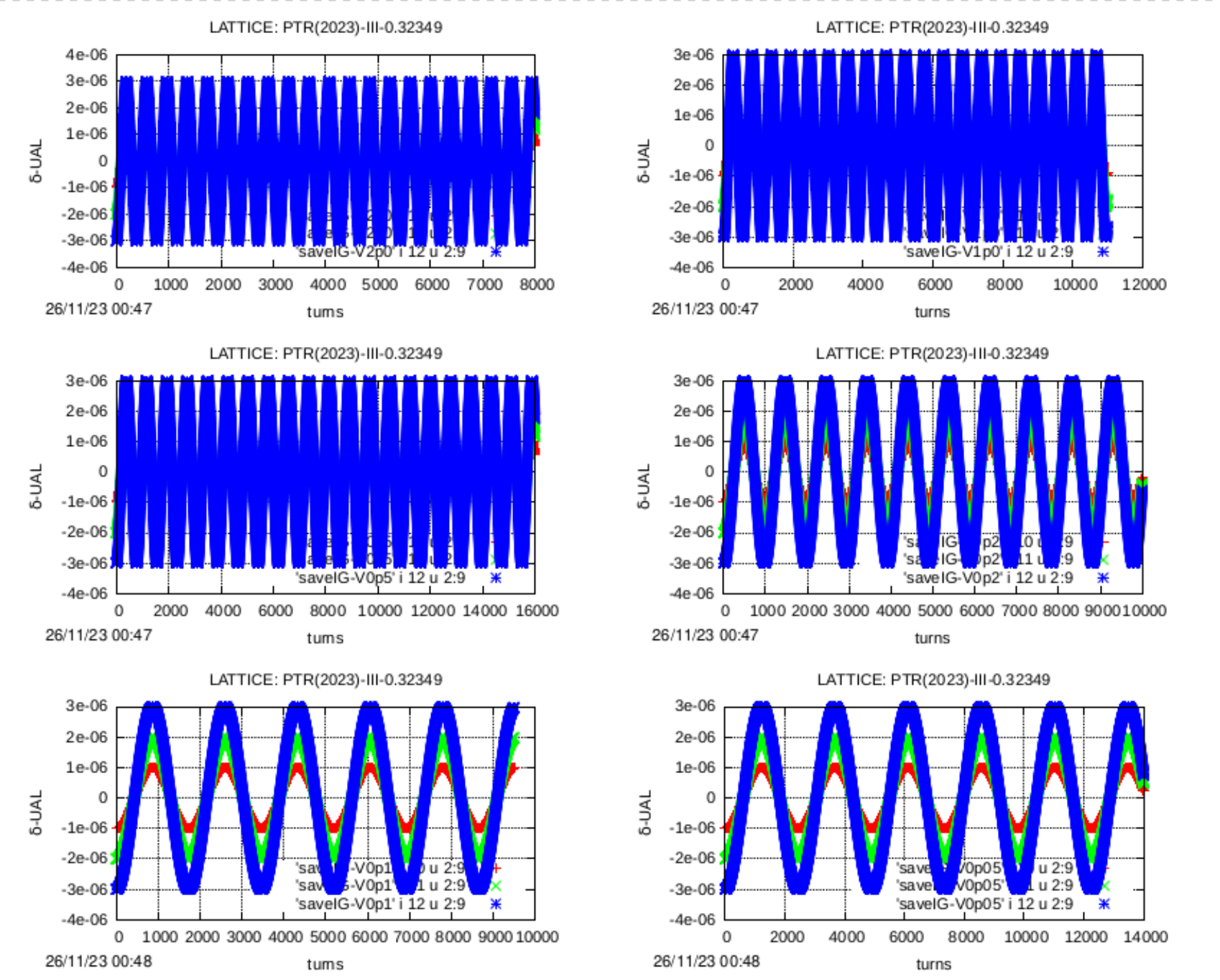}
\caption{\label{fig:delta_vs_turn_V}Fractional offset 
$\delta_{\rm UAL}\equiv\delta{\mathcal{E}}/(p_0c)$
plotted against turn number for
values of RF amplitude $\hat V_{\rm RF}$:
2.0, 1.0, 0.5, 0.2, 0.1, 0.05 [kV]---reading from
left to right then top to bottom; lattice {\tt PTR(2023)-III-0.32349}.
Each of the plots shows three synchrotron oscillation 
amplitudes (with the lower amplitude data largely obscured
by the (blue) large amplitude data).
Starting from the origin with vanishing slopes, the
initial ``momentum offsets'' are 
$\delta_{\rm UAL}$=-0.000001,-0.000002, and -0.000003.
The number of turns tracked in each case is
a (large) multiple of 100, which is also close to
a quarter-integer synchrotron tune advance, aliasing
ambiguities are avoided.  
This provides the synchrotron 
tune values plotted in Figure~\ref{fig:QsVSsqrtVRF}.}
\end{figure*}

\begin{figure*}[hbp]
\includegraphics[scale=0.70]{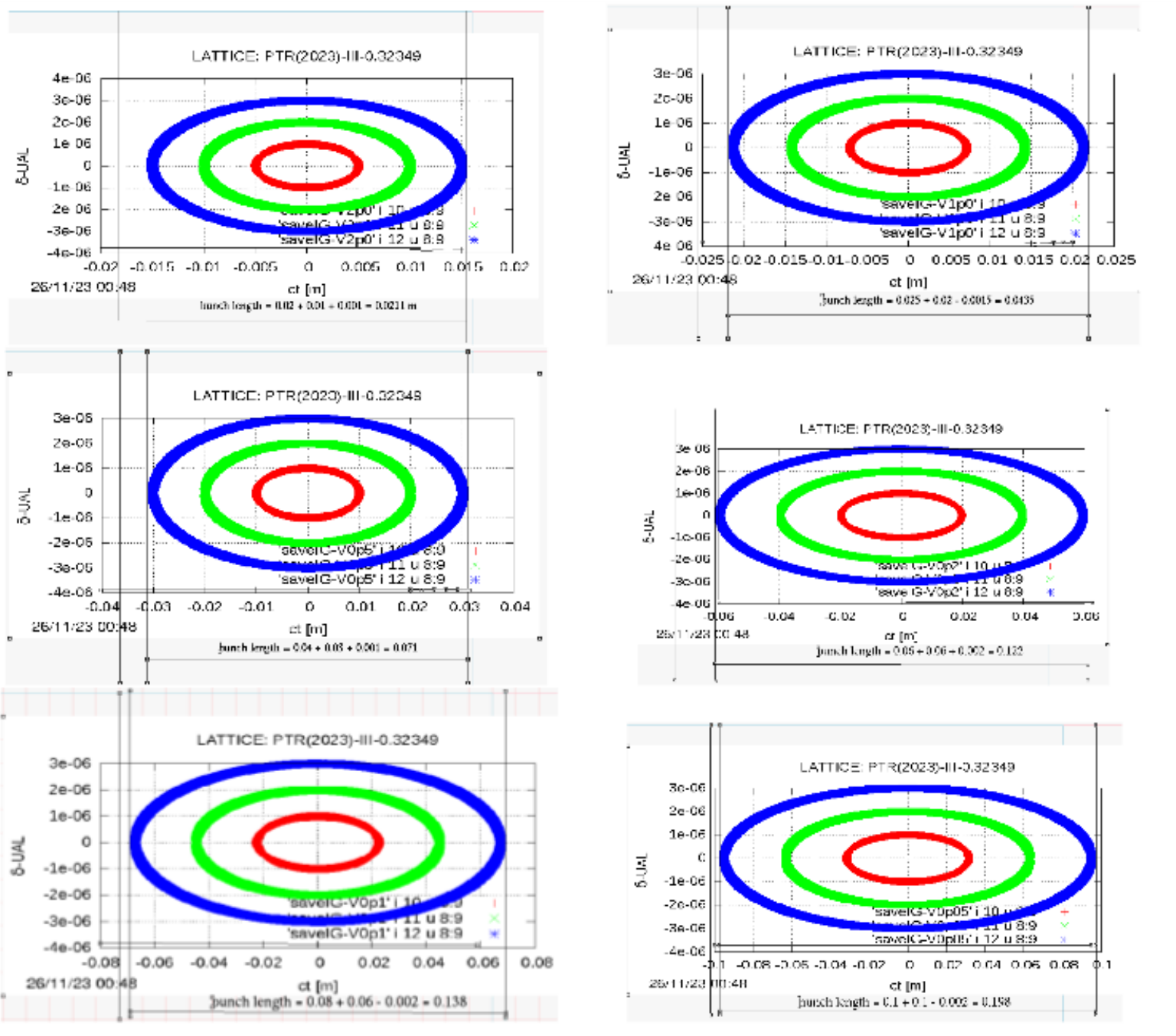}
\caption{\label{fig:delta_vs_ct_V}
Longitudinal phase space plots for the same series of RF voltages,
5.0, 2.0, 1.0, 0.5, 0.2, 0.1, 0.05\,[kV] as in 
Figure~\ref{fig:delta_vs_turn_V}. The horizontal axis is $z$ in meters.
The vertical axis is the fractional offset 
$\delta{\mathcal{E}}/(p_0c)$.}
\end{figure*}

Fractional offset $\delta_{\rm UAL}\equiv\delta{\mathcal{E}}/(p_0c)$ is
plotted against turn number for a range of values of RF amplitude 
$\hat V_{\rm RF}$ in Figure~\ref{fig:delta_vs_turn_V}.  The corresponding 
longitudinal phase space plots are shown in Figure~\ref{fig:delta_vs_ct_V}.

\begin{figure*}[hbp]
\centering
 \includegraphics[scale=1.0]{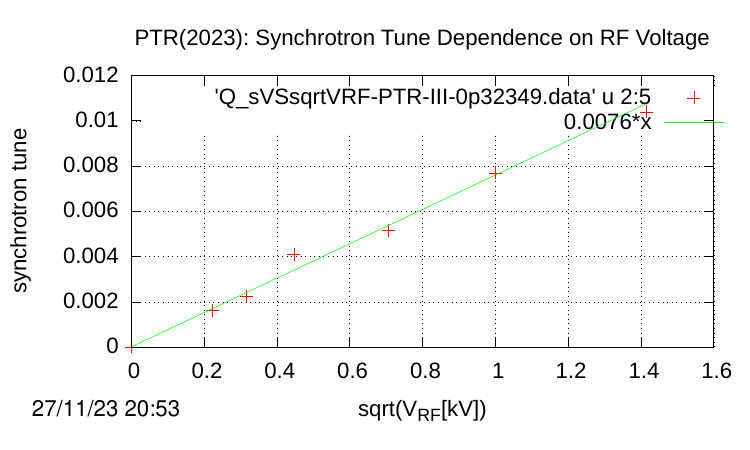}
\caption{\label{fig:QsVSsqrtVRF}Plot of synchrotron tune $Q_s$ 
(obtained by counting periods in plots like those in 
Figure~\ref{fig:delta_vs_turn_V}) versus $\sqrt{V_{\rm RF}[kV]}$.
The fit yields $Q_s= 0.0076\sqrt{V_{\rm RF}[kV]}$.
}
\end{figure*}

\begin{figure*}[hbp]
\centering
\includegraphics[scale=1.0]{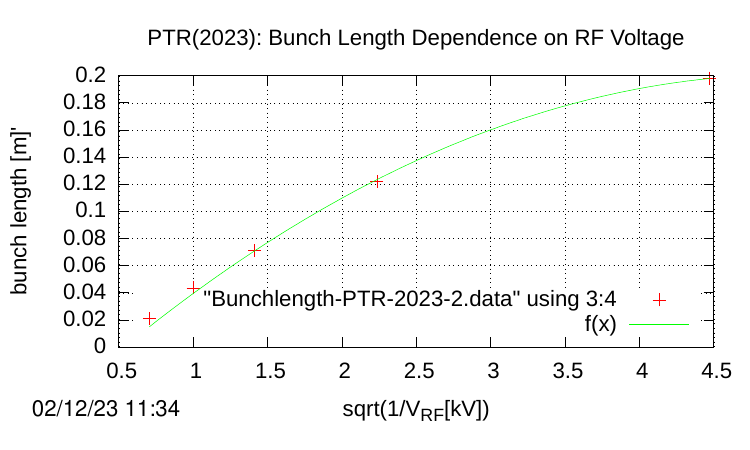}
\caption{\label{fig:BunchlengthVSoneBySqrtVRF}Plot of bunch length
(extremes in plots like those in Figure~\ref{fig:delta_vs_ct_V} for 
$\delta_{\rm UAL}$=0.000003) 
versus $\sqrt{1/V_{\rm RF}[kV]}$.
The best fit in the form ${\rm b.l.} = f(x) = a + b*(x-0.7071) + c*(x-0.7071)**2$, has 
$b = 0.086 +/- 0.0039$,  $c = -0.00996 +/- 0.0011$.  For example, the bunch length
for 1\,KV RF voltage is 4\,cm. 
}
\end{figure*}

From Figures~\ref{fig:delta_vs_turn_V} and \ref{fig:delta_vs_ct_V}
data can be extracted to produce Figure~\ref{fig:QsVSsqrtVRF} and 
Figure~\ref{fig:BunchlengthVSoneBySqrtVRF} 
which plot the synchrotron tune $Q_s$ and the bunch length $\ell_B$
as functions of $V_{\rm RF}$.  Fitting functions are shown in the
captions to the figures. The strict proportionality 
$Q_s\sim\sqrt{V_{\rm RF}}$ is consistent with Eqs.~(\ref{eq:SlipFac.7m}).  
Though the dependence is slightly parabolic, rather than strictly linear,
the bunch length $\ell_B$ is asymptotically  proportional 
to $1/\sqrt{V_{\rm RF}}$.

\section{\bf E.\ Consistent treatment of $g$,$G$ and $g\rightarrow G$}\label{sec:Rational-fractions}\mbox{}

The theoretical importance of anomalous magnetic moments for the experimental investigation of nuclear properties
has been discussed in earlier sections of the present paper.

A chronological listing for NIST values of nuclear physical constants includes, for example, (2014 values), (2010 values) 
(2006 values) (2002 values) (1998 values) (1986 values) (1973 values) (1969 values)\cite{NIST-background}
For example,  ``Atomic Weights and Isotopic Compositions for All Elements''.\cite{NIST-isotope-abundance}


The justification for including this appendix, as explained in reference\ \cite{RT-ICFA}
`` Consistent treatment of $g$,$G$ and $g\rightarrow G$'' is that E\&m storage rings can be utilized as 
``Magnetic Dipole Moment Comparators''. 

In this context, when ultrahigh frequency domain precision is 
required, it is appropriate to have runs long enough for spin orientations to complete an integral number 
of rotations after an integral number of turns.  For this purpose it is appropriate to express the
anomalous MDM as a rational fraction.

\begin{figure*}[hbp]
\caption{\label{fig:Rational-MDM} Rational fraction nuclear MDM $G$-values  } 
\centering
\includegraphics[scale=0.32]{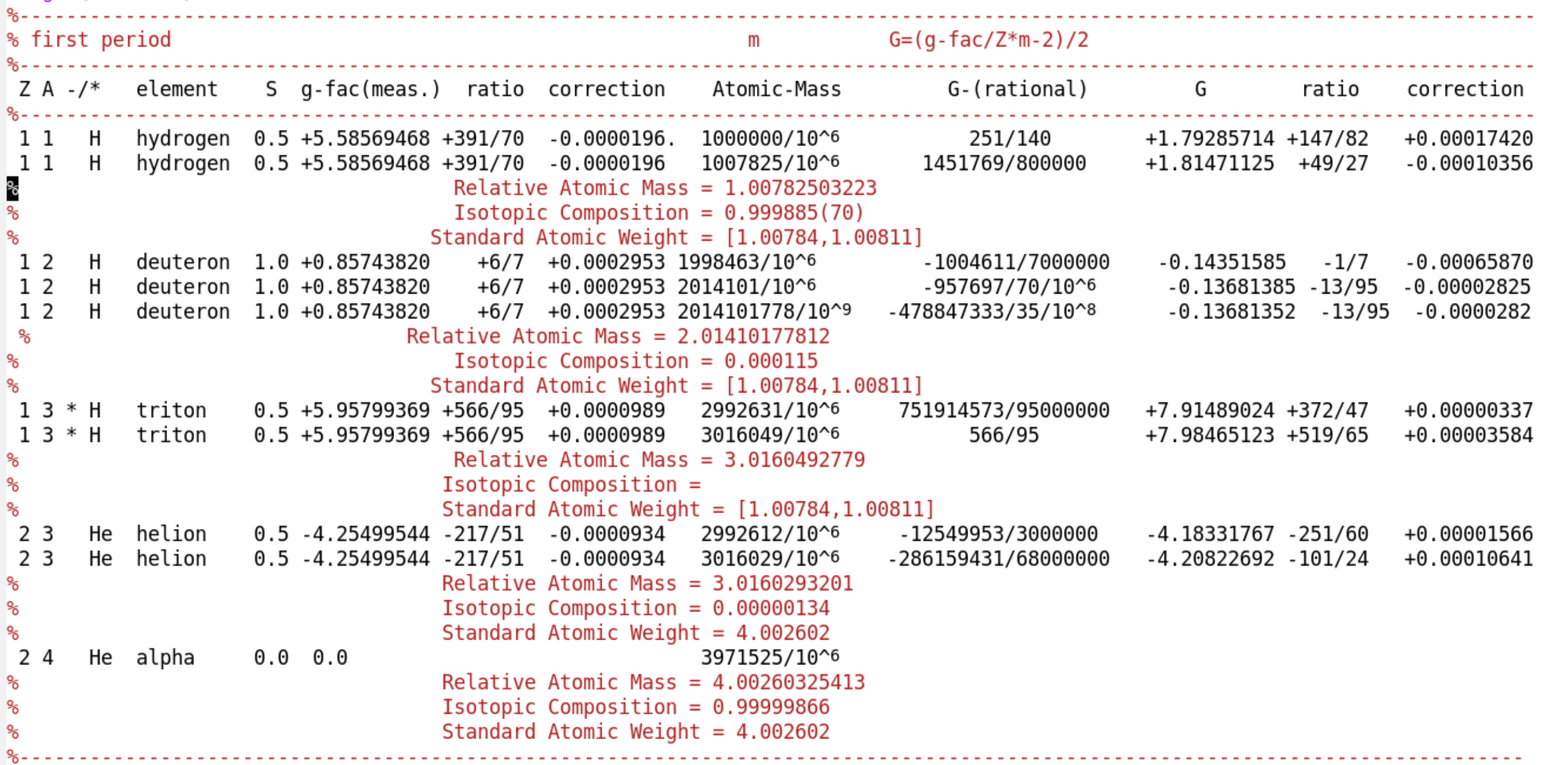}
\includegraphics[scale=0.32]{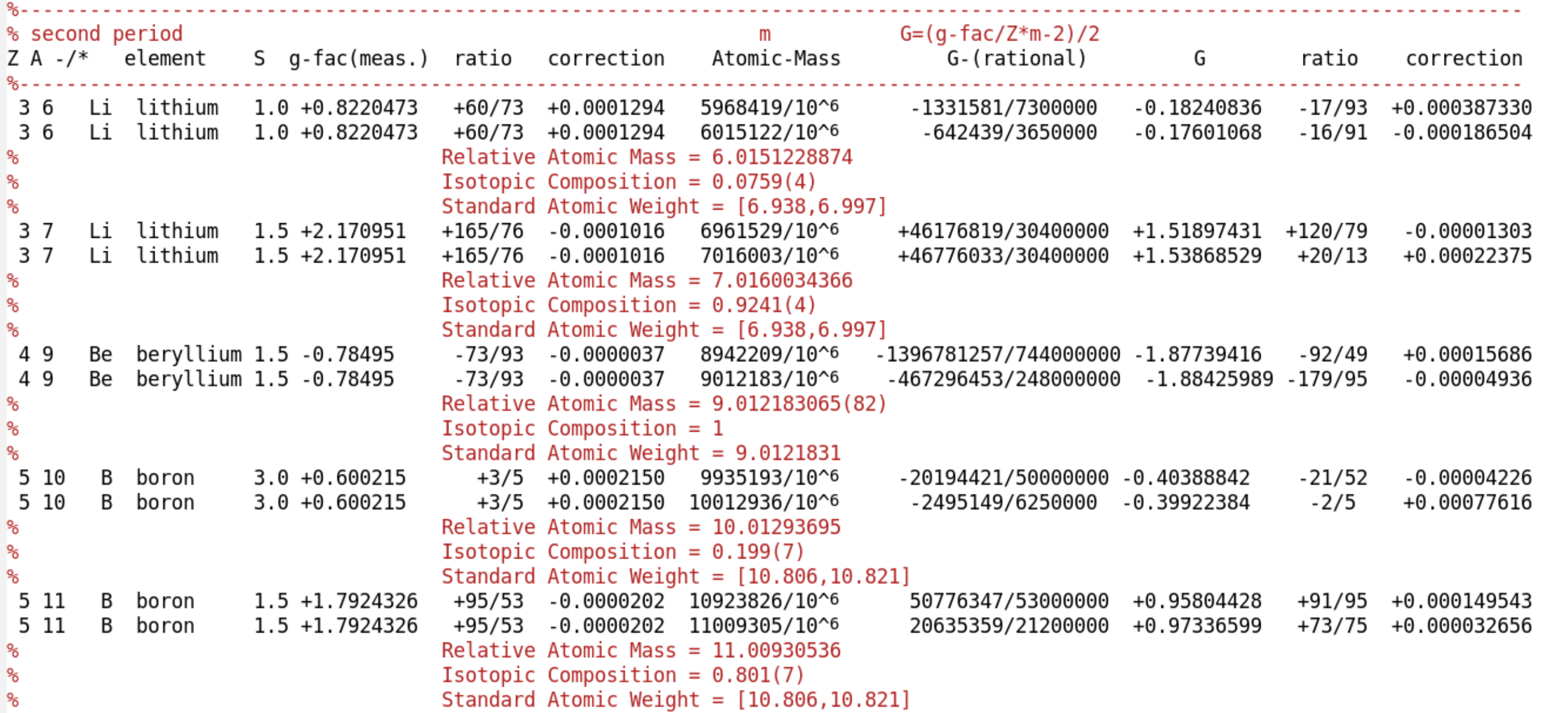}
\includegraphics[scale=0.32]{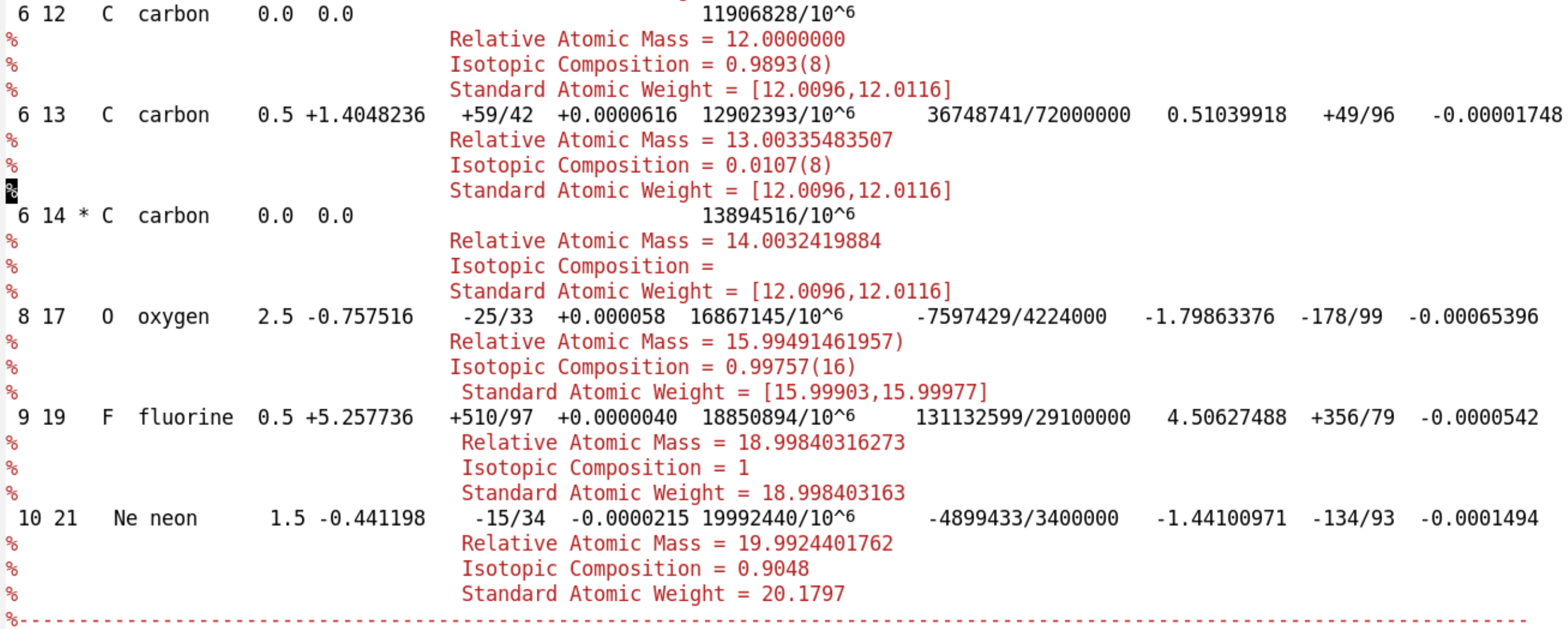}
\includegraphics[scale=0.32]{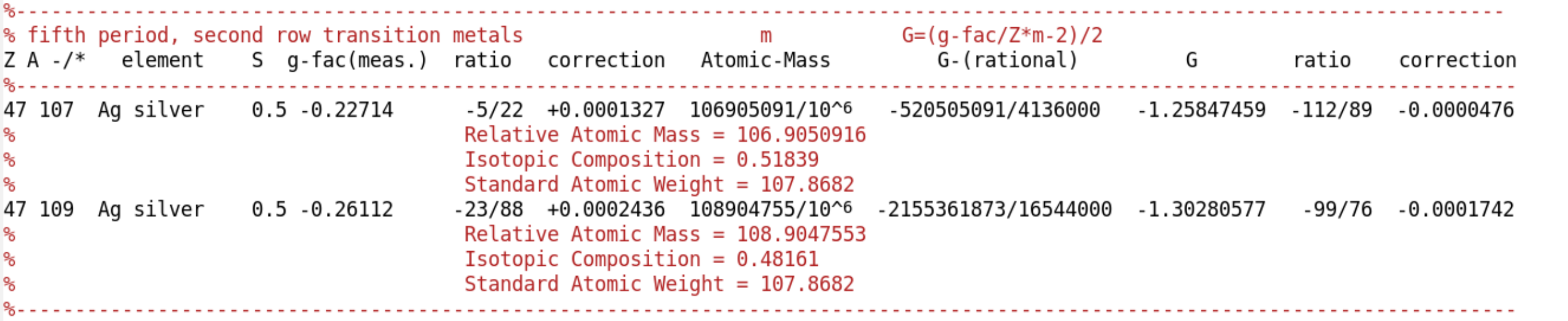}
\end{figure*}

A data table has been assembled from NIST tables and converted to Figure~\ref{fig:Rational-MDM}.  Apart from the
fact that accurate values of the values of anomalous magnetic moments, G, are less systematically
and professionally available available than values of g-factor, g, in spite of the fact that the
values of g and G, are related by a simple formula containing only constants of nature, and measured 
nuclear physical constants.  

Consistent with this paper's emphasis on the importance of nuclear anomalous magnetic moments, this appendix
recommends a procedure, based on integer arithmetic, for most appropriate
conversion from $g$ to $G$, and to provide the conversions for a significant fraction of low mass nuclear 
isotopes.  The conversion $g\ \rightarrow\ G$ uses integer arithmetic and implicates primarily the nuclear isotope 
mass values of $m$ listed in the ``Atomic mass'' column, from NIST tables\cite{NIST-background}\cite{NIST-isotope-abundance}. 

In Figure~\ref{fig:Rational-MDM} the integer entries can be identified by having no decimal points or, better, by being 
represented by ratios of integers.  Integer arithmetic is used in the evaluation of $G=(g\times m/Z-2)/2$, in which $Z$ is
an integer and $m$ is the only measured quantity.  Since $m$ also enters the evaluation $g$ from experimental measurements,
it is important to use the most precise value available and \emph{essential for the same value of $m$ to be
used throughout.} 

The motivation for this is that, along with the obvious requirement that the values of $m$ have to be identical in the 
evaluation of $g$ and $G$, there are no transcendental numbers in the relation between $g$ and $G$.  The importance
of respecting this discipline is apparent in many examples in the table; powers of ten, indicated by exponential notation 
appearing in $m$ column, show up as corresponding trailing zero digits in $G$ denominators.  

Figure~\ref{fig:Rational-MDM} illustrates the ``best approximation'' of $g$ as a ratio of integers $g\approx n/d$, with 
denominator $d<100$, plus a correction provided by the most accurate value of $m$ available.  The maximum value 100 (for example), 
was chosen such that at least the first four digits of the $g$ correction were zero---meaning accuracy, roughly speaking, to 
better than one part in ten thousand.  In numerous cases, for which the bast value for $m$ was unclear, where fractional
abundance in nature is especially significant, or where the rational approximation to $G$ seems surprisingly accurate
(especially the deuteron) more than one row per isotope is given.     

Because the deuteron to proton mass ratio is known to such high accuracy, three determination
are given; the surprising $g-G=1$ at low accuracy is seen, at higher accuracy, to be simple coincidence.  
In many other cases two evaluations are given.  In the cases of lithium, boron, and silver, two stable isotopes are treated.
This provides warning that the purity of samples needs to be taken into account.  

Other than noting that all entries have 
been taken from NIST tables, the sources of entries and their errors are not given in this paper; in fact most values of $m$ 
simply been truncated to seven digits to improve the table's visual uniformity.  Other than canceling common factors in $n/d$
integers are truncated only with the use of integer factors such as $10^6$. 
In every case the three lines providing ``Relative Atomic Mass'', ``Isotopic Composition'', and ``Standard Atomic Weight''have been cut and pasted from NIST 
tables, \cite{NIST-background}\cite{NIST-isotope-abundance}---simple detective work can then identify most of the sources.  
As far as I know there are no reliable tables systematically providing accurate values of $G$.

\section{\bf F.\ Superimposed E\&M bending storage rings}

Commonly storage ring beams are unpolarized, but in the present paper, without loss of generality, it is assumed that 
both beams are, at least to some extent, polarized.

In a superimposed E\&M bending polarized storage ring the circulation direction of a so-called 
``master beam'' (of whatever charge $q_1$) is assumed to be CW or, equivalently,
$p_1>0$. A secondary beam charge $q_2$ is allowed to have either 
sign, and either CW or CCW circulation direction.

Ideally both beam polarizations would be frozen ``globally'' (meaning spin tune $Q_S$ is zero and the angle $\alpha$ 
between polarization vector and momentum is constant everywhere around the ring).  

(Somewhat weaker) ``doubly-frozen'' 
can (and will) be taken to mean that a ``primary beam'' locked to $Q_S=0$, circulates concurrently with a ``secondary'' beam that 
is ``pseudo-frozen'', meaning the spin tune is locked to an unambiguous, exact, rational fraction.  Only if this
rational fraction is zero, would the terminology ``doubly-magic'' be legitimate.

In general, though polarized, neither beam polarization is frozen. 

For brevity one can discuss just electrons (including positrons) 
protons($p$), deuterons($d$), tritons($t$), and helions($h$), or even just $p$ 
and $d$, based on the consideration that most of the apparatus, and all of the 
technology, needed for low energy nuclear physics investigation is presently available,
for example at the COSY laboratory in Juelich, Germany.

These pairings are expected to make spin dependent measurements of unprecedented precision possible.  For any 
arbitrary pairing of particle types ($(p,d),\ (p,e-),\ (\mu,e+),\ (d,h),\ (p,t)$, etc.) continua of such doubly-frozen 
pairings are guaranteed.

A design particle has mass $m>0$ and charge $qe$, with electron charge 
$e>0$ and $q=\pm 1$ (or some other integer). These values produce circular 
motion with radius $r_0>0$, and velocity ${\bf v}=v{\bf\hat z}$, where the motion
is CW (clockwise) for $v>0$ or CCW for $v<0$. With $0<\theta<2\pi$ being 
the cylindrical particle position coordinate around the ring, the angular 
velocity is $d\theta/dt=v/r_0$. 

(In MKS units) $qeE_0$ and $qe\beta c B_0$ are commensurate forces, 
with the magnetic force relatively weakened by a factor $\beta=v/c$ 
because the magnetic Lorentz force is $qe{\bf v}\times{\bf B}$. 
By convention $e$ is the absolute value of the electron charge; where it
appears explicitly, usually as a denominator factor, its purpose in 
MKS formulas is to allow energy factors to be evaluated as electron volts (eV)
in formulas for which the MKS unit of energy is the joule. 
Newton's formula for radius $r_0$ circular motion, expressed in terms of 
momentum and velocity (rather than just velocity, in order to be relativistically valid)
can be expressed using the total force per unit charge in the form
\begin{equation}
\beta\frac{pc}{e} = \Big(E_0 + c\beta B_0\Big)\,qr_0,
\label{eq:CounterCirc.1} 
\end{equation}
Coming from the cross-product Lorentz magnetic force, the factor $q\beta cB_0$
is negative for backward-traveling orbits because the $\beta$ factor 
is negative.

A ``master'' or primary beam travels in the ``forward'', CW direction. 
For the secondary beam, the $\beta$ factor can have either sign.
For $q=1$ and $E_0=0$, formula~(\ref{eq:CounterCirc.1}) reduces to a standard 
accelerator physics ``cB-rho=pc/e'' formula.  For $E_0\ne 0$ the formula 
incorporates the relative ``bending effectiveness'' of $E_0/\beta$ 
compared to $cB_0$.  

As well as fixing the bend radius $r_0$,
this fixes the magnitudes of the electric and magnetic bend field values 
$E_0$ and $B_0$. To begin, we assume the parameters of a frozen spin ``master'',
charge $qe$, particle beam have already been established, including the signs
of the electric and magnetic fields consistent with $\beta_1>0$ and $p_1>0$.  

In general, beams can be traveling either CW or CCW.  For a CCW beam both $p$ and 
$\beta$ have reversed signs, with the effect that the electric force is unchanged, but the 
magnetic force is reversed. The $\beta$ velocity factor can be expressed as
\begin{equation}
\beta = \frac{pc/e}{\sqrt{(pc/e)^2 + (mc^2/e)^2}}.
\label{eq:CounterCirc.2} 
\end{equation}
Eq.~(\ref{eq:CounterCirc.1}) becomes
\begin{equation}
\frac{pc}{e} = \Big(\frac{E_0\sqrt{(pc/e)^2 + (mc^2/e)^2}}{pc/e} + cB_0\Big)qr_0.
\label{eq:CounterCirc.3} 
\end{equation}
Cross-multiplying the denominator factor produces
\begin{equation}
\Big(\frac{pc}{e}\Big)^2 = qE_0r_0\sqrt{(pc/e)^2 + (mc^2/e)^2} + qcB_0r_0\frac{pc}{e}.
\label{eq:CounterCirc.4} 
\end{equation}
To simplify the formulas we make some replacements and alterations, 
starting with
\begin{equation}
pc/e \rightarrow p,
\quad\hbox{and}\quad
m c^2/e\rightarrow m, 
\label{eq:Alterations.1}
\end{equation}
The mass parameter $m$ will be replaced later
by, $m_p$, $m_d$, $m_{\rm tritium}$, $m_e$, etc., as appropriate
for the particular particle types, proton, deuteron, triton, electron, helion, etc..
These changes amount to setting $c=1$ and switching the energy units from joules to electron volts. 
The number of ring and beam parameters can be reduced by forming the combinations 
\begin{equation}
\mathcal{E} = qE_0r_0,
\quad\hbox{and}\quad
\mathcal{B} = qcB_0r_0.
\label{eq:Alterations.2}
\end{equation}
After these changes, the closed orbit equation has become  
\begin{equation}
p_m^4 -2\mathcal{B}p_m^3 + (\mathcal{B}^2-\mathcal{E}^2)p_m^2 - \mathcal{E}^2m^2=0,
\label{eq:AbbrevFieldStrengths.3}
\end{equation}
an equation to be solved for either CW and CCW orbits.  The absence of a term linear in $p_m$
suggests the restoration, using Eq.~(\ref{eq:Alterations.2}), of the explicit form of 
$\mathcal{B}$ in the coefficient of the $p_m^3$ term to produce;
\begin{equation}
\boxed{
p_m^4 - 2cB_0(qr_0)p_m^3 + (\mathcal{B}^2-\mathcal{E}^2)p_m^2 - \mathcal{E}^2m^2 = 0,
\label{eq:AbbrevFieldStrengths.3-rev}}
\end{equation}
Here $c$ has been left explicit when attached to $B_0$ (even though its value is 1) but not to $m^2$ (whose
value is customarily expressed in GeV${}^2$ units). This is to serve as a reminder that, to be commensurate 
with $E_0$ in MKS units, the magnetic field needs to be multiplied by $c$.

Every (correct) table of kinematic parameters in this paper is based on the four roots of this quartic equation,
or, more precisely, \emph{on at least two of its roots being correctly identified as real}.
\footnote{There is a symmetry of Eq.~\ref{eq:AbbrevFieldStrengths.3-rev} which is useful in its application
to storage ring investigation of $\beta$-decay processes.  What makes these weak interaction processes special is
that, because of their small mass, the electrons (of either sign) are always highly relativistic.  In
magnetic storage rings their speeds are always essentially equal to the speed of light.  For this reason it may 
become appropriate to employ a predominantly magnetic e\&M storage ring for the investigation of higher (but still low)
energy $\beta$-decay processes.

To interchange electric an magnetic bending, while changing the sign of the charge, we set $\mathcal{E'}=-qcB_0r_0=-\mathcal{B}$ and $\mathcal{B'}=-\mathcal{E}$,
which amounts to interchanging the fractional bending fractions $\eta_E$ and $\eta_M$.
Eq.~\ref{eq:AbbrevFieldStrengths.3-rev} has then become 
$$p_m^4 + 2E_0(qr_0) p_m^3 + (\mathcal{E'}^2-\mathcal{B'}^2)p_m^2 - \mathcal{B'}^2m^2 = 0 $$
which, except for signs and altered coefficient values, has the same form as Eq.~\ref{eq:AbbrevFieldStrengths.3-rev}\,.
Note, for predominantly magnetic bending, that the magnitude of $E'$ is now perturbatively small, as compared 
to the magnitude of $B'$. Like Eq.~\ref{eq:AbbrevFieldStrengths.3-rev}\,, this equation can be solved in closed form
using some variant of the method discovered by the Italian mathematician Lodovico Ferrari in around 1550, for
example using a MAPLE or MATHEMATICA program, which probably uses the same method, to solve the equation
numerically, with arbitrarily high precision, or to express the LHS as a product of four linear or
quadratic factors, from which real roots can be determined. 
}
Any arbitrarily chosen numerical factor multiplying the product factor $(qr_0)$ can be compensated by
scaling the powers of independent variable $p_m$ by corresponding powers without influencing any implications
of Eq.~(\ref{eq:AbbrevFieldStrengths.3-rev}).  This and other properties can be confirmed by pure reasoning, 
based on the structure of the equation, or by explicit partially-numerical factorization of the left hand side.  

However, noting from  Eq.~\ref{eq:Alterations.2}\, that the structure of the equation has $E_0$ appearing only 
quadratically but $cB_0$ appearing both linearly and quadratically, appreciably complicates the qualitative
interpretation of the equation.

These considerations include some, but not all of the sign ambiguities introduced by the quadratic 
substitutions used in the derivation of Eq.~(\ref{eq:AbbrevFieldStrengths.3-rev}). 
The electric field can still be reversed without altering the set of solutions of the equation. 
Note that this change cannot be compensated by switching the sign of $q$, which also reverses the 
magnetic bending.  

The most significant experimental implication of this is that it is not only positrons, but also electrons,
that can have orbits identical to (necessarily positive) baryons, possibly having the same or opposite 
orientations.

We can contemplate allowing the signs of $E_0$ or $B_0$ to be reversed for experimental purposes, such as interchanging CW and
CCW beams, or replacing positrons by electrons, but only if this can be done 
with sufficiently high reproducability.  \emph{Demonstrating this capability (by promising spin tune measurability
with frequency domain precision) is an important ingredient of this paper.}

\end{document}